\documentclass[aps,prd,preprintnumbers,superscriptaddress,nofootinbib]{revtex4}
\usepackage[dvips]{graphicx}
\usepackage{bm,latexsym,amsmath,amssymb,amsfonts,mathrsfs}
\usepackage{color}
\input{colordvi.tex}

\begin{document}

\title{Effective field theory approach to quasi-single field inflation\\[2mm]
and effects of heavy fields}

\author{Toshifumi~Noumi}
\email[Email: ]{tnoumi"at"hep1.c.u-tokyo.ac.jp}
\affiliation{Institute of Physics, University of Tokyo, Komaba,
Meguro-ku, Tokyo 153-8902, Japan}

\author{Masahide~Yamaguchi}
\email[Email: ]{gucci"at"phys.titech.ac.jp}
\affiliation{Department of Physics, Tokyo Institute of Technology, Tokyo
152-8551, Japan}

\author{Daisuke~Yokoyama}
\email[Email: ]{d.yokoyama"at"th.phys.titech.ac.jp}
\affiliation{Department of Physics, Tokyo Institute of Technology, Tokyo
152-8551, Japan}

\begin{abstract}
We apply the effective field theory approach to quasi-single field inflation,
which contains an additional scalar field with Hubble scale
mass other than inflaton.
Based on the time-dependent spatial
diffeomorphism, which is not broken by the time-dependent background evolution,
the most generic action of quasi-single field inflation is constructed
up to third order fluctuations.
Using the obtained action,
the effects of the additional massive scalar field on
the primordial curvature perturbations are discussed.
In particular,
we calculate the power spectrum and
discuss the momentum-dependence of
three point functions in the squeezed limit
for general settings of quasi-single field inflation.
Our framework can be also applied to inflation models with heavy particles. We make a qualitative
discussion on the effects of heavy particles during inflation and those
of sudden turning trajectory in our framework.
\end{abstract}

\pacs{98.80.Cq}

\preprint{UT-Komaba/12-9}
\preprint{TIT/HEP-625}
\maketitle

\section{Introduction}

Inflation gives the most natural solution to the horizon and the
flatness problems of the big-bang theory as well as generates the
primordial perturbations \cite{oriinf,r2}, whose properties well coincide
with the recent observations of cosmic microwave background anisotropies
like the Wilkinson Microwave Anisotropy Probe
\cite{Komatsu:2010fb}.
Models of inflation can be classified into two
categories with respect to relevant degrees of freedom during inflation:
single-field models and multiple field models.
Recently,
the most general single field inflation model with the second order
equations of motion~\cite{Kobayashi:2011nu} has been
invented in the
context of Horndeski \cite{Horndeski,Charmousis:2011bf} and Galileon theories \cite{Deffayet:2011gz,Kobayashi:2010cm}. Then, the bispectra of primordial
curvature \cite{Mizuno:2010ag,Gao:2011qe} and tensor perturbations
\cite{Gao:2011vs} 
are obtained as well as their powerspectra \cite{Kobayashi:2011nu}.

\medskip
Effective scalar fields are ubiquitous in the
extensions of the standard model of particle physics such as
supergravity and superstring. Then, it is well motivated to
consider multiple field models of inflation. Such multiple field models
are roughly divided into three classes: (i) only one field is light,
while the other fields are very heavy
compared to the Hubble
scale during inflation. Generically, this class virtually falls into the single
field category~\cite{Yamaguchi:2005qm}. However,
it is recently discussed that 
heavy modes can affect the dynamics of light mode
in some particular cases~\cite{Cremonini:2010ua,Achucarro:2010da,Shiu:2011qw,Cespedes:2012hu,Achucarro:2012sm,Pi:2012gf,Gao:2012uq,Saito:2012pd,Gwyn:2012mw}. (ii)
there are multiple light fields, in which isocurvature perturbations are
generated in addition to curvature perturbations. (iii) only one field
is light, while the masses of other fields are comparable to the Hubble
scale during inflation. This class is called quasi-single field
inflation model~\cite{Chen:2009we,Chen:2012ge}.

\medskip
In supergravity,
inflation necessarily involves supersymmetry (SUSY) breaking, whose
effects are transmitted into other scalar fields as Hubble induced
masses \cite{Copeland:1994vg,Baumann:2011nk,McAllister:2012am}.\footnote{The methods to
keep inflaton flat against such SUSY breaking effects are reviewed in
Refs. \cite{Mazumdar:2010sa}.}
Therefore,
quasi-single field inflation is naturally realized in supergravity
and it is well motivated by the model building based on supergravity or inspired by superstring.
Furthermore,
it is known that massive isocurvature modes which
couple to the inflaton and have Hubble scale masses
can give significant impacts on primordial curvature
perturbations.
In the original paper~\cite{Chen:2009we} by Chen and Wang,
it was shown that, for example,
scalar three point functions
take the intermediate shapes between
local and equilateral types.
Based on these backgrounds,
we would like to discuss quasi-single field inflation model in general settings.

\medskip
Recently, effective field theory approach to inflation has been invented in~\cite{Cheung:2007st,Senatore:2010wk,Weinberg:2008hq}, which is based on the symmetry breaking during inflation:
Time diffeomorphism is broken by the time-dependent background evolution during inflation.
Then, based on the unbroken time-dependent spatial diffeomorphism,
the effective action for inflation can be constructed systematically
in unitary gauge,
where inflaton is eaten by graviton and there are no perturbations of inflaton.
By use of
the St\"{u}ckelberg trick, the curvature perturbation can be associated
with the Goldstone boson~$\pi$, which non-linearly realizes the time
diffeomorphism.
The key observation is that the Goldstone $\pi$
could decouple from the metric fluctuations in some parameter region which we call the decoupling regime.
In the decoupling regime,
the dynamics of Goldstone $\pi$ is described by a simplified action,
which does not contain metric perturbations.
As a consequence, calculations of scalar perturbations are also simplified
and seeds of non-Gaussianities become clear.

\medskip
In this paper, we apply this effective field theory approach to
quasi-single field inflation.
First, in unitary gauge, we write down the most general action
invariant under the
time-dependent spatial diffeomorphism
and constructed from graviton and the massive isocurvature mode.
The obtained action is expanded systematically in fluctuations and derivatives around the FRW background.
By the St\"{u}ckelberg trick,
we introduce
the action for Goldstone boson
and carefully discuss its decoupling regime.
Using the action in the decoupling regime,
the power spectrum is calculated in the general setting of quasi-single field inflation. The momentum dependence of scalar three-point function is also discussed in the general setting.
Our framework can be also applied to inflation models with heavy particles. As an application, we
make a qualitative
discussion on the effects of heavy particles during inflation and that
of sudden turning trajectory.

\medskip
The organization of this paper is as follows. In the next section, we
briefly review the effective field theory approach and quasi-single
field inflation. In Sec.\,\ref{section_most_general},
the most general action for quasi-single field inflation is constructed
via effective field theory approach.
The decoupling regime of the obtained action is also discussed.
In Sec.\,\ref{section_PS}, the power spectrum is calculated first in the general setting of quasi-single field inflation with constant mixing couplings.
Then, the effects of sudden turning trajectory on the power spectrum
is qualitatively discussed.
In Sec.\,\ref{section_squeezed},
the momentum dependence of scalar three-point
functions are discussed. Final section is devoted to summary and discussions.
Technical details of the calculation of the power spectrum
are summarized in Appendices.

\section{Effective field theory approach and quasi-single field inflation}
\label{section_review}

In this section we briefly review the effective field theory approach to
inflation developed in~\cite{Cheung:2007st} and the quasi-single field
inflation model proposed in~\cite{Chen:2009we}.

\subsection{Effective field theory approach to inflation}

Inflation is an accelerated cosmic expansion with an approximately
constant Hubble parameter:
\begin{equation}
ds^2=-dt^2+a^2(t)dx^idx^i
\quad
{\rm with}
\quad
H(t)=\frac{\dot{a}}{a}\,,\quad\epsilon=-\frac{\dot{H}}{H^2}\ll1\,.
\end{equation}
It is characterized by
the spontaneous breaking of the time-diffeomorphism:
\begin{equation}
\langle\phi(t,x)\rangle=\phi_0(t)\,,
\end{equation}
where $\phi(t,x)$ is a certain scalar operator. Here we chose the frame in
which the vacuum expectation value of $\phi(t,x)$ is spatially uniform.
Assuming the degrees of freedom relevant to the cosmological
perturbation and invariance under the time-dependent spatial
diffeomorphism, $x^i\to x^{\prime\, i}=x^i+\xi^i(t,x^j)$, which is not
broken by the condensation~$\phi_0(t)$, we can construct the effective
action for inflation.

\medskip
In the simplest case, relevant degrees of freedom are three
physical modes of graviton: two transverse modes and one longitudinal
mode related to the inflaton. 
As discussed in~\cite{Cheung:2007st}, any action of graviton invariant
under the time-dependent spatial diffeomorphism can be written in terms
of the Riemann tensor $R_{\mu\nu\rho\sigma}$, the time-like component of
the metric $g^{00}$, the extrinsic curvature $K_{\mu\nu}$ on
constant-$t$ surfaces, the covariant derivative $\nabla_\mu$, and the
time coordinate~$t$:
\begin{equation}
\label{general_single_1}
S=\int d^4x\sqrt{-g}\,F(R_{\mu\nu\rho\sigma},g^{00},K_{\mu\nu},\nabla_\mu,t)\,,
\end{equation}
where all the free indices inside the function $F$
must be upper $0$'s.
Note that $g^{00}$ should be treated as a scalar
when considering its covariant derivative,
and we can use $\partial_\mu g^{00}$ for example.
The explicit form of the extrinsic curvature $K_{\mu\nu}$ is
\begin{align}
K_{\mu\nu}=h_\mu^\sigma\nabla_\sigma n_\nu
=-\frac{\delta^0_\mu\partial_\nu g^{00}+\delta^0_\nu\partial_\mu g^{00}}{2(-g^{00})^{3/2}}
-\frac{\delta^0_\mu\delta^0_\nu g^{0\sigma}\partial_\sigma g^{00}}{2(-g^{00})^{5/2}}
+\frac{g^{0\rho}(\partial_\mu g_{\rho\nu}+\partial_\nu g_{\rho\mu}-\partial_\rho g_{\mu\nu})}{2(-g^{00})^{1/2}}\,,
\end{align}
where $\displaystyle n_\mu=-\frac{\delta^0_\mu}{\sqrt{-g^{00}}}$ is a unit vector
perpendicular to constant $t$ surfaces
and $h_{\mu\nu}=g_{\mu\nu}+n_\mu n_\nu$
is the induced spatial metric on constant $t$ surfaces.
In~\cite{Cheung:2007st},
it was shown that the action~(\ref{general_single_1})
can be expanded around a given FRW background
as
\begin{equation}
\label{S_grav_unitary}
S=\int d^4x \sqrt{-g}\left[
\frac{1}{2}M_{\rm Pl}^2R
+M_{\rm Pl}^2\dot{H}g^{00}
-M_{\rm Pl}^2(3H^2+\dot{H})
+F^{(2)}(\delta g^{00},\delta K_{\mu\nu},\delta R_{\mu\nu\rho\sigma};\delta^0_\mu,g_{\mu\nu},g^{\mu\nu},\nabla_\mu,t)\right]\,,
\end{equation}
where the function $F^{(2)}$ starts with quadratic terms of the arguments $\delta g^{00}$, $\delta K_{\mu\nu}$, and $\delta R_{\mu\nu\rho\sigma}$
and all the free indices must be upper $0$'s.
The arguments $\delta g^{00}$, $\delta K_{\mu\nu}$, and $\delta R_{\mu\nu\rho\sigma}$
are defined by\footnote{Here and in what follows, we concentrate on the spatially flat FRW background.}
\begin{align}
\delta g^{00}&=g^{00}+1\,,\\
\delta K_{\mu\nu}&=K_{\mu\nu}-Hh_{\mu\nu}\,,\\
\delta R_{\mu\nu\rho\sigma}&=R_{\mu\nu\rho\sigma}-2H^2h_{\mu[\rho}h_{\sigma]\nu}
+(\dot{H}+H^2)(h_{\mu\rho}\delta^0_\nu\delta^0_\sigma+(3\text{ permutations}))\,.
\end{align}
They are covariant under time-dependent spatial-diffeomorphism
and vanish on the FRW background.
Notice that
the action of single field inflation
in the uniform inflaton gauge
can be reproduced
by gauge-fixing
the time-dependent spatial diffeomorphism
as
\begin{equation}
g_{ij}(x)=a^2(t)e^{2\zeta(x)}(e^{\gamma(x)})_{ij}
\quad
{\rm with}
\quad
\gamma_{ii}=\partial_i\gamma_{ij}=0\,,
\end{equation}
where $\zeta(x)$ is the scalar perturbation.

\medskip
For the calculation of correlation functions of the scalar perturbation
$\zeta$, it is convenient to introduce the action for the Goldstone
boson $\pi$ by the St\"uckelberg method.  We perform the following
time-diffeomorphism on the action~(\ref{S_grav_unitary}) in the unitary
gauge:
\begin{align}
\label{unitary_to_pi}
t\to \tilde{t}\,,\quad
x^i\to\tilde{x}^i
\quad{\rm with}
\quad
\tilde{t} +\tilde{\pi}(\tilde{t},\tilde{x})=t\,,\quad
\tilde{x}^{i}=x^i\,.
\end{align}
In general,
the transformation~(\ref{unitary_to_pi}) is realized by the following replacement:
\begin{align}
\nonumber
&\delta^0_\mu\to \delta^0_\mu+\partial_\mu\pi\,,\quad
f(t)\to f(t+\pi)\,,
\quad
\int d^4x \sqrt{-g}\to\int d^4x \sqrt{-g}\,,\\
\label{replacement}
&\nabla_\mu\to\nabla_\mu\,,
\quad g_{\mu\nu}\to g_{\mu\nu}\,,
\quad g^{\mu\nu}\to g^{\mu\nu}\,,
\quad R_{\mu\nu\rho\sigma}\to R_{\mu\nu\rho\sigma}\,,
\end{align}
where we dropped the tilde for simplicity
and $g^{00}$ transforms, for example, as
\begin{align}
g^{00}\to g^{00}+2g^{0\mu}\partial_\mu\pi
+g^{\mu\nu}\partial_\mu\pi\partial_\nu\pi\,.
\end{align}
The transformation rules of $K_{\mu\nu}$ and $h_{\mu\nu}$ also follow
from~(\ref{replacement}) straightforwardly. These procedures lead to the
following action for the Goldstone boson $\pi$:
\begin{align}
\label{action_pi}
S=\int d^4x \sqrt{-g}\left[
\frac{1}{2}M_{\rm Pl}^2R
+M_{\rm Pl}^2\dot{H}(t+\pi)\left(g^{00}+2g^{0\mu}\partial_\mu\pi+g^{\mu\nu}\partial_\mu\pi\partial_\nu\pi\right)-M_{\rm Pl}^2\left(3H^2(t+\pi)+\dot{H}(t+\pi)\right)
+\ldots\right]\,,
\end{align}
where the dots stand for the terms corresponding to $F^{(2)}$.  The
obtained action enjoys the time-diffeomorphism by assigning to $\pi$ the
non-linear transformation rule\footnote{ Here it should be noted that
while the action is invariant under the time-diffeomorphism, it does not
have the shift symmetry $\pi\to\pi+{\rm constant}$ because of the
time-dependent free parameters such as $H(t+\pi)$ or $\dot{H}(t+\pi)$.
Expanding the parameters in $\pi$, we find, for example, that $\pi$ has
a mass term $\sim M_{\rm Pl}^2\dot{H}^2\pi^2$, which is sub-leading in
the slow-roll approximation.  }
\begin{align}
\pi(x)\to\tilde{\pi}(\tilde{x})=\pi(x)-\xi^0(x)
\quad
{\rm with}
\quad
t\to\tilde{t}=t+\xi^0(x)\,,\quad
x^i\to\tilde{x}^i=x^i\,,
\end{align}
and the action in the unitary gauge can be reproduced by gauge-fixing
the time-diffeomorphism as $\pi(x)=0$.

\medskip 
It is important to recognize that, in the action (\ref{action_pi}),
terms with graviton fluctuations have less derivatives than those
without graviton.  Because of this property, it is expected that the
mixing of the Goldstone boson and graviton becomes irrelevant to the
dynamics of the Goldstone boson at a sufficiently high energy scale.
For example, let us consider the following simplest case:
\begin{align}
S=\int d^4x \sqrt{-g}\left[
\frac{1}{2}M_{\rm Pl}^2R
+M_{\rm Pl}^2\dot{H}(t+\pi)\left(g^{00}+2g^{0\mu}\partial_\mu\pi+g^{\mu\nu}\partial_\mu\pi\partial_\nu\pi\right)-M_{\rm Pl}^2\left(3H^2(t+\pi)+\dot{H}(t+\pi)\right)
\right]\,.
\end{align}
In the canonical normalization ($\pi_c\sim M_{\rm Pl}(-\dot{H})^{1/2}\pi$, $\delta g_c^{\mu\nu}\sim M_{\rm Pl} \delta g^{\mu\nu}$),
the mixing term $M_{\rm Pl}^2\,\dot{H}\,\delta g^{0\mu}\,\partial_\mu \pi$ can be written as
\begin{align}
M_{\rm Pl}^2\,\dot{H}\,\delta g^{0\mu}\,\partial_\mu \pi
\sim
(-\dot{H})^{1/2}\,\delta g_c^{0\mu}\,\partial_\mu \pi_c
\sim\epsilon^{1/2}H\,\delta g_c^{0\mu}\,\partial_\mu \pi_c\,,
\end{align}
and it can be neglected in the energy scale $E\gg \epsilon^{1/2}H$.
In other words,
when the slow-roll parameter $\epsilon$ is small,
the mixing becomes irrelevant
inside the horizon.
The dynamics of $\pi$ inside the horizon
are then determined by the following action
in the decoupling limit:
\begin{align}
S\sim\int d^4x \,a^3\Big[
M_{\rm Pl}^2\dot{H}(t+\pi)\Big(-1-2\dot{\pi}-\dot{\pi}^2+\frac{(\partial_i\pi)^2}{a^2}\Big)-M_{\rm Pl}^2(3H^2(t+\pi)+\dot{H}(t+\pi))\Big]\,,
\end{align}
where the metric reduces to the FRW background. More generally,
the mixing of the Goldstone boson and graviton becomes irrelevant inside
the horizon when the free parameters of the action are in some regime
(decoupling regime) as in the slow-roll regime for the above simplest case.
In the decoupling regime, the calculation of scalar correlation
functions also becomes tractable. Taking the spatially flat
gauge\footnote{Strictly speaking, the name of this gauge may be
inadequate because there are still tensor fluctuations and hence the
spatial hypersurface is not exactly flat.}
\begin{equation}
\label{spatially_flat_gauge}
g_{ij}(x)=a^2(t)(e^{\gamma(x)})_{ij}
\quad
{\rm with}
\quad
\gamma_{ii}=\partial_i\gamma_{ij}=0\,,
\end{equation}
the scalar perturbation $\zeta(x)$
is given by $\zeta(x)=-H(t)\pi(x)$ at the linear order,
and the calculation of correlation functions of $\zeta$
reduces to those of $\pi$,
which can be obtained using the simplified action in the decoupling limit.
This kind of simplification in the decoupling regime
is one of the advantages to use the effective field theory approach.
In the next section
we extend this approach to quasi-single field inflation.

\subsection{Quasi-single field inflation}
The original model~\cite{Chen:2009we} of quasi-single field inflation
is described by the following matter action:
\begin{align}
S_{\rm matter}=\int d^4x\sqrt{-g}\Big[-\frac{1}{2}(\tilde{R}+\chi)^2g^{\mu\nu}
\partial_\mu\theta\partial_\nu\theta
-\frac{1}{2}g^{\mu\nu}\partial_\mu\chi\partial_\nu\chi
-V_{\rm sr}(\theta)-V(\chi)
\Big]\,,
\end{align}
where $\theta$ and $\chi$ are the tangential and radial directions of a
circle with radius $\tilde{R}$ and the potential $V_{\rm sr}(\theta)$
along the tangential direction is of slow-roll type. The
homogeneous backgrounds and their equations of motion are given by
\begin{align}
\nonumber
&\theta=\theta_0(t)\,,\quad\chi=\chi_0 \,(\text{constant})\,,\\
\label{eom_grav_QSI}
&3M_{\rm Pl}^2H^2=\frac{1}{2}R^2\dot{\theta}_0^2+V(\chi_0)+V_{\rm sr}(\theta_0)\,,
\quad
-2M_{\rm Pl}^2\dot{H}=R^2\dot{\theta}^2_0\,,
\quad
V^\prime(\chi_0)=R\dot{\theta}_0^2\,,
\quad
R^2\ddot{\theta}_0+3R^2H\dot{\theta}_0+V^\prime_{\rm sr}(\theta_0)=0\,,
\end{align}
where $R=\tilde{R}+\chi_0$.  Expanding the action around the homogeneous
background, it yields the following second order action of the fluctuations
$\delta\theta=\theta-\theta_0$ and $\sigma=\chi-\chi_0$:
\begin{align}
S_{\rm matter}^{(2)}=\int d^4x\sqrt{-g}\Big[-\frac{1}{2}R^2g^{\mu\nu}
\partial_\mu\delta\theta\,\partial_\nu\delta\theta
-\frac{1}{2}g^{\mu\nu}\partial_\mu\sigma\,\partial_\nu\sigma
-R\dot{\theta}_0\,\sigma\partial^0\delta\theta -\frac{1}{2}(V^{\prime\prime}(\chi_0)-\dot{\theta}_0^2)\sigma^2
\Big]\,.
\end{align}
The mixing coupling $\sigma\delta\dot{\theta}$ converts the $\sigma^3$
coupling, for example, into three point functions of $\delta \theta$,
and hence this model can potentially give a large non-Gaussianities.
Furthermore, it is known that the squeezed limit of scalar three point
functions is sensitive to the mass of $\sigma$:
\begin{align}
\lim_{k_3/k_1=k_3/k_2=\kappa\to0}\langle\zeta_{{\bf k}_1}\zeta_{{\bf k}_2}\zeta_{{\bf k}_3}\rangle\propto \kappa^{-3/2-\nu}k_1^{-6}\,,
\end{align}
where $\displaystyle\nu=\sqrt{\frac{9}{4}-\frac{m_\sigma^2}{H^2}}$ and
$m_\sigma^2=V^{\prime\prime}(\chi_0)-R\dot{\theta}_0^2<\frac{9}{4}H$.
As this simple model implies, massive scalar fields of Hubble scale
mass can cause a non-trivial behavior of non-Gaussianities.  In the
following sections we discuss more general setting for quasi-single
field inflation using the effective field theory approach.

\section{Most generic action of quasi-single field inflation}
\label{section_most_general}

In this section, we construct the most generic action of quasi-single
field inflation using the effective field theory approach. After
constructing the action in the unitary gauge first, we derive the
action for the Goldstone boson $\pi$ and discuss its decoupling regime.
Relations between our approach and models in the literatures are
also discussed.

\subsection{Action in the unitary gauge}
\label{sub_quasi_uni}

In the unitary gauge, the relevant degrees of freedom in quasi-single
field inflation are three physical modes of graviton and
an additional scalar field $\sigma$. The typical mass of $\sigma$ is
supposed to be of the order of the Hubble scale during the inflationary
era.  In this subsection, we construct the most generic action invariant
under the time-dependent spatial diffeomorphism from graviton
and the scalar field $\sigma$ up to the third order fluctuations.  Here
it should be noticed that the action constructed in this section can be
applied to any two field models because no conditions on $\sigma$ are
imposed.\footnote{For example, we do not require the shift
symmetry of $\sigma$, $\sigma\to\sigma+{\rm constant}$, which is assumed
in multi-field inflation~\cite{Senatore:2010wk}.  The mass of $\sigma$
is not necessarily of order Hubble scale so that the action constructed
in this section can be applied not only for quasi-single field inflation
but also inflation models with an additional heavy scalar.}

\medskip
Extending the procedures in~\cite{Cheung:2007st} to our case, the most
general action invariant under the time-dependent spatial diffeomorphism
is given by
\begin{equation}
S=\int d^4x\sqrt{-g}\,F(R_{\mu\nu\rho\sigma},g^{00},K_{\mu\nu},\nabla_\mu,t,\sigma)\,,
\end{equation}
and it is expanded around the given FRW background as
\begin{equation}
S=\int d^4x \sqrt{-g}\left[
\frac{1}{2}M_{\rm Pl}^2R
+M_{\rm Pl}^2\dot{H}g^{00}
-M_{\rm Pl}^2(3H^2+\dot{H})
+F^{(2)}(\delta g^{00},\sigma,\delta K_{\mu\nu},\delta R_{\mu\nu\rho\sigma};\delta^0_\mu,g_{\mu\nu},g^{\mu\nu},\nabla_\mu,t)\right]\,,
\end{equation}
where all the free indices inside the functions $F$ and $F^{(2)}$ must
be again upper $0$'s and $F^{(2)}$ starts with quadratic terms of the arguments
$\delta g^{00}$, $\sigma$, $\delta K_{\mu\nu}$, and $\delta
R_{\mu\nu\rho\sigma}$.\footnote{ \label{footnote_by_parts} In
general, sums of terms linear in the fluctuations can be practically
second order. For example, let us consider the term $\int
d^4x\sqrt{-g}\left[f_1(t)\sigma+f_2(t)\partial^0\sigma\right]$.
Although this kind of action seems to be first order apparently,
it turns out to be second order
after taking into account the equation of motion for $\sigma$:
$f_1(t)+\dot{f}_2(t)+3Hf_2(t)=0$.
Then, the function $F^{(2)}$
seems to contain such a combination of linear order terms.
However,
using the relation
\begin{align}
\nonumber
\int d^4x\sqrt{-g}f(t)\partial^0(\ldots)&=
-\int d^4x\sqrt{-g}f(t)\sqrt{-g^{00}}n^\mu\partial_\mu(\ldots)
=\int d^4x\sqrt{-g}\Big(-g^{00}\dot{f}(t)-\frac{1}{2}f(t)\partial^0\ln(-g^{00})+f(t)\sqrt{-g^{00}}K_\mu^\mu\Big)(\ldots)\\
\label{integ_by_parts}
&=\int d^4x\sqrt{-g}\Big(\dot{f}(t)+3Hf(t)+\mathcal{O}(\delta g^{00},\delta K_\mu^\mu)\Big)(\ldots)\,,
\end{align}
we can rewrite it into the second and higher order terms
in $\sigma$, $\delta g^{00}$, and $\delta K_\mu^\mu$.
Similar discussions hold for more general cases
and we conclude that $F^{(2)}$ starts with quadratic terms of $\delta g^{00}$, $\sigma$, $\delta K_{\mu\nu}$, and $\delta
R_{\mu\nu\rho\sigma}$.
See also appendix B in~\cite{Cheung:2007st}. 
}
Then, let us write down
possible terms in the action up to the third order fluctuations.
Schematically,
we write the action in the following way:
\begin{align}
S=S_{\rm grav}+S_{\sigma}+S_{\rm mix}\,,
\end{align}
where the first term $S_{\rm grav}$ in the right hand side denotes terms
constructed from $\delta g^{00}$, $\delta K_{\mu\nu}$, and $\delta
R_{\mu\nu\rho\sigma}$, the second term $S_{\sigma}$ denotes those only
from $\sigma$, and the last term $S_{\rm mix}$ denotes those mixing the
graviton fluctuations and $\sigma$.  As discussed
in~\cite{Cheung:2007st}, the first term $S_{\rm grav}$ can be expanded
as
\begin{align}
\label{S_grav}
S_{\rm grav}=\int d^4x \sqrt{-g}\Bigg[
\frac{1}{2}M_{\rm Pl}^2R
+M_{\rm Pl}^2\dot{H}(t)g^{00}-M_{\rm Pl}^2\left(3H^2(t)+\dot{H}(t)\right)
+\frac{M_2^4(t)}{2!}(\delta g^{00})^2
+\frac{M_3^4(t)}{3!}(\delta g^{00})^3+\ldots\Bigg]\,,
\end{align}
where the dots stand for terms of higher order in the fluctuations or
with more derivatives.  When we rewrite the action in the unitary gauge
in terms of the Goldstone boson~$\pi$, the terms displayed
in~(\ref{S_grav}) are described by $\pi$ and its first order
derivatives. In this paper, we consider the action up to the same order
in derivatives of $\pi$ and $\sigma$.

\medskip
Let us first construct the second order action.
The second order action $S_{\sigma}^{(2)}$ containing
$\sigma$ and its first order derivative can be
written generally as
\begin{equation}
\label{S_sigma^2}
S_{\sigma}^{(2)}=\int d^4x\sqrt{-g}\Big[-\frac{\alpha_1(t)}{2}g^{\mu\nu}\partial_\mu\sigma\partial_\nu\sigma+\frac{\alpha_2(t)}{2}(\partial^0\sigma)^2-\frac{\alpha_3(t)}{2}\sigma^2+\alpha_4(t)\sigma\partial^0\sigma\Big]\,,
\end{equation}
where we note that terms such as $\sigma(\partial^0)^2\sigma$
can be absorbed into other terms up to higher order fluctuations
by integrating by parts.
As discussed in~\cite{Senatore:2010wk},
the second term leads to
a non-trivial sound speed $c_\sigma$ of $\sigma$
given by $c_\sigma^2=\alpha_1/(\alpha_1+\alpha_2)$.
The second order mixing $S_{\rm mix}^{(2)}$
is generally given by
\begin{equation}
\label{second_K}
S_{\rm mix}^{(2)}=\int d^4x\sqrt{-g}\Big[\overline{\beta}_1(t)\delta g^{00}\sigma+\overline{\beta}_2(t)\delta g^{00}\partial^0\sigma+\beta_3(t)\delta K_\mu^\mu\sigma
\Big]\,.
\end{equation}
It is convenient to note the relation
\begin{align}
\label{K_to_partial}
\int d^4x\sqrt{-g}f(t)\delta K_\mu^\mu(\ldots)&=\int d^4x\sqrt{-g}
\Big[f(t)\partial^0(\ldots)
-\Big(\dot{f}(t)+3Hf(t)\Big)(\ldots)
+\frac{\dot{f}(t)}{2}\delta g^{00}(\ldots)+\frac{f(t)}{2}\delta g^{00}\partial^0(\ldots)+\ldots\Big]\,,
\end{align}
which can be obtained using~(\ref{integ_by_parts}) in footnote~\ref{footnote_by_parts} twice.
Here the last dots stand for higher order terms in the fluctuations,
which can be written using $\delta g^{00}$, 
$\partial^0\delta g^{00}$, and $\delta K^\mu_\mu$.
Using this relation~(\ref{K_to_partial}),
$S_{\rm mix}^{(2)}$ in (\ref{second_K}) is rewritten as
\begin{equation}
\label{S_mix^2}
S_{\rm mix}^{(2)}=\int d^4x\sqrt{-g}\Big[\beta_1(t)\delta g^{00}\sigma+\beta_2(t)\delta g^{00}\partial^0\sigma+\beta_3(t)\partial^0\sigma
-(\dot{\beta}_3(t)+3H\beta_3(t))\sigma
\Big]\,,
\end{equation}
where $\beta_1=\overline{\beta}_1+\dot{\beta}_3/2$,
$\beta_2=\overline{\beta}_2+\beta_3/2$, and we dropped higher
order terms constructed from $\sigma$, $\delta g^{00}$,
$\partial^0\delta g^{00}$, and~$\delta K^\mu_\mu$.  In the
following, we employ Eq.~(\ref{S_mix^2}) as a definition of the second
order mixing action $S_{\rm mix}^{(2)}$.  

\medskip
The third order action $S_{\sigma}^{(3)}$ of $\sigma$ is generally given
by
\begin{align}
\label{S_sigma^3}
S_{\sigma}^{(3)}&=\int d^4x\sqrt{-g}\Big[
\gamma_1(t)\sigma^3
+\gamma_2(t)\sigma^2\partial^0\sigma
+\gamma_3(t)\sigma(\partial^0\sigma)^2
+\gamma_4(t)(\partial^0\sigma)^3
+\gamma_5(t)\,\sigma \partial_\mu\sigma\partial^\mu\sigma
+\gamma_6(t)\partial^0\sigma \partial_\mu\sigma\partial^\mu\sigma
\Big]\,,
\end{align}
and the third order mixing $S_{\rm mix}^{(3)}$ is given by
\begin{align}
\nonumber
S_{\rm mix}^{(3)}&=\int d^4x\sqrt{-g}\Big[
\overline{\gamma}_1(t)\delta g^{00}\,\sigma^2
+\overline{\gamma}_2(t)\delta g^{00}\,\sigma\partial^0\sigma
+\overline{\gamma}_3(t)\delta g^{00}(\partial^0\sigma)^2
+\overline{\gamma}_4(t)\partial^0\delta g^{00}\sigma\partial^0\sigma\\
\label{S_mix3}
&\qquad\qquad\qquad\quad
+\overline{\gamma}_5(t)\,\delta g^{00} \partial_\mu\sigma\partial^\mu\sigma
+\overline{\gamma}_6(t)(\delta g^{00})^2\sigma
+\overline{\gamma}_7(t)(\delta g^{00})^2\partial^0\sigma
\Big]\,.
\end{align}
Here, it may be wondered if the terms such as $\delta K^\mu_\mu\,
\sigma^2$, $\delta K^\mu_\mu \,\sigma\partial^0\sigma$, $\delta
R^{00}\sigma^2$, and $\delta K_\mu^\mu \delta g^{00}\,\sigma$ can appear
at the same order in derivatives.  However, they can be absorbed into
other third order terms in~(\ref{S_mix3}) and the second order terms
in~(\ref{S_sigma^2}) and (\ref{S_mix^2}) by integrating by parts as we
did to rewrite (\ref{second_K}) into (\ref{S_mix^2}). The term
proportional to $\delta R\sigma^2$ can also appear but such a term
vanishes in the decoupling limit, so we do not consider it here for
simplicity.

\medskip
To summarize, the most generic action in the unitary gauge can be
written up to the third order fluctuations as follows:
\begin{align}
\label{unitary_action}
S=S_{\rm grav}+S_{\sigma}^{(2)}+S_{\rm mix}^{(2)}+S_{\sigma}^{(3)}+S_{\rm mix}^{(3)}\,,
\end{align}
where $S_{\rm grav}$, $S_{\sigma}^{(2)}$, $S_{\rm mix}^{(2)}$,
$S_{\sigma}^{(3)}$, and $S_{\rm mix}^{(3)}$ are defined in
(\ref{S_grav}), (\ref{S_sigma^2}), (\ref{S_mix^2}), (\ref{S_sigma^3}),
and (\ref{S_mix3}).

\subsection{Ambiguity of the action in the unitary gauge}
In multiple field inflation models,
there are some ambiguities of the action
in the unitary gauge:
there are degrees of freedom
of the field redefinition of $\sigma$
and
time coordinate transformations
vanishing on the background trajectory $\sigma=0$.
Using these degrees of freedom,
it is possible to drop some
terms and simplify the action.
Without loss of generalities,
the action can be written into the following
three normalizations using these ambiguities.\footnote{
Note that it is not possible in general to
impose some of the three conditions at the same time.}

\medskip
\paragraph{{\rm Normalization} $1:$ $\alpha_1+\alpha_2=1$.}
Let us first consider the kinetic term of $\sigma$.
The second order action $S_\sigma^{(2)}$ of $\sigma$
can be expanded up to the second order fluctuations as
\begin{align}
\int d^4x\,a^3\left[\frac{\alpha_\sigma^2}{2}\Big(\dot{\sigma}^2-c_\sigma^2\frac{(\partial_i\sigma)^2}{a^2}
-\frac{\alpha_3-3H\alpha_4-\dot{\alpha}_4}{\alpha_1+\alpha_2}\sigma^2\Big)\right]\,,
\end{align}
where the normalization factor $\alpha_\sigma$ is defined as
$\alpha_\sigma^2=\alpha_1+\alpha_2$.  Although the factor
$\alpha_\sigma$ is time-dependent in general, it can be taken unity
by redefining $\sigma$ as
$\tilde{\sigma}=\alpha_\sigma \sigma$.
Since the derivative of $\sigma$ can be written as
\begin{align}
\partial_\mu \sigma=\alpha_\sigma^{-1} \partial_\mu \tilde{\sigma}
-\delta^0_\mu\,\frac{\dot{\alpha}_\sigma}{\alpha_\sigma^2}\tilde{\sigma}\,,
\end{align}
the action still takes the form~(\ref{S_sigma^2})
after the redefinition.

\medskip
\paragraph{{\rm Normalization} $2:$ $\alpha_4=\beta_3=\gamma_2=0$.}
Using the time coordinate transformation
vanishing on the background trajectory $\sigma=0$,
the following form of interaction terms can be eliminated:
\begin{align}
\label{int_eliminated_gauge}
\int d^4x\sqrt{-g}\Big[
f(t,\sigma)\partial^0\sigma\Big]\,,
\end{align}
where $f(t,\sigma)$ is a function of $t$ and
$\sigma$ but does not contain derivatives of $\sigma$.
As a simple example,
let us consider the action
\begin{align}
\label{norm2_simple}
S=\int d^4x\sqrt{-g}\Big[
\frac{1}{2}M_{\rm Pl}^2R
+M_{\rm Pl}^2\dot{H}(t)g^{00}-M_{\rm Pl}^2\left(3H^2(t)+\dot{H}(t)\right)
+f(\sigma,t)\partial^0\sigma\Big]\,.
\end{align}
Under the time coordinate transformation
\begin{align}
t\to\tilde{t}\quad
{\rm with}
\quad
t=\tilde{t}-\epsilon(\tilde{t},\sigma)\,,
\end{align}
the action~(\ref{norm2_simple}) is transformed into
\begin{align}
\nonumber
S&=\int d^4x\sqrt{-g}\Bigg[
\frac{1}{2}M_{\rm Pl}^2R
+M_{\rm Pl}^2\dot{H}(t-\epsilon)\Big(g^{00}(1-\partial_{t}\epsilon)^2-2(1-\partial_{t}\epsilon)\partial_\sigma\epsilon\,\partial^0\sigma
+(\partial_\sigma\epsilon)^2\partial_\mu\sigma\partial^\mu\sigma\Big)\\
\label{after_transformed_norm2}
&\qquad\qquad\qquad\quad
-M_{\rm Pl}^2\left(3H^2(t-\epsilon)+\dot{H}(t-\epsilon)\right)
+f(\sigma,t-\epsilon)(1-\partial_{t}\epsilon)\partial^0\sigma
-f(\sigma,t-\epsilon)\partial_\sigma\epsilon\,\partial_\mu\sigma\partial^\mu\sigma\Bigg]\,.
\end{align}
Therefore, if we take $\epsilon$ such that
\begin{align}
\label{epsilon_rec}
\partial_\sigma\epsilon=\frac{f(\sigma,t-\epsilon)}{2M_{\rm Pl}^2\dot{H}(t-\epsilon)}\,,
\quad
\epsilon(t,\sigma=0)=0\,,
\end{align}
the action~(\ref{after_transformed_norm2}) reduces to
\begin{align}
\nonumber
S&=\int d^4x\sqrt{-g}\Bigg[
\frac{1}{2}M_{\rm Pl}^2R
-M_{\rm Pl}^2\left(3H^2(t-\epsilon)+\dot{H}(t-\epsilon)\right)\\
&\qquad\qquad\qquad\quad+M_{\rm Pl}^2\dot{H}(t-\epsilon)\Big(g^{00}(1-\partial_{t}\epsilon)^2
-(\partial_\sigma\epsilon)^2\partial_\mu\sigma\partial^\mu\sigma\Big)
-f(\sigma,t-\epsilon)\partial_\sigma\epsilon\,\partial_\mu\sigma\partial^\mu\sigma
\Bigg]\,,
\end{align}
which does not contain interaction terms in the form $f(t,\sigma)\partial^0\sigma$.
Note that the conditions~(\ref{epsilon_rec})
can be always solved at least as an expansion in $\sigma$.
It is straightforward to extend this discussion
to general cases
and we conclude that
interaction terms in the form of~(\ref{int_eliminated_gauge})
can be eliminated
using the time coordinate transformation
vanishing on the background trajectory $\sigma=0$.
In particular,
we can set $\alpha_4=\beta_3=\gamma_2=0$
without loss of generalities.

\medskip
\paragraph{{\rm Normalization} $3:$ $\alpha_1=1$, $\gamma_5=0$.}

The action $S$ in Eq.~(\ref{unitary_action}) contains the following
term:
\begin{align}
\label{normalization3_action}
\int d^4x\sqrt{-g}\Big[
-\frac{1}{2}f^2(t,\sigma)g^{\mu\nu}\partial_\mu\sigma\partial_\nu\sigma\Big]\,,
\end{align}
where $f(t,\sigma)$ is a function of $t$ and $\sigma$, and does not
contain derivatives of $\sigma$.  The function is expanded in $\sigma$
as $f^2(t,\sigma)=\alpha_1(t)-2\gamma_5(t)\sigma+\mathcal{O}(\sigma^2)$.
By the field redefinition,
it is possible to eliminate this kind of derivative couplings
and to rewrite~(\ref{normalization3_action})
into the canonical form of the kinetic term.
Let us define $\tilde{\sigma}$ as
$\tilde{\sigma}=F(t,\sigma)$ such that
\begin{align}
\partial_\sigma F(t,\sigma) = f(t,\sigma)\,,
\quad
F(t,\sigma=0)=0\,.
\end{align}
Since the derivative of $\tilde\sigma$ is given by
\begin{align}
\partial_\mu \tilde\sigma=f \partial_\mu\sigma + \delta^{0}_{\mu}
 \partial_t F\,,
\end{align}
the action can be rewritten as
\begin{align}
\int d^4x\sqrt{-g}\Big[
-\frac{1}{2}g^{\mu\nu}\partial_\mu\tilde{\sigma}\partial_\nu\tilde{\sigma}
+\partial_t F \partial^0\tilde{\sigma}-\frac12 g^{00} 
\left(\partial_t F\right)^2
\Big]\,,
\end{align}
which still takes the form~(\ref{unitary_action}).
Here $F$ should be regarded as a function of $\tilde{\sigma}$ and $t$.
It also should be noticed that, though we can set the function $f$ to be
unity, the terms proportional to $\partial^0 \sigma$ and $g^{00}$
appear.

\subsection{Action for the Goldstone boson and the decoupling regime}
\label{subsection_quasiaction_for_pi}

In this subsection we construct the action for the Goldstone boson $\pi$
and discuss its decoupling regime.
As in the last section,
we perform the time diffeomorphism~(\ref{unitary_to_pi}).
Practically, it is realized by the following replacements:
\begin{align}
\nonumber
&g^{00}\to g^{00}+2\partial^0\pi
+\partial_\mu\pi\partial^\mu\pi\,,
\quad
\partial^0\sigma\to \partial^0\sigma+\partial_\mu\pi\,\partial^\mu\sigma\,,\\
&\sigma\to\sigma\,,
\quad
f(t)\to f(t+\pi)\,,
\quad
\int d^4x \sqrt{-g}\to\int d^4x \sqrt{-g}\,.
\end{align}
With these replacements, $S_{\rm grav}$, $S_{\sigma}^{(2)}$, and $S_{\rm
mix}^{(2)}$ are rewritten as
\begin{align}
\nonumber
S_{\rm grav}&=\int d^4x \sqrt{-g}\Bigg[
\frac{1}{2}M_{\rm Pl}^2R
+M_{\rm Pl}^2\dot{H}(t+\pi)\left(g^{00}+2\partial^0\pi+\partial_\mu\pi\partial^\mu\pi\right)-M_{\rm Pl}^2\left(3H^2(t+\pi)+\dot{H}(t+\pi)\right)\\
\label{S_grav_pi}
&\qquad\qquad\qquad\quad+\frac{M_2^4(t+\pi)}{2!}\left(\delta g^{00}+2\partial^0\pi+\partial_\mu\pi\partial^\mu\pi\right)^2
+\frac{M_3^4(t+\pi)}{3!}\left(\delta g^{00}+2\partial^0\pi+\partial_\mu\pi\partial^\mu\pi\right)^3\Bigg]\,,\\
S_{\sigma}^{(2)}&=\int d^4x\sqrt{-g}\Big[-\frac{\alpha_1(t+\pi)}{2}\partial_\mu\sigma\partial^\mu\sigma+\frac{\alpha_2(t+\pi)}{2}\left(\partial^0\sigma+\partial_\mu\pi\partial^\mu\sigma\right)^2-\frac{\alpha_3(t+\pi)}{2}\sigma^2+\alpha_4(t+\pi)\sigma(\partial^0\sigma+\partial_\mu\pi\,\partial^\mu\sigma)\Big]\,,\\
\nonumber
S_{\rm mix}^{(2)}&=\int d^4x\sqrt{-g}\Big[
\beta_1(t+\pi)\left(\delta g^{00}+2\partial^0\pi+\partial_\mu\pi\partial^\mu\pi\right)\sigma
+\beta_2(t+\pi)\left(\delta g^{00}+2\partial^0\pi+\partial_\mu\pi\partial^\mu\pi\right)
\left(\partial^0\sigma+\partial_\mu\pi\partial^\mu\sigma\right)\\
\label{S_mix2_pi}
&\qquad\qquad\qquad\quad+\beta_3(t+\pi)(\partial^0\sigma+\partial_\mu\pi\partial^\mu\sigma)-\Big(\dot{\beta}_3(t+\pi)+3H(t+\pi)\beta_3(t+\pi)\Big)\sigma
\Big]\,.
\end{align}
The third order actions $S_{\sigma}^{(3)}$ and $S_{\rm mix}^{(3)}$
can be obtained in a similar way.

\medskip In order to discuss the decoupling regime of the action, we
first clarify in which regime graviton fluctuations become irrelevant to
tree-level three point functions of $\pi$. For this purpose, let us take
the spatially flat gauge~(\ref{spatially_flat_gauge}) and use the ADM
decomposition:
\begin{equation}
ds^2=-(N^2-N_iN^i)dt^2+2N_i dx^i dt+a^2(e^{\gamma})_{ij}\,dx^idx^j
\quad
{\rm with}
\quad
\gamma_{ii}=\partial_i\gamma_{ij}=0
\,.
\end{equation}
Here and in what follows we use the spatial metric $h_{ij}=a^2
(e^\gamma)_{ij}$ and its inverse $h^{ij}=a^{-2} (e^{-\gamma})_{ij}$ to
raise or lower the indices of $N^i$.  The inverse metric $g^{\mu\nu}$
are written in terms of $N$, $N^i$, and $h_{ij}$ as
\begin{equation}
g^{00}=-\frac{1}{N^2}\,,
\quad
g^{0i}=g^{i0}=\frac{N^i}{N^2}\,,
\quad
g^{ij}=h^{ij}-\frac{N^iN^j}{N^2}\,.
\end{equation}
In this gauge, there are no second order mixing terms of $\pi$ and
$\gamma_{ij}$ because $\gamma_{ij}$ has two spatial indices and
is transverse-traceless. Then, the tensor fluctuation $\gamma_{ij}$ does
not contribute to tree-level three point functions of $\pi$.
Therefore, possible contributions of graviton fluctuations come only
from the auxiliary fields $\delta N=N-1$ and $N^i$.  As discussed
in~\cite{Maldacena:2002vr}, it is sufficient for the calculation of
three-point functions to solve the constraints up to first order.
Expanding the actions (\ref{S_grav_pi})-(\ref{S_mix2_pi}) up to the
second order in $\pi$, $\delta N$, and $N^i$,
\begin{align}
\nonumber
S_{\rm grav}&=\int d^4x \,a^3\Bigg[
-M_{\rm Pl}^2(3H^2+c_\pi^{-2}\dot{H})\delta N^2
-2M_{\rm Pl}^2H\delta N\partial_i N^i
+M_{\rm Pl}^2\frac{1}{4}N^i\partial_i\partial_j N^j
-M_{\rm Pl}^2\frac{1}{4}N^i\partial^2 N^i\\
&\quad\quad\qquad\qquad-\frac{M_{\rm Pl}^2\dot{H}}{c_\pi^{2}}\Big(\dot{\pi}^2-c_\pi^2\frac{(\partial_i\pi)^2}{a^2}\Big)-3M_{\rm Pl}^2\dot{H}^2\pi^2
+M_{\rm Pl}^2(2c_\pi^{-2}\dot{H}\dot{\pi}-6H\dot{H}\pi)\delta N
+2M_{\rm Pl}^2\dot{H}N^i\partial_i\pi
\Bigg]\,,\\
S_{\sigma}^{(2)}&=\int d^4x\,a^3\Big[\alpha_\sigma^2\Big(\dot{\sigma}^2-c_\sigma^2\frac{(\partial_i\sigma)^2}{a^2}\Big)
-\frac{\alpha_3}{2}\sigma^2
-\alpha_4\sigma\dot{\sigma}\Big]\,,
\\
\nonumber
S_{\rm mix}^{(2)}&=\int d^4x\,a^3\Big[
2\beta_1\left(\delta N-\dot{\pi}\right)\sigma-2\beta_2\left(\delta N-\dot{\pi}\right)\dot{\sigma}
-\delta N\Big(-\beta_3\dot{\sigma}+\dot{\beta}_3\sigma+3H\beta_3\sigma\Big)
+\beta_3N^i\partial_i\sigma\\
&\qquad\qquad\qquad\quad+\beta_3\Big(-\dot{\pi}\dot{\sigma}+\frac{\partial_i\pi\partial_i\sigma}{a^2}\Big)
+\dot{\beta}_3\dot{\pi}\sigma-3\dot{H} \beta_3 \pi \sigma
\Big]\,,
\end{align}
the constraints are solved up to first order as follows:
\begin{align}
\nonumber
\delta N&=-\frac{\dot{H}}{H}\pi-\frac{\beta_3}{2M_{\rm Pl}^2H}\sigma\,,\quad
N^i=a^{-2}\partial_i\psi\\
\label{constraints}
&
{\rm with}
\quad
\psi=a^2\partial^{-2}
\Big(c_\pi^{-2}\frac{\dot{H}}{H^2}(\dot{H}\pi+H\dot{\pi})
+\frac{\beta_1}{M_{\rm Pl}^2H}\sigma
-\frac{\beta_2}{M_{\rm Pl}^2H}\dot{\sigma}
+\frac{\beta_3}{2M_{\rm Pl}^2H}\Big(c_\pi^{-2}\frac{\dot{H}}{H}\sigma+\dot{\sigma}\Big)
-\frac{\dot{\beta}_3}{2M_{\rm Pl}^2H}\sigma
\Big)\,.
\end{align}
Here the sound speed $c_\pi^2$ of $\pi$ are defined as
$c_\pi^2=\dot{H}M_{\rm Pl}^2/(\dot{H}M_{\rm Pl}^2-2M_2^4)$.  The factors
$c_\sigma^2=\alpha_1/(\alpha_1+\alpha_2)$ and
$\alpha_\sigma^2=\alpha_1+\alpha_2$ are the sound speed and the
normalization factor of $\sigma$, respectively.  Using the canonical
normalization
\begin{align}
\pi_c\sim M_{\rm Pl}(-\dot{H})^{1/2}c_{\pi}^{-1}\pi\,,
\quad
\sigma_c\sim\alpha_\sigma\sigma\,,
\quad
\delta N_c\sim M_{\rm Pl}\delta N\,,
\quad
N^i_c\sim M_{\rm Pl}N^i\,,
\end{align}
and redefining the coupling constants $\beta_1$, 
$\beta_2$, and $\beta_3$ correspondingly as
\begin{align}
\beta^c_1\sim\frac{c_\pi}{\alpha_\sigma M_{\rm Pl}(-\dot{H})^{1/2}}\beta_1\,,
\quad
\beta^c_2\sim\frac{c_\pi}{\alpha_\sigma M_{\rm Pl}(-\dot{H})^{1/2}}\beta_2\,,
\quad
\beta^c_3\sim\frac{c_\pi}{\alpha_\sigma M_{\rm Pl}(-\dot{H})^{1/2}}\beta_3\,,
\end{align}
we rewrite the constraints~(\ref{constraints}) as
\begin{align}
\delta N_c&\sim \tilde{\epsilon}^{1/2}\Big(c_{\pi}^2 \pi_c-\frac{1}{2}\beta_3^c\,\sigma_c\Big)\,,\\
N^i_c&\sim\tilde{\epsilon}^{1/2}\frac{\partial_i}{\partial^2}
\Bigg(-\dot{\pi}_c+\frac{1}{2}\tilde{\eta}H\pi_c
+\beta^c_1\sigma_c
-\beta^c_2\Big(\dot{\sigma}_c-\frac{\dot{\alpha}_\sigma}{\alpha_\sigma}\sigma\Big)
-\frac12 \beta_3^c \dot{\sigma}_c
+\frac12 \beta_3^c \sigma_c
\Big( -2\frac{\dot{\alpha}_\sigma}{\alpha_\sigma}\sigma 
+ (c_\pi^2-1)\tilde{\epsilon} H - \frac{\tilde{\eta}}{2}H \Big)
-\frac12 \dot{\beta}_3^c \sigma_c 
\Bigg)\,,
\end{align}
where we have defined
$\displaystyle\tilde{\epsilon}=-c_\pi^{-2}\frac{\dot{H}}{H^2}$ and
$\displaystyle\tilde{\eta}=\frac{\dot{\tilde{\epsilon}}}{\tilde{\epsilon}H}$
in analogy with usual slow-roll parameters $\epsilon$ and $\eta$. It is
manifest that $\delta N_c$ and $N_c^i$ are suppressed by the parameter
$\tilde{\epsilon}^{1/2}$ and contributions from $\delta N_c$ and $N_c^i$
become irrelevant in the limit $\tilde{\epsilon}\to0$.\footnote{ To be
precise, we need to assume that $\tilde{\epsilon}$ is small enough to
vanish when multiplied by other parameters in the calculation.  For
example, we need to assume that $\tilde{\epsilon}\frac{H}{E}\ll1$.  } In
this limit, tree-level three point functions of $\pi$ are determined by
the following action in the decoupling limit:
\begin{align}
\nonumber
S_{\rm grav}&=\int d^4x \,a^3\Big[
-\frac{M_{\rm Pl}^2\dot{H}}{c_\pi^2}\Big(\dot{\pi}^2-c_{\pi}^2\frac{(\partial_i\pi)^2}{a^2}\Big)
-M_{\rm Pl}^2\dot{H}(c_\pi^{-2}-1)\Big(\dot{\pi}^3-\dot{\pi}\frac{(\partial_i\pi)^2}{a^2}\Big)
-\frac{4M_3^4}{3}\dot{\pi}^3\\
&\qquad\qquad\qquad-3M_{\rm Pl}^2\dot{H}^2\pi^2
-\partial_t\Big(\frac{M_{\rm Pl}^2\dot{H}}{c_\pi^2}\Big)\dot{\pi}^2\pi
+M_{\rm Pl}^2\ddot{H}\frac{(\partial_i\pi)^2}{a^2}\pi 
- 3 M_{\rm pl}^2 \dot{H} \ddot{H} \pi^3
\Big]\,,\\
\nonumber
S_{\sigma}^{(2)}&=\int d^4x\,a^3\Big[\frac{\alpha_\sigma^2}{2}\Big(\dot{\sigma}^2-c_\sigma^2\frac{(\partial_i\sigma)^2}{a^2}-\frac{\alpha_3-3H\alpha_4-\dot{\alpha}_4}{\alpha_\sigma^2}\sigma^2\Big)
+\alpha_\sigma^2(1-c_\sigma^2)\Big(\dot{\pi}\dot{\sigma}^2-\dot{\sigma}\frac{(\partial_i\pi\partial_i\sigma)}{a^2}\Big)\\
&\qquad\qquad\qquad\quad
-\alpha_4\sigma\Big(\dot{\pi}\dot{\sigma}-\frac{\partial_i\pi\partial_i\sigma}{a^2}\Big)
+\frac{\dot{(\alpha_\sigma^2)}}{2}\pi\dot{\sigma}^2-\frac{\dot{\left(\alpha_\sigma^2c_\sigma^2\right)}}{2}\pi\frac{(\partial_i\sigma)^2}{a^2}-\frac{\dot{\alpha}_3}{2}\sigma^2\pi
-\dot{\alpha}_4\pi\sigma\dot{\sigma}
\Big]\,,\\
\nonumber
S_{\rm mix}^{(2)}&=\int d^4x\,a^3\Big[
(-2\beta_1+\dot{\beta}_3)\dot{\pi}\sigma
+(2\beta_2-\beta_3)\dot{\pi}\dot{\sigma}
+\beta_3\frac{\partial_i\pi\partial_i\sigma}{a^2}
-3 \dot{H} \beta_3 \pi \sigma
-\beta_1\Big(\dot{\pi}^2-\frac{(\partial_i\pi)^2}{a^2}\Big)\sigma
+(-2\dot{\beta}_1+\ddot{\beta}_3)\pi\dot{\pi}\sigma 
\\
&\qquad\qquad\quad
+3\beta_2\dot{\pi}^2\dot{\sigma}
-2\beta_2\dot{\pi}\frac{\partial_i\pi\partial_i\sigma}{a^2}-
\beta_2\frac{(\partial_i\pi)^2}{a^2}\dot{\sigma}
+(2\dot{\beta}_2-\dot{\beta}_3)\pi\dot{\pi}\dot{\sigma}
+\dot{\beta}_3\pi\Big(\frac{\partial_i\pi\partial_i\sigma}{a^2}\Big)
-3\left(\dot{H}\dot{\beta}_3 + \frac12 \ddot{H} \beta_3 \right)
 \pi^2 \sigma
\Big]\,,\\
S_{\sigma}^{(3)}&=\int d^4x\,a^3\Big[
\Big(\gamma_1+H\gamma_2+\frac{1}{3}\dot{\gamma}_2\Big)\sigma^3
+(\gamma_3-\gamma_5)\sigma\dot{\sigma}^2
+(-\gamma_4+\gamma_6)\dot{\sigma}^3
+\gamma_5\,\sigma \frac{(\partial_i\sigma)^2}{a^2}
-\gamma_6\dot{\sigma} \frac{(\partial_i\sigma)^2}{a^2}
\Big]\,,\\
S_{\rm mix}^{(3)}&=\int d^4x\,a^3\Big[
-2\overline{\gamma}_1\dot{\pi}\sigma^2
+2\overline{\gamma}_2\dot{\pi}\sigma\dot{\sigma}
+2(-\overline{\gamma}_3+\overline{\gamma}_5)\dot{\pi}\dot{\sigma}^2
-2\overline{\gamma}_4\ddot{\pi}\sigma\dot{\sigma}
-2\overline{\gamma}_5\dot{\pi} \frac{(\partial_i\sigma)^2}{a^2}
+4\overline{\gamma}_6\dot{\pi}^2\sigma
-4\overline{\gamma}_7\dot{\pi}^2\dot{\sigma}
\Big]\,.
\end{align}
It should be noticed that
non-trivial cubic interactions appear generically
when the sound speed $c_\sigma$ of $\sigma$ is small,
$\alpha_4$ is non-zero,
or mixing couplings $\beta_1$ and $\beta_2$ exist
as well as the sound speed $c_\pi$ of $\pi$ is small.

\subsection{Examples}
Before closing this section, we clarify the relation between our
approach and models in the literatures. For this purpose, we first
discuss the original model of quasi-single field
inflation~\cite{Chen:2009we}, and then, we investigate the effects
of heavy particles during inflation. At the end of this subsection, a
class of two field models will be considered.

\subsubsection{Original model discussed by Chen and Wang}

As was reviewed in section~\ref{section_review}, the original
model~\cite{Chen:2009we} of quasi-single field inflation is described by
the following matter action:
\begin{align}
S_{\rm matter}=\int d^4x\sqrt{-g}\Big[-\frac{1}{2}(\tilde{R}+\chi)^2g^{\mu\nu}
\partial_\mu\theta\partial_\nu\theta
-\frac{1}{2}g^{\mu\nu}\partial_\mu\chi\partial_\nu\chi
-V_{\rm sr}(\theta)-V(\chi)
\Big]\, .
\end{align}
The homogeneous backgrounds are given by $\theta=\theta_0(t)$ and
$\chi=\chi_0 \,(\text{constant})$, which leads to the action in the
unitary gauge $\delta\theta=\theta-\theta_0=0$,
\begin{align}
\nonumber
S_{\rm matter}&=\int d^4x\sqrt{-g}\Big[-\frac{1}{2}(R+\sigma)^2g^{00}
\dot{\theta}_0^2
-\frac{1}{2}g^{\mu\nu}\partial_\mu\sigma\,\partial_\nu\sigma
-V_{\rm sr}(\theta_0)-V(\chi_0+\sigma)
\Big]
\\
\nonumber
&=\int d^4x\sqrt{-g}\Big[-\frac{1}{2}R^2\dot{\theta}_0^2\,g^{00}
-(V_{\rm sr}(\theta_0)+V(\chi_0))
-\frac{1}{2}g^{\mu\nu}\partial_\mu\sigma\,\partial_\nu\sigma
-\frac{1}{2}\Big(V^{\prime\prime}(\chi_0)-\dot{\theta}_0^2\Big)\sigma^2\\
\label{quasi_unitary_1}
&\qquad\qquad\qquad\quad
-R\,\dot{\theta}_0^2\,\delta g^{00}\sigma
-\frac{V^{\prime\prime\prime}(\chi_0)}{3!}\sigma^3
-\frac{1}{2}
\dot{\theta}_0^2\,\delta g^{00}\sigma^2
+\mathcal{O}(\sigma^4)
\Big]\,,
\end{align}
where $R=\tilde{R}+\chi_0$, $\sigma=\chi-\chi_0$ and we used the background equations of
motion.
Using the equations of motion (\ref{eom_grav_QSI}),
the action (\ref{quasi_unitary_1})
can be written in terms of the Hubble parameter $H$ as
\begin{align}
\nonumber
S_{\rm matter}&=\int d^4x\sqrt{-g}\Big[M_{\rm Pl}^2\dot{H}g^{00}
-M_{\rm Pl}^2(3H^2+\dot{H})
-\frac{1}{2}g^{\mu\nu}\partial_\mu\sigma\,\partial_\nu\sigma
-\frac{1}{2}\Big(V^{\prime\prime}(\chi_0)+\frac{2M_{\rm Pl}^2\dot{H}}{R^2}\Big) \sigma^2\\
&\qquad\qquad\qquad\quad
+\frac{2M_{\rm Pl}^2\dot{H}}{R}\,\delta g^{00}\sigma
-\frac{V^{\prime\prime\prime}(\chi_0)}{3!}\sigma^3
+\frac{M_{\rm Pl}^2\dot{H}}{R^2}\,\delta g^{00}\sigma^2
+\mathcal{O}(\delta\sigma^4)
\Big]\,,
\end{align}
which corresponds to the following parameters in our framework,
\begin{align}
\alpha_1=1\,,
\quad
\alpha_3=V^{\prime\prime}(\chi_0)+\frac{2M_{\rm Pl}^2\dot{H}}{R^2}\,,
\quad
\beta_1=\frac{2M_{\rm Pl}^2\dot{H}}{R}\,,
\quad
\gamma_1=-\frac{1}{3!}V^{\prime\prime\prime}(\chi_0)\,,
\quad
\overline{\gamma}_1=\frac{M_{\rm Pl}^2\dot{H}}{R^2}\,,
\quad
(\,{\rm others}\,)=0\,.
\end{align}

\subsubsection{Effects of heavy particles}
\label{subsubsection_heavy}

Recently, it is argued that the existence of heavy particles can cause a
non-trivial sound speed of effective single field
inflation~\cite{Cremonini:2010ua,Achucarro:2010da,Shiu:2011qw,Cespedes:2012hu,Achucarro:2012sm,Pi:2012gf,Gao:2012uq,Saito:2012pd,Gwyn:2012mw}.
As was mentioned earlier, our framework is also applicable for such
inflation models with heavy particles.
In the following, we give a simple explanation
for the effects of heavy fields.

\medskip
Let us start from the following simplest case:
\begin{align}
S=\int d^4x \sqrt{-g}\left[
\frac{1}{2}M_{\rm Pl}^2R
+M_{\rm Pl}^2\dot{H}g^{00}
-M_{\rm Pl}^2(3H^2+\dot{H})
-\frac{1}{2}g^{\mu\nu}\partial_\mu\sigma\,\partial_\nu\sigma
-\frac{m^2}{2}\sigma^2+\beta\, \delta g^{00}\sigma\right]\,.
\end{align}
Here $m$ is the mass of $\sigma$ and $\beta$ is the mixing coupling
between the adiabatic mode and the massive particle $\sigma$.
We assume that
the mass of
$\sigma$ is much larger than the Hubble scale during inflation, $m\gg
H$,
and the time-dependence of $\beta$ is negligible compared
to the mass $m$.
In such a regime, the kinetic term of $\sigma$ becomes irrelevant
and the dynamics is determined by
\begin{align}
S\sim\int d^4x \sqrt{-g}\left[
\frac{1}{2}M_{\rm Pl}^2R
+M_{\rm Pl}^2\dot{H}g^{00}
-M_{\rm Pl}^2(3H^2+\dot{H})
-\frac{m^2}{2}\Big(\sigma-\frac{\beta}{m^2}\delta g^{00}\Big)^2
+\frac{\beta^2}{2m^2}(\delta g^{00})^2\right]\,,
\end{align}
which implies that the perturbation $\sigma$ quickly responds to
the variation of the adiabatic mode $\delta g^{00}$.  Integrating out the massive
particle $\sigma$, we obtain the following effective action for
single-field inflation:
\begin{align}
S_{\rm eff}=\int d^4x \sqrt{-g}\left[
\frac{1}{2}M_{\rm Pl}^2R
+M_{\rm Pl}^2\dot{H}g^{00}
-M_{\rm Pl}^2(3H^2+\dot{H})
+\frac{\beta^2}{2m^2}(\delta g^{00})^2\right]\,.
\end{align}
In particular,
the last term gives the following non-trivial sound speed:
\begin{align}
c_\pi^2=\frac{-\dot{H}M_{\rm Pl}^2}{-\dot{H}M_{\rm Pl}^2+2\beta^2/m^2}\,,
\end{align}
which reproduces the result in~\cite{Cremonini:2010ua,Achucarro:2010da,Shiu:2011qw,Cespedes:2012hu,Achucarro:2012sm,Pi:2012gf,Gao:2012uq,Saito:2012pd,Gwyn:2012mw}.
Note that
it is obvious in our approach
that the effective action
contains the $(\delta g^{00})^2$ interaction:
our result explains not only the effective sound speed
but also non-trivial cubic effective interactions
of the Goldstone boson $\pi$
associated with the $(\delta g^{00})^2$ term.

\medskip
The above discussions can be extended to
more general settings.
Let us consider the following action with more generic mixing couplings:
\begin{align}
\nonumber
S&=\int d^4x \sqrt{-g}\left[
\frac{1}{2}M_{\rm Pl}^2R
+M_{\rm Pl}^2\dot{H}g^{00}
-M_{\rm Pl}^2(3H^2+\dot{H})\right.\\
\label{S_heavy_2}
&\quad\qquad\qquad\qquad\left.
-\frac{1}{2}g^{\mu\nu}\partial_\mu\sigma\,\partial_\nu\sigma
-\frac{m^2}{2}\sigma^2+\beta_1\delta g^{00}\sigma+\beta_2\delta g^{00}\partial^0\sigma+\beta_3\partial^0\sigma
-(\dot{\beta}_3+3H\beta_3)\sigma\right]\,.
\end{align}
When the mass of
$\sigma$ is much larger than the Hubble scale during inflation, $m\gg
H$,
and the time-dependence of $\beta_i$'s is negligible compared
to the mass $m$,
the low energy effective action can be obtained
via the following procedure (see appendix~\ref{app_heavy}
for more detailed discussions):
\begin{enumerate}
\item Drop the kinetic term of heavy fields.
\item Eliminate derivatives of heavy fields by partial integrals.
\item Complete square the Lagrangian and integrate out heavy fields.
\end{enumerate}
To perform the second step,
it is convenient to introduce the following relations, which follow from
the formulae~(\ref{integ_by_parts}) and (\ref{K_to_partial}):
\begin{align}
\int d^4x \sqrt{-g}\,f(t)\delta g^{00}\partial^0\sigma
&=\int d^4x \sqrt{-g}\left[(\dot{f}(t)+3H\,f(t))\delta g^{00}\sigma
-f(t)\partial^0\delta g^{00}\,\sigma+\ldots\right]\,,\\
\int d^4x \sqrt{-g}\left[f(t)\partial^0\sigma
-(\dot{f}(t)+3H\,f(t))\sigma\right]
&=\int d^4x \sqrt{-g}\left[f(t)\delta K_\mu^\mu\sigma-\frac{\dot{f}(t)}{2}\delta g^{00}\sigma-\frac{f(t)}{2}\delta g^{00}\partial^0\sigma
+\ldots\right]\,,
\end{align}
where dots stand for higher order terms in $\delta g^{00}$, $\delta K^\mu_\mu$, and their derivatives.
From these relations,
it follows that
\begin{align}
\nonumber
&\quad\int d^4x \sqrt{-g}\left[\beta_1\delta g^{00}\sigma+\beta_2\delta g^{00}\partial^0\sigma+\beta_3\partial^0\sigma
-(\dot{\beta}_3+3H\beta_3)\sigma\right]\\
\nonumber
&=\quad\int d^4x \sqrt{-g}\left[\Big(\beta_1-\frac{\dot{\beta}_3}{2}\Big)\delta g^{00}\sigma+\Big(\beta_2-\frac{\beta_3}{2}\Big)\delta g^{00}\partial^0\sigma+\beta_3\,\delta K_\mu^\mu\,\sigma+\ldots\right]\\
&=\quad\int d^4x \sqrt{-g}\left[\Big(\beta_1+\dot{\beta}_2+3H\beta_2-\dot{\beta}_3-\frac{3}{2}H\beta_3\Big)\delta g^{00}\sigma
-\Big(\beta_2-\frac{\beta_3}{2}\Big)\partial^0\delta g^{00}\,\sigma+\beta_3\,\delta K_\mu^\mu\,\sigma+\ldots\right]\,.
\end{align}
Note that the first equality is the same as the relation used
to rewrite (\ref{second_K}) into the form of~(\ref{S_mix^2}) plus higher order terms.
Then, it is straightforward to
obtain the following effective action for
single-field inflation
using the above prescription:
\begin{align}
\nonumber
S_{\rm eff}&=\int d^4x \sqrt{-g}\Bigg[
\frac{1}{2}M_{\rm Pl}^2R
+M_{\rm Pl}^2\dot{H}g^{00}
-M_{\rm Pl}^2(3H^2+\dot{H})\\
&\qquad\qquad\qquad\quad
+\frac{1}{2m^2}\left(\Big(\beta_1+\dot{\beta}_2+3H\beta_2-\dot{\beta}_3-\frac{3}{2}H\beta_3\Big)\delta g^{00}
-\Big(\beta_2-\frac{\beta_3}{2}\Big)\partial^0\delta g^{00}+\beta_3\,\delta K_\mu^\mu\right)^2
+\ldots
\Bigg]\,,
\end{align}
where dots stand for higher order terms in $\delta g^{00}$,
$\delta K^\mu_\mu$, and their derivatives.
It turns out that
interactions such as $(\partial^0\delta g^{00})^2$
and $(\delta K^\mu_\mu)^2$ appear
in the effective action
as well as the $(\delta g^{00})^2$ interaction.

\subsubsection{A class of two-field models}

Let us then consider a class of two-field models described by the
following matter action:
\begin{align}
\label{two_fields}
S_{\rm matter}=\int d^4x\sqrt{-g}\Big[
-\frac{1}{2}\gamma_{ab}(\phi^a)g^{\mu\nu}\partial_\mu\phi^a\partial_\nu\phi^b
-V(\phi^a)
\Big]
\quad
(a=1,2)\,,
\end{align}
where $\gamma_{ab}(\phi^a)$ is the metric on the field space.  This
class of models were carefully studied in~\cite{GrootNibbelink:2001qt}
and recently discussed in~\cite{Shiu:2011qw} to investigate effects of
massive particles during inflation.  Suppose that the trajectory of the
homogeneous background fields $\phi_0^a(t)$ is on a curve
$\phi^a=\bar{\phi}^a(\lambda)$, and the background fields are given by
$\phi_0^a(t)=\bar{\phi}(\lambda_0(t))$.  We also assume that
$\dot{\lambda}>0$.  Defining the coordinates $(\lambda,\sigma)$ of the
field space such that the curve $\sigma=0$ coincides with the trajectory
curve, the fields $\lambda(x)$ and $\sigma(x)$ describe the adiabatic
mode and the isocurvature mode, respectively.  There are still
many choices of the coordinates or degrees of freedom of the field
redefinition.  In the following, we consider two types of basis of the
fields and discuss their properties in our framework.

\medskip
\paragraph{Orthogonal basis}

We first consider the orthogonal-basis.
We can always take the coordinate $(\lambda,\sigma)$ such that
\begin{align}
\gamma_{\lambda\sigma}=\gamma_{\sigma\lambda}=0\,.
\end{align}
By the field redefinition of $\lambda$,
we further require $\gamma_{\lambda\lambda}(\lambda,\sigma=0)=1$.
In this basis,
the matter action~(\ref{two_fields}) is given by
\begin{align}
S_{\rm matter}=\int d^4x\sqrt{-g}\Big[
-\frac{1}{2}\gamma_{\lambda\lambda}(\lambda,\sigma)g^{\mu\nu}\partial_\mu\lambda\partial_\nu\lambda
-\frac{1}{2}\gamma_{\sigma\sigma}(\lambda,\sigma)g^{\mu\nu}\partial_\mu\sigma\partial_\nu\sigma
-V(\lambda,\sigma)
\Big]\,,
\end{align}
and
it can be expanded in the unitary gauge $\delta\lambda=\lambda-\lambda_0=0$ up to the third order fluctuations
as
\begin{align}
\nonumber
S_{\rm matter}&=\int d^4x\sqrt{-g}\Big[
-\frac{1}{2}\dot{\lambda}_0^2\,g^{00}
-\frac{1}{2}\dot{\lambda}_0^2(\gamma_{\lambda\lambda})^\prime_0
\delta g^{00}\sigma
+\frac{1}{4}\dot{\lambda}_0^2(\gamma_{\lambda\lambda})^{\prime\prime}_0
\sigma^2
-\frac{1}{4}\dot{\lambda}_0^2(\gamma_{\lambda\lambda})_0^{\prime\prime}
\delta g^{00}\sigma^2
+\frac{1}{12}\dot{\lambda}_0^2(\gamma_{\lambda\lambda})^{\prime\prime\prime}_0
\sigma^3\\
\label{orthog_1}
&\qquad\qquad\qquad\quad
-\frac{1}{2}(\gamma_{\sigma\sigma})_0\,g^{\mu\nu}\partial_\mu\sigma\partial_\nu\sigma
-\frac{1}{2}(\gamma_{\sigma\sigma})^\prime_0\,\sigma\,g^{\mu\nu}\partial_\mu\sigma\partial_\nu\sigma
-V_0
-\frac{1}{2}V_0^{\prime\prime}\sigma^2
-\frac{1}{6}V_0^{\prime\prime\prime}\sigma^3
\Big]\,,
\end{align}
where we have used the equation of motion for $\sigma$.  In this
subsection, we write derivatives of the metric and the potential
evaluated at the classical value, for example, as
$(\gamma_{\lambda\lambda})_0^\prime=\partial_\sigma\gamma_{\lambda\lambda}(\lambda,\sigma)|_{\lambda=\lambda_0,\sigma=0}$
and $V_0^\prime=\partial_\sigma
V(\lambda,\sigma)|_{\lambda=\lambda_0,\sigma=0}$.  Using the equations
of motion for graviton,
\begin{align}
M_{\rm Pl}^2\dot{H}=-\frac{1}{2}\dot{\lambda}_0^2\,,
\quad
M_{\rm Pl}^2(3H^2+\dot{H})=V_0\,,
\end{align}
we can rewrite~(\ref{orthog_1}) as
\begin{align}
\nonumber
S_{\rm matter}&=\int d^4x\sqrt{-g}\Big[
M_{\rm Pl}^2g^{00}\dot{H}
-M_{\rm Pl}^2(3H^2+\dot{H})
-\frac{1}{2}(\gamma_{\sigma\sigma})_0\,g^{\mu\nu}\partial_\mu\sigma\partial_\nu\sigma
-\frac{1}{2}\Big(V_0^{\prime\prime}+M_{\rm Pl}^2\dot{H}(\gamma_{\lambda\lambda})_0^{\prime\prime}\Big)\sigma^2
\\
&\quad
+M_{\rm Pl}^2\dot{H}
(\gamma_{\lambda\lambda})_0^\prime
\delta g^{00}\sigma
-\frac{1}{2}(\gamma_{\sigma\sigma})^\prime_0\,\sigma\,g^{\mu\nu}\partial_\mu\sigma\partial_\nu\sigma
-\frac{1}{6}\left(V_0^{\prime\prime\prime}+M_{\rm Pl}^2\dot{H}(\gamma_{\lambda\lambda})_0^{\prime\prime\prime}\right)\sigma^3
+\frac{1}{2}M_{\rm Pl}^2\dot{H}(\gamma_{\lambda\lambda})^{\prime\prime}_0
\delta g^{00}\sigma^2
\Big]\,,
\end{align}
which is described in our framework as
\begin{align}
\nonumber
&\alpha_1=(\gamma_{\sigma\sigma})_0\,,
\quad
\alpha_3=V_0^{\prime\prime}+M_{\rm Pl}^2\dot{H}(\gamma_{\lambda\lambda})_0^{\prime\prime}\,,
\quad
\beta_1=M_{\rm Pl}^2\dot{H}
(\gamma_{\lambda\lambda})_0^\prime\,,
\\
&\gamma_1=-\frac{1}{6}\left(V_0^{\prime\prime\prime}+M_{\rm Pl}^2\dot{H}(\gamma_{\lambda\lambda})_0^{\prime\prime\prime}\right)\,,
\quad
\gamma_5=-\frac{1}{2}(\gamma_{\sigma\sigma})^\prime_0\,,
\quad
\overline{\gamma}_1=\frac{1}{2}M_{\rm Pl}^2\dot{H}(\gamma_{\lambda\lambda})^{\prime\prime}_0\,,
\quad
(\text{ others })=0\,.
\end{align}
We notice that this basis corresponds to
the second normalization of $\sigma$
in section~\ref{sub_quasi_uni}.

\medskip
\paragraph{Canonical $\sigma$ basis}

We then choose the coordinate $(\lambda,\sigma)$ such that
\begin{align}
\gamma_{\sigma\sigma}(\lambda,\sigma)=1\,,
\quad
\gamma_{\lambda\lambda}(\lambda,\sigma=0)=1\,,
\end{align}
where the normalization of $\sigma$ is always canonical
and that of $\lambda$ is canonical only on the trajectory curve.
In this basis,
the matter action~(\ref{two_fields}) is given by
\begin{align}
S_{\rm matter}=\int d^4x\sqrt{-g}\Big[
-\frac{1}{2}\gamma_{\lambda\lambda}(\lambda,\sigma)g^{\mu\nu}\partial_\mu\lambda\partial_\nu\lambda
-\gamma_{\lambda\sigma}(\lambda,\sigma)g^{\mu\nu}\partial_\mu\lambda\partial_\nu\sigma
-\frac{1}{2}g^{\mu\nu}\partial_\mu\sigma\partial_\nu\sigma
-V(\lambda,\sigma)
\Big]\,,
\end{align}
and
it can be expanded in the unitary gauge $\delta\lambda=\lambda-\lambda_0=0$ up to the third order fluctuations
as
\begin{align}
\nonumber
S_{\rm matter}&=\int d^4x\sqrt{-g}\Big[
-\frac{1}{2}\dot{\lambda}_0^2\,g^{00}
+\frac{1}{2} \dot{\lambda}_0^2(\gamma_{\lambda\lambda})^\prime_0 \,\sigma
-\frac{1}{2}\dot{\lambda}_0^2(\gamma_{\lambda\lambda})_0^\prime
\delta g^{00}\sigma
+\frac{1}{4}\dot{\lambda}_0^2(\gamma_{\lambda\lambda})_0^{\prime\prime}
\sigma^2
-\frac{1}{4}\dot{\lambda}_0^2(\gamma_{\lambda\lambda})_0^{\prime\prime}
\delta g^{00}\sigma^2
+\frac{1}{12}\dot{\lambda}_0^2(\gamma_{\lambda\lambda})^{\prime\prime}_0
\sigma^3\\
\nonumber
&\qquad\qquad\qquad\quad
-\dot{\lambda}_0(\gamma_{\lambda\sigma})_0\partial^0\sigma
-\dot{\lambda}_0(\gamma_{\lambda\sigma})^\prime_0\sigma\partial^0\sigma
-\frac{1}{2}\dot{\lambda}_0(\gamma_{\lambda\sigma})^{\prime\prime}_0\sigma^2\partial^0\sigma
-\frac{1}{2}g^{\mu\nu}\partial_\mu\sigma\partial_\nu\sigma\\
&\qquad\qquad\qquad\quad
-V_0
-V_0^\prime \sigma
-\frac{1}{2}V_0^{\prime\prime}\sigma^2
-\frac{1}{6}V_0^{\prime\prime\prime}\sigma^3
\Big]\,,
\end{align}
where
the equation of motion for $\sigma$
implies $V_0^\prime=\frac{1}{2} \dot{\lambda}_0^2(\gamma_{\lambda\lambda})^\prime_0 -\Big(3H\dot{\lambda}_0(\gamma_{\lambda\sigma})_0+
\ddot{\lambda}_0(\gamma_{\lambda\sigma})_0+
\dot{\lambda}_0\,\partial_t(\gamma_{\lambda\sigma})_0\Big)$.
In terms of the Hubble parameter $H$,
we can rewrite it as
\begin{align}
\nonumber
S_{\rm matter}&=\int d^4x\sqrt{-g}\Big[
M_{\rm Pl}^2g^{00}\dot{H}
-M_{\rm Pl}^2(3H^2+\dot{H})
-\sqrt{2}M_{\rm Pl}(-\dot{H})^{1/2}(\gamma_{\lambda\sigma})_0\partial^0\sigma
-(V_0^\prime + M_{\rm Pl}^2\dot{H} (\gamma_{\lambda\lambda})^\prime_0 )\sigma\\
\nonumber
&\qquad\qquad\quad\quad-\frac{1}{2}g^{\mu\nu}\partial_\mu\sigma\partial_\nu\sigma
-\frac{1}{2}\Big(V_0^{\prime\prime}+M_{\rm Pl}^2\dot{H}(\gamma_{\lambda\lambda})^{\prime\prime}_0\Big)\sigma^2
-\sqrt{2}M_{\rm Pl}(-\dot{H})^{1/2}(\gamma_{\lambda\sigma})^\prime_0\sigma\partial^0\sigma
+M_{\rm Pl}^2\dot{H}
(\gamma_{\lambda\lambda})^\prime_0
\delta g^{00}\sigma\\
&\qquad\qquad\quad\quad
-\frac{1}{6}\left(V_0^{\prime\prime\prime}+M_{\rm Pl}\dot{H}(\gamma_{\lambda\lambda})_0^{\prime\prime\prime}\right)\sigma^3
-\frac{\sqrt{2}}{2}M_{\rm Pl}(-\dot{H})^{1/2}(\gamma_{\lambda\sigma})_0^{\prime\prime}\sigma^2\partial^0\sigma
+\frac{1}{2}M_{\rm Pl}^2\dot{H}(\gamma_{\lambda\lambda})^{\prime\prime}_0\delta g^{00} \sigma^2
\Big]\,,
\end{align}
which is described in our framework as
\begin{align}
\nonumber
&\alpha_1=1\,,
\quad
\alpha_3=V_0^{\prime\prime}+M_{\rm Pl}^2\dot{H}(\gamma_{\lambda\lambda})^{\prime\prime}_0\,,
\quad
\alpha_4=-\sqrt{2}M_{\rm Pl}(-\dot{H})^{1/2}(\gamma_{\lambda\sigma})^\prime_0\,,
\quad
\beta_1=M_{\rm Pl}^2\dot{H}
(\gamma_{\lambda\lambda})_0^\prime\,,\\
\nonumber
&\beta_3=-\sqrt{2}M_{\rm Pl}(-\dot{H})^{1/2}(\gamma_{\lambda\sigma})_0\,,
\quad
\gamma_1=-\frac{1}{6}\left(V_0^{\prime\prime\prime}+M_{\rm Pl}\dot{H}(\gamma_{\lambda\lambda})_0^{\prime\prime\prime}\right)\,,
\quad
\gamma_2=-\frac{\sqrt{2}}{2}M_{\rm Pl}(-\dot{H})^{1/2}(\gamma_{\lambda\sigma})_0^{\prime\prime}\,,\\
&\overline{\gamma}_1=\frac{1}{2}M_{\rm Pl}^2\dot{H}(\gamma_{\lambda\lambda})_0^{\prime\prime}\,,
\quad
(\text{ others })=0\,.
\end{align}
The above action apparently includes terms linear in $\sigma$. However,
it can be easily shown that such terms turn out to be more than second
order after integration by parts.
We notice that this basis corresponds to the third normalization of
$\sigma$ in section~\ref{sub_quasi_uni}.

\section{Power spectrum}
\label{section_PS}
In this section we calculate the power spectrum for a class of
quasi-single field inflation models in our framework.  In the following,
we take the decoupling limit $\tilde{\epsilon}\to0$ and use the action
for the Goldstone boson $\pi$. It is also assumed that the background
trajectory satisfies $\epsilon\ll1$.  Up to the second order
fluctuations, the action for $\pi$ and $\sigma$ takes the following form
in the decoupling limit:
\begin{align}
\label{second_action_decoupled}
S&=\int d^4x \,a^3\Big[
\frac{\alpha_\pi^2}{2}
\Big(\dot{\pi}^2-c_{\pi}^2\frac{(\partial_i\pi)^2}{a^2}\Big)
+\frac{\alpha_\sigma^2}{2}\Big(\dot{\sigma}^2-c_\sigma^2\frac{(\partial_i\sigma)^2}{a^2}-m_\sigma^2\sigma^2\Big)
+\tilde{\beta}_1\dot{\pi}\sigma
+\tilde{\beta}_2\dot{\pi}\dot{\sigma}
+\tilde{\beta}_3c_\pi^2\frac{\partial_i\pi\partial_i\sigma}{a^2}
\Big]\,,
\end{align}
where we have dropped sub-leading terms $\sim M_{\rm
Pl}^2\dot{H}\pi^2$ and $\sim \dot{H}\pi\sigma$ in the regime
$\epsilon\ll1$, and have defined
\begin{align}
\label{second_parameters}
\alpha_\pi^2=-\frac{2M_{\rm Pl}^2\dot{H}}{c_\pi^2}\,,\quad
m_\sigma^2=\frac{\alpha_3-3H\alpha_4-\dot{\alpha}_4}{\alpha_\sigma^2}\,,
\quad
\tilde{\beta}_1=-2\beta_1+\dot{\beta}_3\,,
\quad
\tilde{\beta}_2=2\beta_2-\beta_3\,,
\quad
\tilde{\beta}_3=c_\pi^{-2}\beta_3\,.
\end{align}
The corresponding Hamiltonian
is given by
\begin{align}
\nonumber
H&=\int d^3x
\left[
\frac{1}{2a^3}\left(
\alpha_\pi^{-2}P_\pi^2+\alpha_\sigma^{-2}P_\sigma^2\right)
+\frac{a^3}{2}\Big(\alpha_\pi^2c_{\pi}^2\frac{(\partial_i\pi)^2}{a^2}
+\alpha_\sigma^2 c_\sigma^2\frac{(\partial_i\sigma)^2}{a^2}
+\alpha_\sigma^2m_\sigma^2\sigma^2\Big)
\right.\\
\label{H_before_int}
&\quad\qquad\qquad\left.
-\alpha_\pi^{-2}\tilde{\beta}_1P_\pi\,\sigma
-a^{-3}\alpha_\pi^{-2}\alpha_\sigma^{-2}\tilde{\beta}_2P_\pi P_\sigma-\tilde{\beta}_3a\,c_\pi^2\partial_i\pi\partial_i\sigma
+\frac{1}{2a^3}\alpha_\pi^{-4}\alpha_\sigma^{-2}\tilde{\beta}_2^2P_\pi^2
+\ldots\right]\,,
\end{align}
where $P_\pi$ and $P_\sigma$
are canonical momentum variables conjugate to
$\pi$ and $\sigma$, respectively.
In this paper, we assume that the mixing couplings can be treated as perturbations
and calculate the power spectrum up to the second order
in the mixing couplings $\tilde{\beta}_i$'s.
The dots in~(\ref{H_before_int}) stand
for term irrelevant to the power spectrum at this order,
and therefore, we drop them
in the following.

\medskip
Let us then calculate the power spectrum using the in-in formalism.
Before going to concrete models, we first introduce general expressions for
the power spectrum. 
Choosing the free part $H_{\rm free}$
and the interaction part $H_{\rm int}$
of the Hamiltonian as
\begin{align}
H&=H_{\rm free}+H_{\rm int}\,,\\
H_{\rm free}&
=\int d^3x\,\mathcal{H}_{\rm free}
=\int d^3x\left[
\frac{1}{2a^3}\left(
\alpha_\pi^{-2}P_\pi^2+\alpha_\sigma^{-2}P_\sigma^2\right)
+\frac{a^3}{2}\Big(\alpha_\pi^2c_{\pi}^2\frac{(\partial_i\pi)^2}{a^2}
+\alpha_\sigma^2 c_\sigma^2\frac{(\partial_i\sigma)^2}{a^2}
+\alpha_\sigma^2m_\sigma^2\sigma^2\Big)
\right]\,,\\
H_{\rm int}&
=\int d^3x\,\mathcal{H}_{\rm int}
=\int d^3x
\left[
-\alpha_\pi^{-2}\tilde{\beta}_1P_\pi\,\sigma
-a^{-3}\alpha_\pi^{-2}\alpha_\sigma^{-2}\tilde{\beta}_2P_\pi P_\sigma
-\tilde{\beta}_3a\,c_\pi^2\partial_i\pi\partial_i\sigma
+\frac{1}{2a^3}\alpha_\pi^{-4}\alpha_\sigma^{-2}\tilde{\beta}_2^2P_\pi^2\right]\,,
\end{align}
the dynamics of canonical variables in the interaction picture
are determined by
\begin{align}
\dot{\pi}=\frac{\partial \mathcal{H}_{\rm free}}{\partial P_\pi}
=a^{-3}\alpha^{-2}_\pi P_\pi\,,&
\quad
\dot{\sigma}=\frac{\partial \mathcal{H}_{\rm free}}{\partial P_\sigma}
=a^{-3}\alpha^{-2}_\sigma P_\sigma\,,\\
\dot{P}_\pi=-\frac{\partial \mathcal{H}_{\rm free}}{\partial \pi}=\alpha_\pi^2c_\pi^2a\partial_i^2\pi\,,&
\quad
\dot{P}_\sigma=-\frac{\partial \mathcal{H}_{\rm free}}{\partial \sigma}
=\alpha_\sigma^2(c_\sigma^2a\partial_i^2\sigma-a^3m_\sigma^2\sigma)\,.
\end{align}
The fields $\pi$ and
$\sigma$ are then expanded in the Fourier space as
\begin{align}
\pi_{\bf k}=u_{k}\, a_{\bf k}+u^\ast_{k}\, a^\dagger_{-{\bf k}}\,,
\quad
\sigma_{\bf k}=v_{k}\, b_{\bf k}+v^\ast_{k}\, b^\dagger_{-{\bf k}}
\end{align}
with the standard commutation relations
\begin{align}
\label{creation_annihilation}
[a_{\bf k},a^\dagger_{\bf k^\prime}]=(2\pi)^3\delta^{(3)}({\bf k}-{\bf k}^\prime)\,,
\quad
[b_{\bf k},b^\dagger_{\bf k^\prime}]=(2\pi)^3\delta^{(3)}({\bf k}-{\bf k}^\prime)\,.
\end{align}
Here the mode functions $u_{k}$ and $v_{k}$ satisfy the equations of
motion in the free theory and depend on $k=|{\bf k}|$:
\begin{align}
\label{eom_no_app}
\ddot{u}_{ k}+H(3-2\epsilon+\tilde{\eta})\dot{u}_{ k}+\frac{c_\pi^2\,k^2}{a^2}u_{ k}=0\,,
\quad
\ddot{v}_{ k}+H(3+2\,\delta_\alpha)\dot{v}_{k}+\Big(m_\sigma^2+\frac{c_{\sigma}^2\,k^2}{a^2}\Big)v_{ k}=0
\quad{\rm with}
\quad
\delta_\alpha=\frac{\dot{\alpha}_\sigma}{H\alpha_\sigma}\,.
\end{align}
Their normalization follows from
\begin{align}
\label{normalization_mode}
\alpha_\pi^2\,a^3\left(u_{k}\,\dot{u}_{k}^\ast-\dot{u}_{k}\,u^\ast_{k}\right)=
2M_{\rm Pl}^2H^2\tilde{\epsilon}\,a^3\left(u_{k}\,\dot{u}_{k}^\ast-\dot{u}_{k}\,u^\ast_{k}\right)=i\,,
\quad
\alpha_\sigma^2\,a^3\left(v_{k}\,\dot{v}_{k}^\ast-\dot{v}_{k}\,v^\ast_{k}\right)=i\,.
\end{align}
Using these expressions,
the Hamiltonian in the interaction picture
can be written as
\begin{align}
H_{\rm int}(t)=\int \frac{d^3{\bf k}}{(2\pi)^3}a^3(t)
\Big[-\tilde{\beta}_1\dot{\pi}_{\bf k}\,\sigma_{\bf -k}
-\tilde{\beta}_2\dot{\pi}_{\bf k}\,\dot{\sigma}_{\bf -k}
-\tilde{\beta}_3c_\pi^2\frac{k^2}{a^2}\pi_{\bf k}\,\sigma_{\bf -k}
+\frac{1}{2\alpha_\sigma^2}\tilde{\beta}_2^2\dot{\pi}_{\bf k}\,\dot{\pi}_{\bf -k}
\Big](t)\,.
\end{align}
Then,
the expectation value of $\pi_{\bf k}(t)\pi_{\bf k^\prime}(t)$
is calculated as
\begin{align}
\nonumber
\langle\pi_{\bf k}(t)\pi_{\bf k^\prime}(t)\rangle
&=\langle0|
\left[\bar{T}\exp\Big(i\int_{t_0}^tdt^\prime H_{\rm int}(t^\prime)\Big)\right]
\pi_{\bf k}(t)\pi_{\bf k^\prime}(t)
\left[T\exp\Big(-i\int_{t_0}^tdt^\prime H_{\rm int}(t^\prime)\Big)\right]
|0\rangle\\
\nonumber
&=\langle0|
\pi_{\bf k}(t)\pi_{\bf k^\prime}(t)
|0\rangle
-2{\rm Re}\left[i\int_{t_0}^tdt_1\langle0|
\pi_{\bf k}(t)\pi_{\bf k^\prime}(t)
H_{\rm int}(t_1)|0\rangle\right]\\
\nonumber
&\quad+\int_{t_0}^td\tilde{t}_1\int_{t_0}^tdt_1\langle0|
 H_{\rm int}(\tilde{t}_1)
\pi_{\bf k}(t)\pi_{\bf k^\prime}(t)
H_{\rm int}(t_1)
|0\rangle\\
\label{ps_H}
&\quad-2{\rm Re}\left[\int_{t_0}^tdt_1\int_{t_0}^{t_1}dt_2\langle0|
\pi_{\bf k}(t)\pi_{\bf k^\prime}(t)
H_{\rm int}(t_1)H_{\rm int}(t_2)|0\rangle\right]
+\ldots\,,
\end{align}
where the dots stand for the higher order terms
in the couplings.
In terms of the mode functions and the couplings,
the general form of the two point function~(\ref{ps_H})
is given up to the leading order corrections from the mixing couplings~by
\begin{align}
\langle\pi_{\bf k}(t)\pi_{\bf k^\prime}(t)\rangle&=(2\pi)^3\delta^{(3)}({\bf k}+{\bf k}^\prime)u^\ast_{k}(t)u_{k}(t)
\Big[1+\mathcal{C}_1+\mathcal{C}_2+\mathcal{C}_3\Big]\,,
\end{align}
where $\mathcal{C}_1$, $\mathcal{C}_2$, and $\mathcal{C}_3$ are defined by
\begin{align}
\label{C_1_general}
\mathcal{C}_1&=2\left|\int_{t_0}^tdt_1\,a^3\Big[
\tilde{\beta}_1\,\dot{u}_kv_k+
\tilde{\beta}_2\,\dot{u}_k\dot{v}_k+
\tilde{\beta}_3c_\pi^2\frac{k^2}{a^2}\,u_kv_k
\Big](t_1)\right|^2\,,\\
\nonumber
\mathcal{C}_2&=-4{\rm Re}\Bigg[\frac{u_k^2(t)}{|u_k(t)|^2}
\int_{t_0}^tdt_1\,a^3\Big[
\tilde{\beta}_1\,\dot{u}^\ast_kv_k+
\tilde{\beta}_2\,\dot{u}^\ast_k\dot{v}_k+
\tilde{\beta}_3c_\pi^2\frac{k^2}{a^2}\,u^\ast_kv_k
\Big](t_1)\\
\label{C_2_general}
&\qquad\qquad\qquad\qquad\qquad\times
\int_{t_0}^{t_1}dt_2
\,a^3\Big[
\tilde{\beta}_1\,\dot{u}^\ast_kv^\ast_k+
\tilde{\beta}_2\,\dot{u}^\ast_k\dot{v}^\ast_k+
\tilde{\beta}_3c_\pi^2\frac{k^2}{a^2}\,u^\ast_kv^\ast_k
\Big](t_2)
\Bigg]\,,\\
\mathcal{C}_3&=2{\rm Re}\Bigg[
-i\frac{u_k^2(t)}{|u_k(t)|^2}\int_{t_0}^tdt_1\,a^3\,
\alpha_\sigma^{-2}\tilde{\beta}_2^2\,\dot{u}_k^\ast{}^2
(t_1)\Bigg]\,.
\end{align}
It is also convenient to rewrite
$\mathcal{C}=\mathcal{C}_1+\mathcal{C}_2+\mathcal{C}_3$ as follows:
\begin{align}
\nonumber
\mathcal{C}&=4{\rm Re}\Bigg[
\int_{t_0}^tdt_1\,a^3\left[
\Big(\tilde{\beta}_1\,\dot{u}_kv_k+
\tilde{\beta}_2\,\dot{u}_k\dot{v}_k+
\tilde{\beta}_3c_\pi^2\frac{k^2}{a^2}\,u_kv_k\Big)-
\frac{u_k^2(t)}{|u_k(t)|^2}\Big(\tilde{\beta}_1\,\dot{u}^\ast_kv_k+
\tilde{\beta}_2\,\dot{u}^\ast_k\dot{v}_k+
\tilde{\beta}_3c_\pi^2\frac{k^2}{a^2}\,u^\ast_kv_k\Big)
\right](t_1)\\
\nonumber
&\qquad\qquad\qquad\qquad\times
\int_{t_0}^{t_1}dt_2
\,a^3\Big[
\tilde{\beta}_1\,\dot{u}^\ast_kv^\ast_k+
\tilde{\beta}_2\,\dot{u}^\ast_k\dot{v}^\ast_k+
\tilde{\beta}_3c_\pi^2\frac{k^2}{a^2}\,u^\ast_kv^\ast_k
\Big](t_2)
\Bigg]\\
\label{C_general}
&\quad+2{\rm Re}\Bigg[
-i\frac{u_k^2(t)}{|u_k(t)|^2}\int_{t_0}^tdt_1\,a^3\,
\alpha_\sigma^{-2}\tilde{\beta}_2^2\,\dot{u}_k^\ast{}^2
(t_1)\Bigg]\,.
\end{align}
Since the scalar perturbation
$\zeta$ is given at the linear order by $\zeta=-H\pi$,
we obtain
the expectation value of $\zeta_{\bf k}(t)\zeta_{\bf k^\prime}(t)$~as
\begin{align}
\langle\zeta_{\bf k}(t)\zeta_{\bf k^\prime}(t)\rangle&=(2\pi)^3\delta^{(3)}
({\bf k}+{\bf k}^\prime)
\frac{2\pi^2}{k^3}
\mathcal{P}_\zeta(k)\,,
\end{align}
where the power spectrum $\mathcal{P}_\zeta(k)$ is given by
\begin{align}
\label{C_to_P}
\mathcal{P}_\zeta(k)&=\frac{H^2(t)k^3}{2\pi^2}u^\ast_{\bf k}(t)u_{\bf k}(t)
\left(1+\mathcal{C}\right)\,.
\end{align}
The factor $\mathcal{C}$ can be considered as a deviation factor from single field inflation.
Here it should be noticed that in the derivation of the above general expression we
assumed only $\epsilon\ll1$, $\tilde{\epsilon}\ll1$, and the
perturbativity of the mixing couplings.
In principle, we can calculate
the power spectrum for any models satisfying those three conditions.
Note that, as pointed out in~\cite{Chen:2009we}, the perturbativity of
mixing couplings is justified
even in the case when they are large only
in a sufficiently short period of time,
which is realized for example
by the sudden turning background
trajectory~\cite{Achucarro:2010da,Shiu:2011qw,Cespedes:2012hu,Achucarro:2012sm,Gao:2012uq,Saito:2012pd}.
In
the rest of this section, we first perform the calculation concretely
for the case when the time-dependence of mixing couplings is
irrelevant (constant turning trajectory), and then we discuss the
qualitative features of the case when the mixing couplings are large in
a sufficiently short period of time (sudden turning trajectory).

\subsection{Constant turning trajectory}

In this subsection, the power spectrum is calculated in the case that
the time-dependence of mixing couplings is irrelevant.  To make
the calculation tractable, we assume that the time-dependence of
$\epsilon$, $\tilde{\epsilon}$, $\alpha_\sigma$, $c_\sigma$, and
$m_\sigma$ is negligible and we use the de-Sitter approximation.  In
this approximation, the equations of motion~(\ref{eom_no_app}) for the
mode functions $u_k$ and $v_k$ can be written as
\begin{align}
\label{eom_app}
u_{k}^{\prime\prime}-\frac{2}{\tau}u_{k}^\prime+c_\pi^2\,k^2u_{ k}=0\,,
\qquad
v_{k}^{\prime\prime}-\frac{2}{\tau}v_{k}^\prime+c_{\sigma}^2\,k^2v_{ k}+\frac{m_\sigma^2}{H^2\tau^{2}}v_k=0\,,
\end{align}
where the conformal time $d\tau=a^{-1}dt$
is given by $\tau=-1/(aH)$ in the de-Sitter approximation
and the primes denote derivatives with respect to $\tau$.
The equations~(\ref{eom_app}) can be solved as follows:
\begin{align}
u_k&=\frac{1}{2M_{\rm Pl}\tilde{\epsilon}^{1/2}(c_\pi k)^{3/2}}(1+ic_\pi k\tau)e^{-ic_\pi k\tau}
=\frac{1}{2M_{\rm Pl}\tilde{\epsilon}^{1/2}(c_\pi k)^{3/2}}(1-i x)e^{ ix}\,,\\
v_k&=-ie^{\frac{i}{2}\pi\nu+\frac{i}{4}\pi}\frac{\sqrt{\pi}H}{2\alpha_\sigma}
(-\tau)^{3/2}H_\nu^{(1)}(-c_\sigma k\tau)
=-ie^{\frac{i}{2}\pi\nu+\frac{i}{4}\pi}\frac{\sqrt{\pi}H}{2\alpha_\sigma (c_\pi k)^{3/2}}
x^{3/2}H_\nu^{(1)}(r_s x)\,,
\end{align}
where $x=-c_\pi k\tau$, $r_s=c_\sigma/c_\pi$, and we chose the Bunch-Davies vacuum
for $\pi$ and $\sigma$.
The function
$H_\nu^{(1)}=J_\nu+iY_\nu$ is the Hankel function
and $\nu$ is defined as
\begin{align}
\nu=\sqrt{\frac{9}{4}-\frac{m_\sigma^2}{H^2}}\quad{\rm for}
\quad m_\sigma<\frac{3}{2}H\,,
\quad\quad
\nu=i\sqrt{\frac{m_\sigma^2}{H^2}-\frac{9}{4}}\quad{\rm for}
\quad m_\sigma>\frac{3}{2}H\,.
\end{align}
The time derivatives of $u_k$ and $v_k$
are given by
\begin{align}
\dot{u}_k&=-H\tau u_k^\prime=-\frac{H}{2M_{\rm Pl}\tilde{\epsilon}^{1/2}(c_\pi k)^{3/2}} x^2e^{i x}\,,\\
\dot{v}_k&=-H\tau v_k^\prime=
ie^{\frac{i}{2}\pi\nu+\frac{i}{4}\pi}\frac{\sqrt{\pi}H^2}{2\alpha_\sigma (c_\pi k)^{3/2}}
x^{3/2}
\Big((3/2-\nu)H_\nu^{(1)}(r_s x)+
(r_s x)H_{\nu-1}^{(1)}(r_s x)\Big)\,,
\end{align}
where we used the identity $z\partial_zH_\nu^{(1)}=zH_{\nu-1}^{(1)}-\nu H_\nu^{(1)}$.
Therefore,
the factor $\mathcal{C}$ defined in (\ref{C_general})
takes the form
\begin{align}
\nonumber
\mathcal{C}&=e^{\frac{i}{2}\pi(\nu-\nu^\ast)}\frac{\pi}{4\alpha^2_\sigma }\frac{c_\pi^2 }{M_{\rm Pl}^2(-\dot{H})}\\
\nonumber
&\quad\times{\rm Re}\Bigg[
\int_0^\infty dx_1\Big[
\Big(\frac{\tilde{\beta}_1}{H}+(\nu-3/2)^\ast\tilde{\beta}_2-\tilde{\beta}_3\Big)\Big(A_1(x)^\ast-\widetilde{A}_1(x_1)\Big)-\tilde{\beta}_2
\Big(A_2(x)^\ast-\widetilde{A}_2(x_1)\Big)
+\tilde{\beta}_3\Big(A_3(x)^\ast-\widetilde{A}_3(x_1)\Big)
\Big]\\
\nonumber
&\qquad\qquad\qquad\qquad\qquad\qquad\qquad\quad
\times\int_{x_1}^\infty dx_2\Big[\Big(\frac{\tilde{\beta}_1}{H}+(\nu-3/2)\tilde{\beta}_2-\tilde{\beta}_3\Big)A_1(x_2)-\tilde{\beta}_2A_2(x_2)+\tilde{\beta}_3A_3(x_2)\Big]
\Bigg]\\
\label{before_const_beta}
&\quad-\frac{c_\pi^2}{2M_{\rm Pl}^2(-\dot{H})}{\rm Re}\left[i\int_0^\infty dx\,\alpha_\sigma^{-2}\tilde{\beta}_2^2e^{-2ix}\right]\,,
\end{align}
where we set $t_0=-\infty$ and $t=\infty$, 
and we defined
\begin{align}
\nonumber
&A_1(x)=x^{-1/2}e^{ i x}H_\nu^{(1)}(r_s x)\,,
\quad
A_2(x)= r_sx^{1/2}e^{ i x}H_{\nu-1}^{(1)}(r_s x)\,,
\quad
A_3(x)= i x^{1/2}e^{ i x}
H_\nu^{(1)}(r_s x)\,,\\
&\widetilde{A}_1(x)=x^{-1/2}e^{ i x}H_{\nu^\ast}^{(2)}(r_s x)\,,
\quad
\widetilde{A}_2(x)=r_sx^{1/2}e^{ i x}H_{\nu^\ast-1}^{(2)}(r_s x)\,,
\quad
\widetilde{A}_3(x)=ix^{1/2}e^{ i x}H_{\nu^\ast}^{(2)}(r_s x)\,.
\end{align}
When the time-dependence of mixing couplings is irrelevant,
mixing couplings $\tilde{\beta}$'s
can be evaluated
at the time of
horizon-crossing $\tau=-(c_\pi k)^{-1}$.  In such a case,
$\mathcal{C}$ is given by
\begin{align}
\nonumber
\mathcal{C}&=\frac{c_\pi^2 }{\alpha^2_\sigma M_{\rm Pl}^2(-\dot{H})}
{\rm Re}\Bigg[\Big|\tilde{\beta}_1/H+(\nu-3/2)\tilde{\beta}_2-\tilde{\beta}_3\Big|^2\mathcal{I}_{11}
+(\tilde{\beta}_1/H+(\nu^\ast-3/2)\tilde{\beta}_2-\tilde{\beta}_3)(-\tilde{\beta}_2\,
\mathcal{I}_{12}+\tilde{\beta}_3\,\mathcal{I}_{13})\\
\nonumber
&\quad\qquad\qquad\qquad\qquad\qquad\qquad-\tilde{\beta}_2\Big((\tilde{\beta}_1/H+(\nu-3/2)\tilde{\beta}_2-\tilde{\beta}_3)\mathcal{I}_{21}-\tilde{\beta}_2\,
\mathcal{I}_{22}+\tilde{\beta}_3\,\mathcal{I}_{23}\Big)\\
\label{C_almost_const_simple}
&\quad\qquad\qquad\qquad\qquad\qquad\qquad+\tilde{\beta}_3\Big((\tilde{\beta}_1/H+(\nu-3/2)\tilde{\beta}_2-\tilde{\beta}_3)\mathcal{I}_{31}-\tilde{\beta}_2\,
\mathcal{I}_{32}+\tilde{\beta}_3\,\mathcal{I}_{33}\Big)
-\frac{1}{4}\tilde{\beta}_2^2\Bigg]\,,
\end{align}
where
$\mathcal{I}_{ij}$'s are integrals defined by
\begin{align}
\label{I_ij}
\mathcal{I}_{ij}=\frac{\pi}{4}e^{\frac{i}{2}\pi(\nu-\nu^\ast)}
\int_0^\infty dx_1 \Big(A_i(x_1)^\ast-\widetilde{A}_i(x_1)\Big)
\int_{x_1}^\infty dx_2\,A_j(x_2)\,.
\end{align}
The last term in~(\ref{before_const_beta})
was calculated using the $i\epsilon$-prescription as
\begin{align}
-\frac{c_\pi^2}{2M_{\rm Pl}^2(-\dot{H})}{\rm Re}\left[i\int_0^\infty dx\,\alpha_\sigma^{-2}\tilde{\beta}_2^2e^{-2ix}\right]
=\frac{1}{4}\frac{c_\pi^2}{\alpha_\sigma^2M_{\rm Pl}^2(-\dot{H})}\tilde{\beta}_2^2{\rm Re}\left[e^{-2ix}\right]_0^\infty=-\frac{1}{4}\frac{c_\pi^2}{\alpha_\sigma^2M_{\rm Pl}^2(-\dot{H})}\tilde{\beta}_2^2\,.
\end{align}
We then have
\begin{align}
\mathcal{C}&=\frac{c_\pi^2 }{\alpha^2_\sigma M_{\rm Pl}^2(-\dot{H})}
\left(
\frac{\tilde{\beta}_1^2}{H^2}\,\mathcal{C}_{11}
+\tilde{\beta}_2^2\,\mathcal{C}_{22}
+\tilde{\beta}_3^2\,\mathcal{C}_{33}
+\frac{\tilde{\beta}_1}{H}\,\tilde{\beta}_2\,\mathcal{C}_{12}
+\frac{\tilde{\beta}_1}{H}\,\tilde{\beta}_3\,\mathcal{C}_{13}
+\tilde{\beta}_2\,\tilde{\beta}_3\,\mathcal{C}_{23}\right)\,,
\end{align}
and $\mathcal{C}_{ij}$'s are given by
\begin{align}
\nonumber
&\mathcal{C}_{11}={\rm Re}\Big[\,\mathcal{I}_{11}\,\Big]\,,\quad
\mathcal{C}_{22}={\rm Re}
\left[|\nu-3/2|^2\,\mathcal{I}_{11}-(\nu^\ast-3/2)\,\mathcal{I}_{12}-(\nu-3/2)\,\mathcal{I}_{21}+\mathcal{I}_{22}
-\frac{1}{4}
\right]\,,\\
\nonumber
&\mathcal{C}_{33}={\rm Re}\left[\,\mathcal{I}_{11}+\mathcal{I}_{33}-\mathcal{I}_{13}-\mathcal{I}_{31}\,\right]\,,
\quad
\mathcal{C}_{12}={\rm Re}\left[(\nu+\nu^\ast-3)\,\mathcal{I}_{11}-\mathcal{I}_{12}-\mathcal{I}_{21}\right]\,,
\quad\mathcal{C}_{13}={\rm Re}\left[-2\,\mathcal{I}_{11}+\mathcal{I}_{13}+\mathcal{I}_{31}\right]\,,
\\[1mm]
\label{C_ij_def}
&\mathcal{C}_{23}={\rm Re}\left[-(\nu+\nu^\ast-3)\,\mathcal{I}_{11}+\mathcal{I}_{12}+\mathcal{I}_{21}+(\nu^\ast-3/2)\,\mathcal{I}_{13}
+(\nu-3/2)\,\mathcal{I}_{31}-\mathcal{I}_{23}-\mathcal{I}_{32}\right]\,.
\end{align}
Here the explicit form of the power spectrum is
\begin{align}
\mathcal{P}_\zeta(k)&=\frac{H^2}{8\pi^2M^2_{\rm Pl}\epsilon\,c_\pi }\left[1+\frac{c_\pi^2 }{\alpha^2_\sigma M_{\rm Pl}^2(-\dot{H})}
\left(
\frac{\tilde{\beta}_1^2}{H^2}\,\mathcal{C}_{11}
+\tilde{\beta}_2^2\,\mathcal{C}_{22}
+\tilde{\beta}_3^2\,\mathcal{C}_{33}
+\frac{\tilde{\beta}_1}{H}\,\tilde{\beta}_2\,\mathcal{C}_{12}
+\frac{\tilde{\beta}_1}{H}\,\tilde{\beta}_3\,\mathcal{C}_{13}
+\tilde{\beta}_2\,\tilde{\beta}_3\,\mathcal{C}_{23}\right)\right]\,,
\end{align}
where $H$, $\epsilon$, $c_\pi$, $\alpha_\sigma$, $\tilde{\beta}_i$, and $r_s$
are evaluated at the time of horizon-crossing.
Then,
the calculation of the power spectrum
reduces to the evaluation
of
${\rm Re}[\,\mathcal{I}_{ij}+\mathcal{I}_{ji}]$,
${\rm Im}[\,\mathcal{I}_{12}-\mathcal{I}_{21}]$,
and
${\rm Im}[\,\mathcal{I}_{13}-\mathcal{I}_{31}]$.

\begin{figure}[t]
	  		\begin{center}
   				\includegraphics[width=140mm]{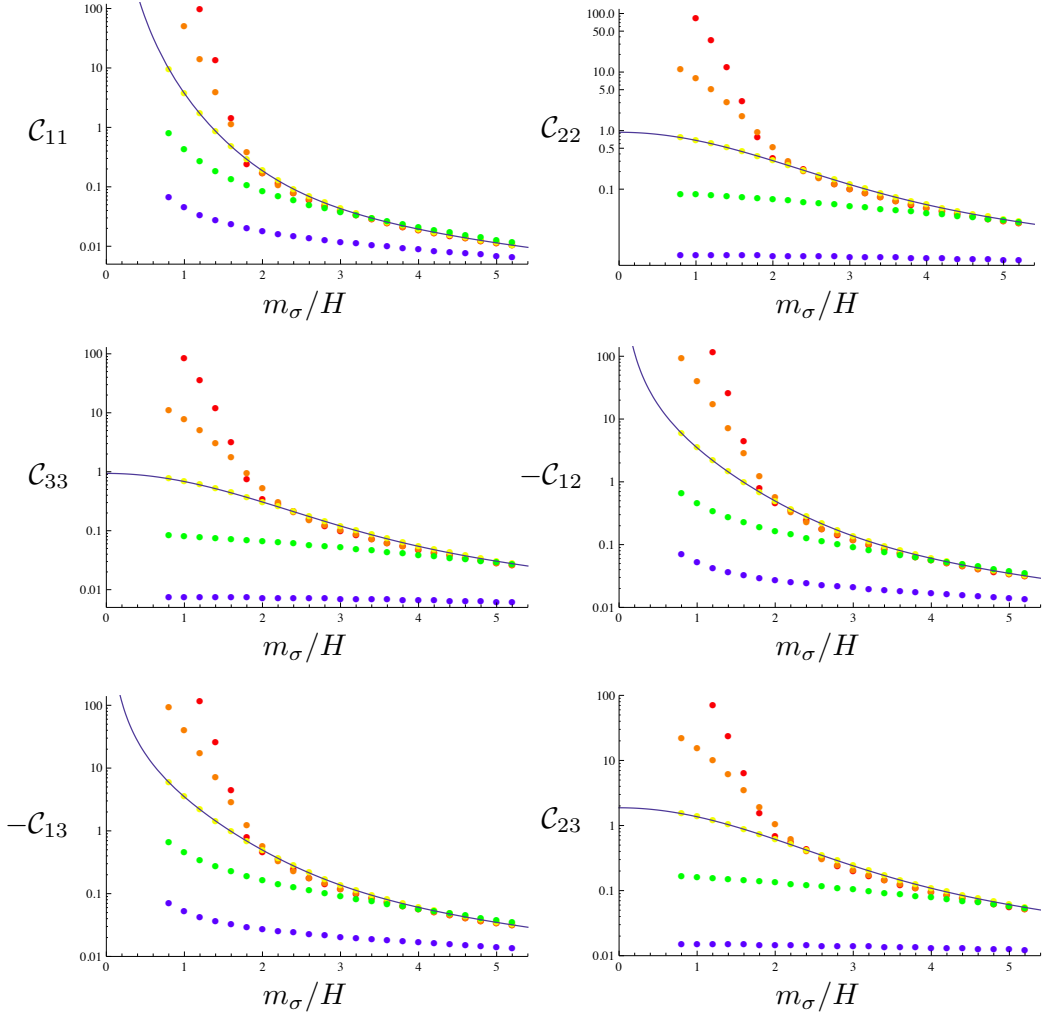}
  			\end{center}
			\vspace{-5mm}
  				\caption{$\mathcal{C}_{ij}$'s for fixed $r_s=c_\sigma/c_\pi$.
				The dots are numerical results for $r_s=0.1$ (red), $0.3$ (orange), $1$ (yellow), $3$ (green), and $10$ (blue). The curves are analytic results for $r_s=1$ obtained in the next subsection.
								}
  		 	\label{fig:PS_m}
	\end{figure}

\begin{figure}[t]
	  		\begin{center}
   				\includegraphics[width=140mm]{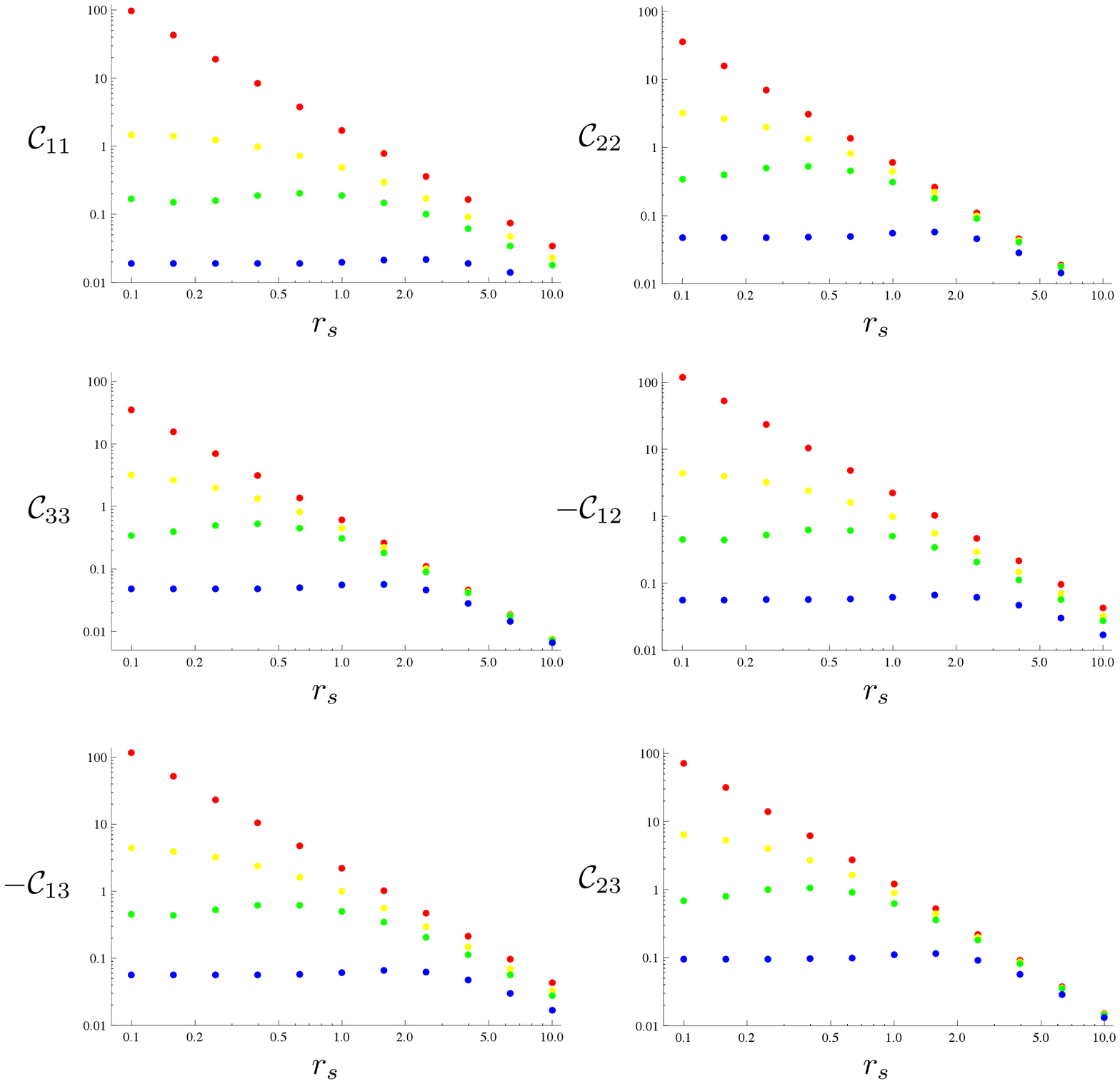}
  			\end{center}
			\vspace{-5mm}
  				\caption{$\mathcal{C}_{ij}$'s for fixed $m_\sigma$.
				The dots are numerical results for $m_\sigma/H=1.2$ (red), $1.6$ (yellow), $2.0$ (green), and $4.0$ (blue).
								}
  		 	\label{fig:PS_r}
	\end{figure}

\medskip
As is understood from the definition,
$\mathcal{I}_{ij}$'s and $\mathcal{C}_{ij}$'s are
functions of $m_\sigma$ and $r_s=c_\sigma/c_\pi$.
For general value of $r_s$,
it is difficult to perform the integrals analytically
and we performed numerical calculations by contour deformations (see appendix~\ref{app_numerical} for details).
For the special case $r_s=1$,
however, it is possible to perform the integrals $\mathcal{I}_{ij}$'s analytically
by extending the results in~\cite{Chen:2012ge},
and the results are summarized in appendix~\ref{app_analytic}.
In such a way,
$\mathcal{C}_{ij}$'s are calculated
and the obtained results are summarized in figure~\ref{fig:PS_m} and figure~\ref{fig:PS_r}.
We find that they monotonically decrease in $m_\sigma$ for fixed $r_s$,
but they are not monotonic for fixed $m_\sigma$.
As is discussed in section~\ref{subsubsection_heavy},
the effects of mixing interactions appear in the form
of $\beta_i^2/m_\sigma^2$
in the heavy mass limit
and it is implied that $\mathcal{C}_{ij}\sim1/m_\sigma^2$ for large $m_\sigma$,
which is consistent with our results in this subsection.
Therefore,
the power spectrum is not affected by heavy particles
unless the mixing couplings are comparable to the mass of heavy particles~\cite{Cremonini:2010ua,Achucarro:2010da,Shiu:2011qw,Cespedes:2012hu,Achucarro:2012sm,Pi:2012gf,Gao:2012uq,Saito:2012pd,Gwyn:2012mw}.

\subsection{Qualitative features of sudden turning trajectory}

\begin{figure}[t]
\begin{center}
 \includegraphics[width=150mm]{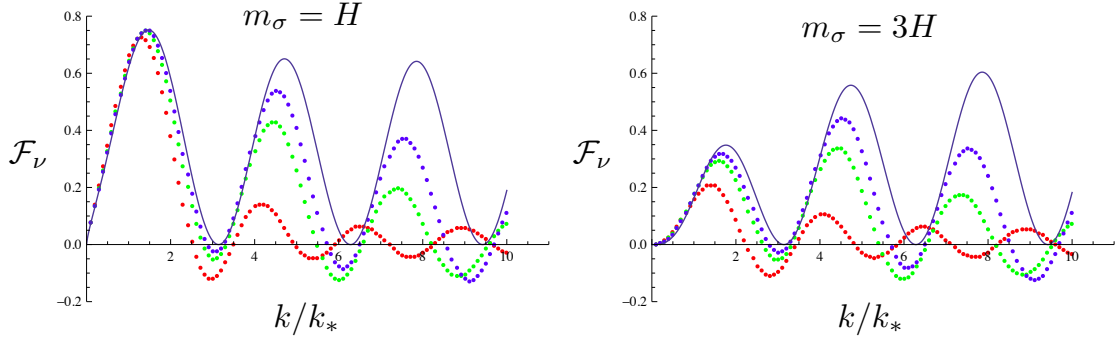}
\end{center}
\vspace{-5mm}
\caption{$\mathcal{F}_\nu$ for fixed $c_\pi k_\ast\Delta\tau$ and
$c_\pi=c_\sigma$}.  The left figure is for $m_\sigma=H$ and the
right one is for $m_\sigma=3H$.  The dots are numerical results for
$c_\pi k_\ast\Delta\tau=5$ (blue), $10$ (green), and $15$ (red). The
curves are analytic results in the limit of $c_\pi k_\ast\Delta\tau\to0$
(delta function limit).   \label{fig:FNx}
\end{figure}

In this subsection, we discuss qualitative features of the
case when the mixing couplings are large in a sufficiently short period
of time (sudden turning trajectory).  As a simplest example, let us first
consider
the case when the $\tilde{\beta}_1$ coupling
is the only relevant mixing coupling and it is proportional to a delta function:
\begin{align}
\label{delta_coupling}
\tilde{\beta}_1=\beta_{\ast }\,\delta(t-t_\ast)=\beta_{\ast }a^{-1}(t_\ast)\,\delta(\tau-\tau_\ast)\,,
\quad
\tilde{\beta}_2=\tilde{\beta}_3=0\,,
\end{align}
where $t_\ast$ is the time of sudden turning,
$\tau_\ast$ is the corresponding conformal time,
and the mode $k_\ast$ crossing the horizon at $t=t_\ast$
is given by $k_\ast=-1/(c_\pi\tau_\ast)$.
We assume that $\beta_\ast$ is small enough to be treated perturbatively.
In this case,
the deviation of the power spectrum from that of single field inflation
can be written as\footnote{
In our calculation
we have treated the mixing coupling $\beta_\ast$
as an interaction.
In such an interaction picture,
it is manifest that the deviation
from single field inflation
starts from the second order in $\beta_\ast$.
On the other hand,
it is also possible to treat
the mixing coupling
as a part of the kinetic and mass terms.
In that picture,
the commutation relations~(\ref{creation_annihilation})
of creation and annihilation operators
are affected by the mixing
as well as the mode functions $u_k$ and $v_k$
are modified.
However,
in some literatures,
these modifications
are not taken into
account adequately
and
the deviation from single field inflation
is calculated to start from the first order in~$\beta_\ast$.}
\begin{align}
\label{sharp_C1C2}
\mathcal{C}&=
2\beta_\ast^2\,{\rm Re}\,\Big[a^6\,(\dot{u}_k^\ast-\dot{u}_k)\,\dot{u}_k\,v_k \,v_k^\ast(t_\ast)\Big]\,,
\end{align}
which is related to the power spectrum by (\ref{C_to_P}).
For simplicity,
suppose that time-dependence of $m_\sigma$, $\alpha_\sigma$, $c_\sigma$, and $c_\pi$
is negligible at the time of sudden turning
and the mode functions $u_k(t_\ast)$ and $v_k(t_\ast)$ are given by
\begin{align}
u_k(t_\ast)&=\frac{1}{2M_{\rm Pl}\tilde{\epsilon}^{1/2}(c_\pi k)^{3/2}}(1-i x_\ast)e^{ ix_\ast}\,,\\
v_k(t_\ast)&
=-ie^{\frac{i}{2}\pi\nu+\frac{i}{4}\pi}\frac{\sqrt{\pi}H}{2\alpha_\sigma (c_\pi k)^{3/2}}
x_\ast^{3/2}H_\nu^{(1)}(r_s x_\ast)\,,
\end{align}
where $x_\ast=k/k_\ast$ 
and parameters such as $m_\sigma$ are evaluated at the time of
sudden turning.
Then, (\ref{sharp_C1C2}) is given by
\begin{align}
\mathcal{C}&=\beta_\ast^2\,\frac{\pi}{4\alpha_\sigma^2}\frac{c_\pi^2}{M_{\rm Pl}^2(-\dot{H})}
\mathcal{F}_\nu(x_\ast)
\quad
{\rm with}
\quad
\mathcal{F}_\nu(x_\ast)=e^{\frac{i}{2}\pi(\nu-\nu^\ast)}x_\ast\sin^2x_\ast|H_\nu^{(1)}(r_s x_\ast)|^2\,.
\end{align}
For $k\ll k_\ast$ or $x_\ast\ll1$,
the asymptotic behavior of $\mathcal{F}_\nu(x_\ast)$ is
\begin{align}
\mathcal{F}_\nu(x_\ast)&\sim
\left\{\begin{array}{ll}\displaystyle\frac{4^\nu}{\pi^2}\Gamma(\nu)^2\,
r_s^{-2\nu}x_\ast^{3-2\nu} &\displaystyle \quad{\rm for}\quad m_\sigma(t_\ast)<\frac{3}{2}H\,, \\[4mm]
\displaystyle \frac{1}{\pi^2}
\left|e^{\frac{i}{2}\pi \nu}2^\nu\Gamma(\nu)(r_sx_\ast)^{-\nu}+e^{-\frac{i}{2}\pi \nu}2^{-\nu}\Gamma(-\nu)(r_sx_\ast)^{\nu}\right|^2
x_\ast^{3} &\displaystyle\quad {\rm for}\quad m_\sigma(t_\ast)>\frac{3}{2}H\,,
\end{array}\right.
\end{align}
which is consistent to the intuition that modes outside the horizon at
$t=t_\ast$ are not much affected by the sudden turning.  For $ k\gg k_\ast$ or
$x_\ast\gg1$, it reduces to
\begin{align}
\mathcal{F}_\nu(x_\ast)&\sim \frac{2}{\pi}r_s^{-1}\sin^2 x_\ast=\frac{1}{\pi}r_s^{-1}(1-\cos 2 x_\ast)\,.
\end{align}
This kind of oscillating behavior was also found in~\cite{Shiu:2011qw,Cespedes:2012hu,Achucarro:2012sm,Gao:2012uq}.
The turning trajectory generically oscillates
around the turning point
and the mixing couplings at the turning point
become more regular than delta functions.
In such a case, it is expected that the oscillating behavior
of short modes $c_\pi k\gtrsim -1/\tau_\ast$
begins damping at some scale characterized by the oscillation of the trajectory.
Let us next confirm such a behavior explicitly for a concrete
example with a finite width of turning.
We consider the following  $\tilde{\beta}_1$ profile:
\begin{align}
\tilde{\beta}_1=\left\{\begin{array}{cc}\beta_\ast(-H\tau_\ast)(\Delta \tau)^{-1}
&\quad{\rm for}\quad \tau_\ast-\frac{\Delta\tau}{2}<\tau<\tau_\ast+\frac{\Delta\tau}{2}\,,
\\[2mm]
0&{\rm otherwise}\,,\end{array}\right.
\end{align}
where we normalized $\beta_\ast$ so that it reproduces the coupling
(\ref{delta_coupling}) in the limit $\Delta\tau\to0$.  For this class of
couplings, we define $\mathcal{F}_\nu$ by
\begin{align}
\mathcal{C}&=\beta_\ast^2\,\frac{\pi}{4\alpha_\sigma^2}\frac{c_\pi^2}{M_{\rm Pl}^2(-\dot{H})}
\mathcal{F}_\nu(x_\ast)\,.
\end{align}
Since it is difficult to calculate $\mathcal{F}_\nu$ analytically,
we can confirm the expected damping behavior of $\mathcal{F}_\nu$ by
numerical calculations and the results are given in
figure~\ref{fig:FNx}.
To summarize,
$\mathcal{C}$ vanishes in the long mode limit $c_\pi k\ll -1/\tau_\ast$, oscillates for
short modes $c_\pi k\gtrsim -1/\tau_\ast$, and damps at some scale
characterized by the oscillating trajectory around the turning point.

\medskip 
It is straightforward to extend the above discussions to the case with
non-vanishing $\tilde{\beta}_2$ and $\tilde{\beta}_3$.
We
generically expect to find a similar behavior of $\mathcal{C}$:
vanishing in the long mode limit $c_\pi k\ll -1/\tau_\ast$, oscillating for
short modes $c_\pi k\gtrsim -1/\tau_\ast$, and damping at some scale
characterized by the oscillating trajectory around the turning point.

\section{Three point functions in the squeezed limit}
\label{section_squeezed}

In this section, we discuss the momentum dependence of three point
functions in the squeezed limit.  We take the decoupling limit and
assume that time dependence of $\alpha$'s, $\beta$'s, $\gamma$'s, and
$\dot{H}$ is negligible. Under these assumptions, the second order action
is given in~(\ref{second_action_decoupled}) and
(\ref{second_parameters}), and three point vertices are given by
\begin{align}
\nonumber
S^{(3)}&=\int d^4x\, a^3\Bigg[-\Big(M_{\rm Pl}^2\dot{H}(c_\pi^{-2}-1)+\frac{4M_3^4}{3}\Big)\dot{\pi}^3
+M_{\rm Pl}^2\dot{H}(c_\pi^{-2}-1)\dot{\pi}\frac{(\partial_i\pi)^2}{a^2}
+(-\beta_1+4\overline{\gamma}_6)\dot{\pi}^2\sigma
+(3\beta_2-4\overline{\gamma}_7)\dot{\pi}^2\dot{\sigma}
\\
\nonumber
&\qquad\qquad\qquad
-2\beta_2\dot{\pi}\frac{\partial_i\pi\partial_i\sigma}{a^2}
+\beta_1\frac{(\partial_i\pi)^2}{a^2}\sigma
-\beta_2\frac{(\partial_i\pi)^2}{a^2}\dot{\sigma}
-2\overline{\gamma}_1\dot{\pi}\sigma^2
+(-\alpha_4+2\overline{\gamma}_2)\dot{\pi}\sigma\dot{\sigma}
\\
\nonumber
&\qquad\qquad\qquad
+\Big(\alpha_\sigma^2(1-c_\sigma^2)-2\overline{\gamma}_3+2\overline{\gamma}_5\Big)\dot{\pi}\dot{\sigma}^2-2\overline{\gamma}_5\dot{\pi}\frac{(\partial_i\sigma)^2}{a^2}
+\alpha_4\frac{\partial_i\pi\partial_i\sigma}{a^2}\sigma
-\alpha_\sigma^2(1-c_\sigma^2)\frac{\partial_i\pi\partial_i\sigma}{a^2}\dot{\sigma}
-2\overline{\gamma}_4\ddot{\pi}\sigma\dot{\sigma}\\
\label{three_point_vertices}
&\qquad\qquad\qquad
+\gamma_1\sigma^3
-\gamma_2\sigma^2\dot{\sigma}
+(\gamma_3-\gamma_5)\sigma\dot{\sigma}^2
+\gamma_5\sigma\frac{(\partial_i\sigma)^2}{a^2}
+(-\gamma_4+\gamma_6)\dot{\sigma}^3
-\gamma_6\dot{\sigma}\frac{(\partial_i\sigma)^2}{a^2}
\Bigg]\,.
\end{align}
In the following,
we discuss what kind of momentum dependence in the squeezed limit appears
from the above three point vertices.\footnote{
For the calculation of three point functions using the in-in formalism,
it is necessary to obtain the Hamiltonian description of the system.
Since the $\ddot{\pi}\sigma\dot{\sigma}$ coupling
contains the second order derivative of $\pi$,
careful discussions are required
when it is relevant
in the action~(\ref{three_point_vertices})
and calculate using the Hamiltonian formalism in the interaction picture.
In this paper,
we do not consider such a situation for simplicity
and concentrate on other cubic couplings.
It should be also noted that
the form of interactions in the Hamiltonian formalism
in the interaction picture does not coincide with
minus that in the Lagrangian formalism
because the action contains derivative interactions.
Correspondingly, the coefficients
of cubic couplings in the Hamiltonian
are changed from minus those in the action~(\ref{three_point_vertices}).}

\begin{figure}[t]
	  		\begin{center}
   				\includegraphics[width=160mm]{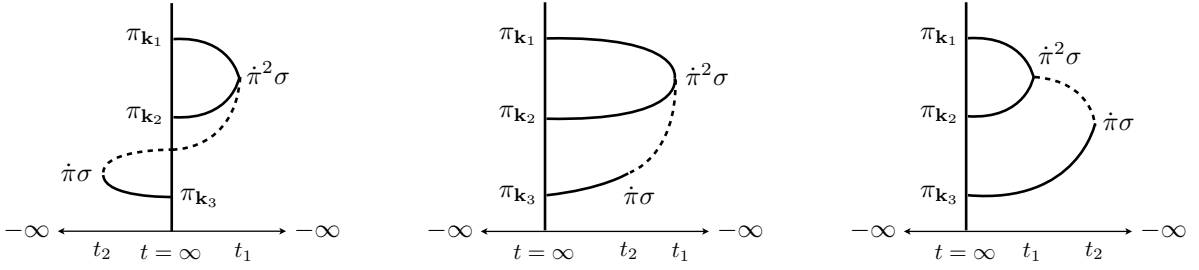}
  			\end{center}
			\vspace{-5mm}
  				\caption{Feynman diagrams for the first term of (\ref{before_squeeze}). The solid lines denote the propagations of $\pi$ and the dotted lines denote those of $\sigma$.
								}
  		 	\label{fig:squeezed}
	\end{figure}

\medskip
As an example,
let us start from the case when the mixing coupling $\tilde{\beta}_1\dot{\pi}\sigma$
and the three point vertex $4\bar{\gamma}_6\dot{\pi}^2\sigma$
are relevant.
In this case, the three point function of $\pi$ takes the form
\begin{align}
\nonumber
&\langle \pi_{{\bf k}_1}(t)\pi_{{\bf k}_2}(t)\pi_{{\bf k}_3}(t)\rangle\\
\nonumber
&\ni \frac{\tilde{\beta}_1\,\bar{\gamma}_6\,(2\pi)^3\delta^{(3)}({\bf k}_1+{\bf k}_2+{\bf k}_3)}{M_{\rm Pl}^3\tilde{\epsilon}^{3/2}(c_\pi k_1)^{3/2}(c_\pi k_2)^{3/2}(c_\pi k_3)^{3/2}}
{\rm Re}\Bigg[
\int_{-\infty}^\infty dt_1\,a^3\,\dot{u}^\ast_{k_1}(t_1)\dot{u}^\ast_{k_2}(t_1)
\\
\nonumber
&\quad\times\left\{ v^\ast_{k_3}(t_1)\int_{-\infty}^\infty dt_2\,a^3\,\dot{u}_{k_3}(t_2)v_{k_3}(t_2)-v^\ast_{k_3}(t_1)\int_{t_1}^\infty dt_2\,a^3\,\dot{u}^\ast_{k_3}(t_2)v_{k_3}(t_2)
-v_{k_3}(t_1)\int_{-\infty}^{t_1} dt_2\,a^3\,\dot{u}_{k_3}^\ast(t_2)v^\ast_{k_3}(t_2) \right\}\Bigg]\\
\label{before_squeeze}
&\quad+( \text{2 permutations})\,.
\end{align}
Here there are three terms in the curly brackets,
and they correspond to the Feynman diagrams in figure \ref{fig:squeezed}, respectively.
In the squeezed limit,
$k_1=k_2=k$ and $k_3/k=\kappa\ll1$,
the long mode $k_3$ crosses the horizon much earlier than
the short modes $k_1$ and $k_2$.
Since the interactions are considered to be relevant around the horizon,
it is expected that the relevant contribution arises from
$(\tau_1,\tau_2)\sim (-1/k_1,-1/k_3)$,
and therefore,
the middle term in the curly brackets becomes irrelevant
in the squeezed limit.
In fact,
we can confirm this expectation explicitly from the expression~(\ref{before_squeeze}),
and the integrals
in the curly brackets can be written at the leading order in $\kappa$ as
\begin{align}
\int_{-\infty}^\infty dt_1\,a^3\,\dot{u}^\ast_{k_1}(t_1)\dot{u}^\ast_{k_2}(t_1)
\,2i\,{\rm Im}\left[v^\ast_{k_3}(t_1)\int_{-\infty}^\infty dt_2\,a^3\,\dot{u}_{k_3}(t_2)v_{k_3}(t_2)\right]\,.
\end{align}
We notice that the $t_2$-integral is $k_3$-independent and
$k_3$-dependence of the total integral appears only via
$v^\ast_{k_3}(t_1)$ originated from the three-point vertex.  Then, the
momentum dependence of the first term in (\ref{before_squeeze}) is given
by
\begin{align}
\label{momentum_dep_first}
\text{the first term in (\ref{before_squeeze}) }\propto \left\{\begin{array}{ll}\kappa^{-3/2-\nu} k^{-6}
&
\displaystyle\quad{\rm for}\quad m_\sigma<\frac{3}{2}H\,,\\[3mm]
\kappa^{-3/2}k^{-6}\sin[i\nu\log \kappa+\delta_\nu]&
\displaystyle\quad{\rm for}\quad m_\sigma>\frac{3}{2}H\,,
\end{array}\right.
\end{align}
where $\delta_\nu$ is a $\nu$-dependent phase factor.  We note that the
only information necessary to derive~(\ref{momentum_dep_first}) was the
fact that the field of momentum $k_3$ in the three point vertex takes
the form $\sigma$.  For the other two permutation terms in (\ref{before_squeeze}),
the $k_3$-dependence of the integral is determined by $u_{k_3}^\ast$
originated from the three point vertex, and their momentum dependence is
given by
\begin{align}
\text{the other two permutation terms in (\ref{before_squeeze})
}\propto \kappa^{-1}k^{-6}\,.
\end{align}
Then, the first term dominates for small $\kappa$.
Therefore, when the mixing coupling $\tilde{\beta}_1$ and
three point coupling $\overline{\gamma}_6$ are relevant,
the momentum dependence of scalar three point functions in the squeezed limit
is given by
\begin{align}
\lim_{k_3/k_1=k_3/k_2=\kappa\to0}\langle \zeta_{{\bf k}_1}\zeta_{{\bf k}_2}\zeta_{{\bf k}_3}\rangle \propto \left\{\begin{array}{ll}\kappa^{-3/2-\nu} k^{-6}
&
\displaystyle\quad{\rm for}\quad m_\sigma<\frac{3}{2}H\,,\\[3mm]
\kappa^{-3/2}k^{-6}\sin[i\nu\log \kappa+\delta_\nu]&
\displaystyle\quad{\rm for}\quad m_\sigma>\frac{3}{2}H\,.
\end{array}\right.
\end{align}

\medskip
It is straightforward to extend the above discussion for more general cases.
First, for general $\tilde{\beta}_i$'s, we can show that when the mixing couplings convert $\pi$ of momentum $k_3=\kappa k$ to $\sigma$,
the $t_2$-integral becomes $\kappa$-independent in the limit $\kappa\ll1$ and
the three point vertex determines
the $\kappa$-dependence of the diagram.
Second,
the only information necessary to obtain the momentum dependence
is the form of the field of momentum $k_3$ in the vertex.
In the previous examples,
when it takes the form $\sigma$ in the vertex
the diagram was proportional to $\kappa^{-3/2-\nu}$ or $\kappa^{-3/2}\sin[i\nu\log \kappa+\delta_\nu]$,
and it was proportional to $\kappa^{-1}$ for $\dot{\pi}$.
More generally, we can obtain the relations in Table\,I between momentum dependence of the diagram and the form of the field of momentum $k_3$ in the three point vertex.
\begin{table}
\begin{center}
$\begin{array}{c|c} \text{\quad form of the field of momentum $k_3$\quad} & \text{momentum dependence of the diagram} \\[1mm]
\hline
\\[-4mm]
\pi & \kappa^{-3}k^{-6} \\[1mm]
\hline
\\[-4mm] \dot{\pi} & \kappa^{-1}k^{-6}
\\[1mm]
\hline
\\[-3mm] \displaystyle\frac{\partial_i\pi}{a} & \kappa^{-2}k^{-6}
\\[3mm]
\hline
\\[-3mm] \sigma,\,\dot{\sigma} &
\left\{\begin{array}{ll}\kappa^{-3/2-\nu} k^{-6}
&
\displaystyle\quad{\rm for}\quad m_\sigma<\frac{3}{2}H\\[3mm]
\kappa^{-3/2}k^{-6}\sin[i\nu\log \kappa+\delta_\nu]&
\displaystyle\quad{\rm for}\quad m_\sigma>\frac{3}{2}H
\end{array}\right.
\\[7mm]
\hline
\\[-3mm] \displaystyle\frac{\partial_i\sigma}{a} &
\left\{\begin{array}{ll}\kappa^{-1/2-\nu} k^{-6}
&
\displaystyle\quad{\rm for}\quad m_\sigma<\frac{3}{2}H\\[3mm]
\kappa^{-1/2}k^{-6}\sin[i\nu\log \kappa+\delta_\nu]&
\displaystyle\quad{\rm for}\quad m_\sigma>\frac{3}{2}H
\end{array}\right.
  \end{array}$
\end{center}
\label{table_1}
\caption{Momentum dependence of the diagram}
\end{table}
Here it should be noted that the $\nu$-dependent phase factor
$\delta_\nu$ depends on the details of mixing couplings. Finally, as
discussed in the previous example, momentum dependence of the
contribution from each vertex is now identified for $\kappa\ll1$
so that it is
straightforward to obtain momentum dependence of the contribution to
scalar three point functions from each three point vertex displayed
in~(\ref{three_point_vertices}). The results are summarized in Table\,II.
\begin{table}
\begin{center}
$\begin{array}{c|c} \text{\quad three point vertices\quad} & \text{\quad momentum dependence\quad} \\[1mm]
\hline
\\[-3mm]
\displaystyle\dot{\pi}^3,\,\dot{\pi}\frac{(\partial_i\pi)^2}{a^2} & \kappa^{-1}k^{-6} \\[3mm]
\hline
\\[-4mm] \begin{array}{c}\dot{\pi}^2\sigma,\,\dot{\pi}^2\dot{\sigma},\,
\dot{\pi}\sigma^2,\,\dot{\pi}\sigma\dot{\sigma},\,\dot{\pi}\dot{\sigma}^2,\\[2mm]
\displaystyle\sigma^3,\,\sigma^2\dot{\sigma},\,\sigma\dot{\sigma}^2,\,\sigma\frac{(\partial_i\sigma)^2}{a^2},\,\dot{\sigma}^3,\,\dot{\sigma}\frac{(\partial_i\sigma)^2}{a^2}
\end{array}
 & \left\{\begin{array}{ll}\kappa^{-3/2-\nu} k^{-6}
&
\displaystyle\quad{\rm for}\quad m_\sigma<\frac{3}{2}H\\[3mm]
\kappa^{-3/2}k^{-6}\sin[i\nu\log \kappa+\delta_\nu]&
\displaystyle\quad{\rm for}\quad m_\sigma>\frac{3}{2}H
\end{array}\right.
\\[7mm]
\hline
\\[-3mm] \displaystyle \dot{\pi}\frac{\partial_i\pi\partial_i\sigma}{a^2} & \kappa^{-2}k^{-6}
\\[3mm]
\hline
\\[-3mm]
\displaystyle\frac{(\partial_i\pi)^2}{a^2}\sigma,\,\frac{(\partial_i\pi)^2}{a^2}\dot{\sigma},\,
\frac{\partial_i\pi\partial_i\sigma}{a^2}\sigma,\,\frac{\partial_i\pi\partial_i\sigma}{a^2}\dot{\sigma} &
\left\{\begin{array}{ll}\kappa^{-3/2-\nu} k^{-6}
&
\displaystyle\quad{\rm for}\quad m_\sigma<\sqrt{2}H\\[3mm]
\kappa^{-2}k^{-6}&
\displaystyle\quad{\rm for}\quad m_\sigma>\sqrt{2}H
\end{array}\right.
\\[6mm]
\hline
\\[-3mm] \displaystyle\dot{\pi}\frac{(\partial_i\sigma)^2}{a^2} &
\left\{\begin{array}{ll}\kappa^{-1/2-\nu} k^{-6}
&
\displaystyle\quad{\rm for}\quad m_\sigma<\sqrt{2}H\\[3mm]
\kappa^{-1}k^{-6}&
\displaystyle\quad{\rm for}\quad m_\sigma>\sqrt{2}H
\end{array}\right.
  \end{array}$
\end{center}
\label{table_2}
\caption{Three point vertices and momentum dependence}
\end{table}
Here note that
although the contribution from the $\displaystyle\dot{\pi}\frac{(\partial_i\pi)^2}{a^2}$ vertex
seems to be proportional to $\kappa^{-2}k^{-6}$ apparently,
explicit calculations show that this kind of leading contribution
vanishes and the three point functions begin with
terms proportional to $\kappa^{-1}k^{-6}$.

\medskip
As we have seen, the momentum dependence of scalar three point functions
in the squeezed limit has robust information about mass of $\sigma$ and
three point vertices.

\section{Summary and discussion}

In this paper we discussed quasi-single field inflation
using the effective field theory approach.
We first constructed the most generic action in
the unitary gauge based on the unbroken
time-dependent spatial diffeomorphism,
and then constructed the action
for the Goldstone boson by the St\"{u}ckelberg method.
Its decoupling regime was also discussed carefully,
and the action in the decoupling regime
implies that
non-trivial cubic interactions generically appear
and non-negligible non-Gaussianities can arise
when the sound speed $c_\sigma$ of $\sigma$ is small,
$\alpha_4$ is non-zero,
or mixing couplings $\beta_1$ and $\beta_2$ exist
as well as the sound speed $c_\pi$ of $\pi$ is small.
Using the obtained action,
two classes of concrete models were discussed:
the constant turning trajectory and the sudden turning trajectory.

\medskip
In the constant turning case,
we first calculated the power spectrum of scalar perturbations
numerically for general values of $r_s=c_\sigma/c_\pi$
and analytically for the special case $r_s=1$.
We then discussed the momentum dependence of scalar three point functions in the squeezed limit for general settings of quasi-single field inflation.
It was shown that the momentum dependence is determined only from the cubic interactions and the cubic interactions were classified into five classes.
The three point functions in the squeezed limit take the intermediate shapes between local and equilateral types
when the mixing couplings are relevant,
and this kind of momentum dependence characterizes quasi-single field inflation.
Recently in~\cite{Sefusatti:2012ye},
the detectability of such a momentum dependence was discussed
for some cases.
It would be interesting to discuss the detectability of
the momentum dependence in the form of
$\kappa^{-3/2}\sin[i\nu\log \kappa+\delta_\nu]$,
which arises in the second class with $m_\sigma>\frac{3}{2}H$.
It is also important to calculate the full bi-spectrum for general settings of quasi-single field inflation.

\medskip
In the sudden turning case,
we made a qualitative discussion of the power spectrum.
It was found that the deviation factor $\mathcal{C}$
from single field inflation
vanishes for long modes $c_\pi k\ll -1/\tau_\ast$, oscillates for
short modes $c_\pi k\gtrsim -1/\tau_\ast$, and damps at some scale
characterized by the oscillating trajectory around the turning point.
Since our framework makes the contributions from the mixing couplings clear,
it would be useful to discuss more on the sudden turning trajectory.

\medskip
Our framework can be considered as a starting point
for systematic discussions on multiple field models,
and there would be a lot of applications
such as those mentioned above. 
We hope to report our progress in these directions elsewhere.

\acknowledgments
We would like to thank Mitsuhiro Kato for useful discussions.
The work of T.N. was supported in part by JSPS Grant-in-Aid for JSPS Fellows.
The work of M.Y. was supported in part by the Grant-in-Aid for Scientific
Research No.~21740187 and the Grant-in-Aid for Scientific
Research on Innovative Areas No.~24111706.  D.Y. acknowledges the
financial support from the Global Center of Excellence Program by MEXT,
Japan through the ``Nanoscience and Quantum Physics'' Project of the
Tokyo Institute of Technology.


\appendix
\section{Integrating out heavy fields}
\label{app_heavy}
In section~\ref{subsubsection_heavy},
we discussed effects of heavy fields
by the following procedure:
\begin{enumerate}
\item Drop the kinetic term of heavy fields.
\item Eliminate derivatives of heavy fields by partial integrals.
\item Complete square the Lagrangian and integrate out heavy fields.
\end{enumerate}
In this appendix
we make some comments and careful discussions
on the first and second steps of this procedure.\footnote{See also~\cite{Gwyn:2012mw} for related discussions.}

\subsection{Role of kinetic term}
We first discuss the procedure to integrate out
heavy fields using the following simple model:
\begin{align}
\label{1d_model}
S&=\int dt \left[\frac{1}{2}\left(\dot{\sigma}^2-m^2\sigma^2\right)
+\sigma f[\phi_i(t);t]\right]\,,
\end{align}
where $f[\phi_i(t);t]$ is a function of light fields $\phi_i$'s and time $t$.
Before discussing the cosmological perturbation,
let us recall the case when we calculate correlation functions in the momentum space
and the total energy of the system is conserved.
Suppose that
the typical energy scale $E$ of external states
is much smaller than the mass of the heavy field:
$E\ll m$.
Then, the typical energy scale of internal $\phi_i$'s and $\sigma$
is also of the order $E$
and much smaller than $m$
because of the energy conservation.
In such a case,
the kinetic term $\frac{1}{2}\dot{\sigma}^2$ becomes irrelevant
compared to the mass term
because
it can be written
as $\frac{1}{2}\dot{\sigma}^2\sim \frac{1}{2}E^2\sigma^2\ll\frac{1}{2}m^2\sigma^2$.
We therefore drop the kinetic term of the heavy field $\sigma$
and obtain
\begin{align}
S&=\int dt \left[-\frac{1}{2}m^2\sigma^2
+\sigma f[\phi_i(t);t]\right]
=\int dt \left[-\frac{m^2}{2}\left(\sigma
-\frac{1}{m^2}f[\phi_i(t);t]\right)^2+\frac{1}{2m^2}\big(f[\phi_i(t);t]\big)^2\right]\,,
\end{align}
which reduces to the following effective action after integrating out $\sigma$:
\begin{align}
\label{app_eff1.5}
S_{\rm eff}&=\frac{1}{2m^2}\int dt\,\big(f[\phi_i(t);t]\big)^2\,.
\end{align}
In this way,
the first step of the procedure is justified
in the momentum space when the momentum conservation holds.

\medskip
In the cosmological perturbation,
we calculate correlation functions in the real time coordinate.
Furthermore, the time translation is broken by the time-dependent background.
As an illustrative toy model for the cosmological perturbations,
let us next consider to calculate correlation functions
of the model~(\ref{1d_model}) in the real time coordinate space.
In such a case,
$\dot{\sigma}$ may seem to behave as $\dot{\sigma}\sim m\sigma$
because the heavy field $\sigma$ oscillates like $\sigma\sim e^{imt}$ on shell:
it may not be so obvious whether the kinetic term of $\sigma$ can be neglected.
To clarify this point, let us make a careful discussion
without neglecting the kinetic term of $\sigma$.
We first rewrite the kinetic term and the mass term as
\begin{align}
\frac{1}{2}\int dt \left(\dot{\sigma}^2-m^2\sigma^2\right)=-\frac{1}{2}\int dt_1\int dt_2
\,\mathcal{K}(t_1,t_2)\sigma(t_1)\sigma(t_2)
\quad
{\rm with}
\quad
\mathcal{K}(t_1,t_2)=m^2\delta(t_1-t_2)+\delta^{\prime\prime}(t_1-t_2)\,,
\end{align}
and introduce the inverse $\mathcal{P}(t_1,t_2)$
of the kinetic operator $\mathcal{K}(t_1,t_2)$ satisfying
\begin{align}
\label{inverse_kinetic}
\int dt \mathcal{P}(t_1,t)\mathcal{K}(t,t_2)=
\int dt \mathcal{K}(t_1,t)\mathcal{P}(t,t_2)=\delta(t_1-t_2)\,.
\end{align}
We then rewrite the action as
\begin{align}
\nonumber
S&=-\frac{1}{2}\int dt_1dt_2\,\mathcal{K}(t_1,t_2)
\Big(\sigma(t_1)-\int dt_1^\prime \mathcal{P}(t_1^\prime,t_1)f[\phi_i(t_1^\prime);t_1^\prime]\Big)
\Big(\sigma(t_2)-\int dt_2^\prime \mathcal{P}(t_2,t_2^\prime)f[\phi_i(t_2^\prime);t_2^\prime]\Big)\\
&\quad+\frac{1}{2}\int dt_1dt_2\,\mathcal{P}(t_1,t_2)
f[\phi_i(t_1);t_1]
f[\phi_i(t_2);t_2]\,.
\end{align}
Integrating out $\sigma$,
the following effective action is obtained:
\begin{align}
\label{app_eff1}
S_{\rm eff}&=\frac{1}{2}\int dt_1dt_2\,\mathcal{P}(t_1,t_2)
f[\phi_i(t_1);t_1]
f[\phi_i(t_2);t_2]\,.
\end{align}
Here note that no approximations are used so far.
Let us then consider the property of $\mathcal{P}(t_1,t_2)$
when $\sigma$ is heavy.
Since the condition~(\ref{inverse_kinetic}) can be rephrased as
\begin{align}
\partial_{t_1}^2\mathcal{P}(t_1,t_2)+m^2\mathcal{P}(t_1,t_2)=\partial_{t_2}^2\mathcal{P}(t_1,t_2)+m^2\mathcal{P}(t_1,t_2)=\delta(t_1-t_2)\,,
\end{align}
$\mathcal{P}(t_1,t_2)$ can be expanded in $1/m^2$ as
\begin{align}
\nonumber
\mathcal{P}(t_1,t_2)&=\frac{1}{m^2}\delta(t_1-t_2)-\frac{1}{m^2}\partial_{t_1}^2\mathcal{P}(t_1,t_2)\\
\nonumber
&=\frac{1}{m^2}\delta(t_1-t_2)-\frac{1}{m^4}\delta^{\prime\prime}(t_1-t_2)
+\frac{1}{m^4}\partial_{t_1}^4\mathcal{P}(t_1,t_2)\\
\nonumber
&=\ldots\\
\label{P_expanded_in_m}
&=\frac{1}{m^2}\delta (t_1-t_2)+\frac{1}{m^2}\sum_{n=1}^\infty\frac{(-1)^n}{m^{2n}}\delta^{(2n)}(t_1-t_2)\,.
\end{align}
Using this expression, the effective action~(\ref{app_eff1})
can be written as
\begin{align}
\label{app_eff2}
S_{\rm eff}=\frac{1}{2m^2}
\int dt \left[\big(f[\phi_i(t);t]\big)^2
+\sum_{n=1}^\infty f[\phi_i(t);t]\frac{(-1)^n}{m^{2n}}\frac{d^{2n}}{dt^{2n}}f[\phi_i(t);t]\right]\,.
\end{align}
Therefore,
if the time-dependence of $f[\phi_i(t);t]$ is negligible compared to the mass $m$ of $\sigma$,
or in other words,
if the fields $\phi_i$'s are light and the explicit time-dependence
of $f[\phi_i(t);t]$ is irrelevant compared to $m$,
the second term in~(\ref{app_eff2}) is negligible
and the effective action reduces to
\begin{align}
\label{app_eff3}
S_{\rm eff}&=\frac{1}{2m^2}\int dt \big(f[\phi_i(t);t]\big)^2\,,
\end{align}
which coincides with the result~(\ref{app_eff1.5}) obtained by
dropping the kinetic term of $\sigma$.

\medskip
To summarize,
the kinetic term $\frac{1}{2}\dot{\sigma}^2$
can be neglected when the time-dependence of $f[\phi_i(t);t]$
is negligible compared to the mass $m$ of the heavy field $\sigma$.
It would be notable that,
when we neglect the kinetic term $\frac{1}{2}\dot{\sigma}^2$,
the kinetic operator $\mathcal{K}(t_1,t_2)$
takes the form $\mathcal{K}(t_1,t_2)=m^2\delta(t_1-t_2)$
and its inverse $\mathcal{P}(t_1,t_2)$
is given by
\begin{align}
\label{singular_P}
\mathcal{P}(t_1,t_2)=\frac{1}{m^2}\delta(t_1-t_2)\,,
\end{align}
which coincides with
the first term in~(\ref{P_expanded_in_m}).
Therefore, it can be considered that
the delta-function like behavior of $\mathcal{P}(t_1,t_2)$
originates from the mass term
and the kinetic term
plays a role to regularize the singular behavior
by reproducing the second term in~(\ref{P_expanded_in_m}).

\subsection{Derivative coupling, partial integral, and Hamiltonian formalism}
Let us next consider the second step of the procedure
using the following two actions:
\begin{align}
\label{app_eff_der_action}
S&=\int dt \left[\frac{1}{2}\left(\dot{\sigma}^2-m^2\sigma^2\right)
+\dot{\sigma} g[\phi_i(t);t]\right]\,,\\
\label{app_eff_partial_action}
S^\prime&=\int dt \left[\frac{1}{2}\left(\dot{\sigma}^2-m^2\sigma^2\right)
-\sigma \frac{d}{dt}g[\phi_i(t);t]\right]\,,
\end{align}
where
$g[\phi_i(t);t]$ is a function of light fields $\phi_i$'s
and time $t$.
Since~(\ref{app_eff_der_action}) and~(\ref{app_eff_partial_action}) are related to each other by partial integrals,
they are expected to describe the same dynamics.
In particular,
they are expected to reproduce the same effective theory in the low energy regime.
However,
it may be wondered that
the mixing term in~(\ref{app_eff_der_action})
becomes relevant when $\sigma$ is heavy because $\dot{\sigma}\sim m\sigma$ on shell
and that the low energy dynamics can be different from
those of~(\ref{app_eff_partial_action}).
In this subsection
we would like to clarify this point
and show that (\ref{app_eff_der_action}) and~(\ref{app_eff_partial_action})
describe the same dynamics as is expected from the partial integral perspective.

\medskip
Similarly to the previous discussions,
the action~(\ref{app_eff_der_action})
can be written in terms of $\mathcal{K}(t_1,t_2)$ and $\mathcal{P}(t_1,t_2)$
as
\begin{align}
\nonumber
S&=-\frac{1}{2}\int dt_1dt_2\,\mathcal{K}(t_1,t_2)
\Big(\sigma(t_1)-\int dt_1^\prime \big[\partial_{t_1^\prime}\mathcal{P}(t_1^\prime,t_1)\big]g[\phi_i(t_1^\prime);t_1^\prime]\Big)
\Big(\sigma(t_2)-\int dt_2^\prime \big[\partial_{t_2^\prime}\mathcal{P}(t_2,t_2^\prime)\big]g[\phi_i(t_2^\prime);t_2^\prime]\Big)\\
&\quad+\frac{1}{2}\int dt_1dt_2\,
\big[\partial_{t_1}\partial_{t_2}
\mathcal{P}(t_1,t_2)\big]
g[\phi_i(t_1);t_1]
g\big[\phi_i(t_2);t_2]\,,
\end{align}
and we obtain the following effective action
after integrating out $\sigma$:
\begin{align}
\label{eff_der_1}
S_{\rm eff}=\frac{1}{2}\int dt_1dt_2\,
\big[\partial_{t_1}\partial_{t_2}
\mathcal{P}(t_1,t_2)\big]
g\big[\phi_i(t_1);t_1\big]
g\big[\phi_i(t_2);t_2\big]\,.
\end{align}
It follows from the expression~(\ref{P_expanded_in_m}) of $\mathcal{P}(t_1,t_2)$
that
\begin{align}
\nonumber
\partial_{t_1}\partial_{t_2}\mathcal{P}(t_1,t_2)
&=\partial_{t_1}\partial_{t_2}\left[\frac{1}{m^2}\delta (t_1-t_2)+\frac{1}{m^2}\sum_{n=1}^\infty\frac{(-1)^n}{m^{2n}}\delta^{(2n)}(t_1-t_2)\right]\\
&=\sum_{n=1}^\infty\frac{(-1)^n}{m^{2n}}\delta^{(2n)}(t_1-t_2)\,,
\end{align}
and therefore,
the effective action~(\ref{eff_der_1}) can be written as
\begin{align}
\nonumber
S_{\rm eff}&=\frac{1}{2}\int dt\,
\sum_{n=1}^\infty \frac{(-1)^n}{m^{2n}}
g\big[t,\phi_i(t)\big]
\frac{d^{2n}}{dt^{2n}}g\big[\phi_i(t);t\big]\\
&=\frac{1}{2m^2}\int dt\left[
\Big(\frac{d}{dt}g\big[\phi_i(t);t\big]\Big)^2+
\sum_{n=1}^\infty \frac{(-1)^n}{m^{2n}}
\Big(\frac{d}{dt}g\big[\phi_i(t);t\big]\Big)
\frac{d^{2n}}{dt^{2n}}\Big(\frac{d}{dt}g\big[\phi_i(t);t\big]\Big)\right]\,,
\end{align}
which is nothing but
the effective action for (\ref{app_eff_partial_action})
obtained by applying the previous result~(\ref{app_eff2}).
We therefore conclude that
the actions~(\ref{app_eff_der_action}) and~(\ref{app_eff_partial_action})
describe the same dynamics.

\medskip
Then, what was wrong in the naive expectation
that the mixing term in~(\ref{app_eff_der_action})
becomes relevant for heavy $\sigma$?
To answer this question,
it may be instructive
to reconsider the above discussion
using the following concrete form of $\mathcal{P}(t_1,t_2)$:
\begin{align}
\label{P_Feynman}
\mathcal{P}(t_1,t_2)=\frac{1}{2im}\left[\theta(t_1-t_2)e^{im(t_1-t_2)}+\theta(t_2-t_1)e^{-im(t_1-t_2)}\right]\,,
\end{align}
which is essentially the same as the Feynman propagator.
The property of the effective action~(\ref{eff_der_1})
is determined by $\partial_{t_1}\partial_{t_2}\mathcal{P}(t_1,t_2)$
and it is given for the choice~(\ref{P_Feynman}) by
\begin{align}
\label{12P}
\partial_{t_1}\partial_{t_2}\mathcal{P}(t_1,t_2)
=m^2\mathcal{P}(t_1,t_2)-\delta(t_1-t_2)\,.
\end{align}
Here
the first term is obtained by taking derivatives of
$e^{\pm imt}$ originated from mode functions of $\sigma$
and the factor $m^2$ is the expected one
from the observation that $\dot{\sigma}\sim m\sigma$.
An important point is that
we also have the second term obtained by taking derivatives
of both of step functions and $e^{\pm imt}$.
Because of this second term,
the leading order term in the $1/m^2$ expansions of $m^2\mathcal{P}(t_1,t_2)$
is canceled out as
\begin{align}
\nonumber
\partial_{t_1}\partial_{t_2}\mathcal{P}(t_1,t_2)
&=m^2\frac{1}{m^2}\left[\delta(t_1-t_2)+\sum_{n=1}^\infty\frac{(-1)^n}{m^{2n}}\delta^{(2n)}(t_1-t_2)\right]-\delta(t_1-t_2)\\
&=\sum_{n=1}^\infty\frac{(-1)^n}{m^{2n}}\delta^{(2n)}(t_1-t_2)\,,
\end{align}
and the naive expectation that the interaction is enhanced by the mass of $\sigma$
turns out to be wrong.
The lesson is that
it is important to take into account derivatives of step functions
in the Feynman propagator appropriately
when we discuss derivative interactions:
the mass factor naively expected from time derivatives of massive fields
can be canceled out.

\medskip
It would be also notable
that similar situations appear in the Hamiltonian formalism
in the interaction picture.
Let us consider the Hamiltonian system
corresponding to~(\ref{app_eff_der_action}).
For simplicity,
suppose that $g[\phi_i(t);t]$ does not
depend on time derivatives of $\phi_i$'s.
Then, the momentum conjugate $P_\sigma$ of $\sigma$
and the Hamiltonian $H$ are given by
\begin{align}
P_\sigma&=\dot{\sigma}+g[\phi_i(t);t]\,,\\
\nonumber
H&=\dot{\sigma}P_\sigma-\left[\frac{1}{2}\dot{\sigma}^2-\frac{1}{2}m^2\sigma^2+\dot{\sigma}g[\phi_i(t);t]\right]\\
&=\frac{1}{2}P_\sigma^2+\frac{1}{2}m^2\sigma^2-P_\sigma g[\phi_i(t);t]+\frac{1}{2}\big(g[\phi_i(t);t]\big)^2\,.
\end{align}
Choosing the free part $H_{\rm free}$
and the interaction part $H_{\rm int}$
of the Hamiltonian
as
\begin{align}
H_{\rm free}&=\frac{1}{2}P_\sigma^2+\frac{1}{2}m^2\sigma^2\,,\\
H_{\rm int}&=-P_\sigma g[\phi_i(t);t]+\frac{1}{2}\big(g[\phi_i(t);t]\big)^2\,,
\end{align}
the dynamics of canonical variables in the interaction picture
are determined by
\begin{align}
\dot{\sigma}=\frac{\partial H_{\rm free}}{\partial P_\sigma}
= P_\sigma\,,
\quad
\dot{P}_\sigma=-\frac{\partial H_{\rm free}}{\partial \sigma}
=-m\sigma\,,
\end{align}
and we have
\begin{align}
H_{\rm free}&=\frac{1}{2}\dot{\sigma}^2+\frac{1}{2}m^2\sigma^2\,,\\
\label{app_H_int}
H_{\rm int}&=-\dot{\sigma} g[\phi_i(t);t]+\frac{1}{2}\big(g[\phi_i(t);t]\big)^2\,.
\end{align}
Note that
the first term in~(\ref{app_H_int}) corresponds to the interaction part
of the Lagrangian
and $H_{\rm int}$ has the additional second term $\frac{1}{2}\big(g[\phi_i(t);t]\big)^2$
as is usual in systems with derivative interactions.
Since $\dot{\sigma}\sim m\sigma$,
the first term in~(\ref{app_H_int}) is enhanced by
the mass $m$ of $\sigma$
and it may be wondered that
the interaction of the system becomes relevant when $\sigma$ is heavy.
However,
it turns out that
the enhancement is canceled out by
the second term $\frac{1}{2}\big(g[\phi_i(t);t]\big)^2$
just as the second term in~(\ref{12P})
cancels out the leading order in the $1/m^2$ expansions of the first term in~(\ref{12P}):
the additional second term in~(\ref{app_H_int})
plays an important role.

\subsection{Extension to cosmological perturbation}
The above discussions can be extended
to the cosmological perturbation straightforwardly.
Let us consider the following action:
\begin{align}
\int d^4x\sqrt{-g}\left[-\frac{1}{2}g^{\mu\nu}\partial_\mu\sigma\partial_\nu\sigma-\frac{m^2}{2}\sigma^2+\sigma f[\phi_i(x);x]\right]\,.
\end{align}
The kinetic term and the mass term of $\sigma$
can be written as
\begin{align}
S_{\rm free}=-\frac{1}{2}\int d^4x_1\sqrt{-g}(x_1)\int d^4x_2\sqrt{-g}(x_2)\,\sigma(x_1)\mathcal{K}(x_1,x_2)
\sigma(x_2)
\quad
{\rm with}
\quad
\mathcal{K}(x_1,x_2)=(m^2-\Box)\frac{\delta^4(x_1-x_2)}{\sqrt{-g}}\,,
\end{align}
where $\Box=g^{\mu\nu}\nabla_\mu\nabla_\nu$
is the covariant d'Alembertian operator and $\nabla_\mu$ is the covariant derivative.
Introducing the inverse $\mathcal{P}(x_1,x_2)$ of $\mathcal{K}(x_1,x_2)$,
\begin{align}
\label{cond_P_cos}
\int d^4x\sqrt{-g}\,\mathcal{K}(x_1,x)\mathcal{P}(x,x_2)
=\int d^4x\sqrt{-g}\,\mathcal{P}(x_1,x)\mathcal{K}(x,x_2)
=\frac{\delta^4(x_1-x_2)}{\sqrt{-g}}\,,
\end{align}
the following effective action is obtained
after integrating out $\sigma$:\footnote{
Since the kinetic operator $\mathcal{K}(x_1,x_2)$ contains metric perturbations,
non-trivial $\log \det \mathcal{K}$ contributions
arise from the Gaussian path integrals of $\sigma$.
However,
we do not consider these contributions for simplicity
because they do not appear as long as
tree-level perturbations are discussed
around a given background spacetime.}
\begin{align}
S_{\rm eff}&=\frac{1}{2}\int d^4x_1\sqrt{-g}\int d^4x_2\sqrt{-g}\mathcal{P}(x_1,x_2)
f[\phi_i(x_1);x_1]f[\phi_i(x_2);x_2]\,.
\end{align}
Just as in the previous toy models,
the conditions~(\ref{cond_P_cos})
can be rephrased as
\begin{align}
(m^2-\Box_1) \mathcal{P}(x_1,x_2)
=(m^2-\Box_2) \mathcal{P}(x_1,x_2)
=\frac{\delta^4(x_1-x_2)}{\sqrt{-g}}\,,
\end{align}
and $\mathcal{P}(x_1,x_2)$ can be expanded in $1/m^2$ as
\begin{align}
\nonumber
\mathcal{P}(x_1,x_2)
&=\frac{1}{m^2}\frac{\delta^4(x_1-x_2)}{\sqrt{-g}}
+\frac{\Box_1}{m^2}\mathcal{P}(x_1,x_2)\\
\nonumber
&=\ldots\\
&=\frac{1}{m^2}\frac{\delta^4(x_1-x_2)}{\sqrt{-g}}+\frac{1}{m^2}\sum_{n=1}^\infty\left(\frac{\Box_1}{m^2}\right)^n\frac{\delta^4(x_1-x_2)}{\sqrt{-g}}\,.
\end{align}
Then,
the effective action can be written as
\begin{align}
S_{\rm eff}&=\frac{1}{2m^2}\int d^4x\sqrt{-g}
\left[\big(f[\phi_i(x);x]\big)^2+\sum_{n=1}^\infty
f[\phi_i(x);x]\left(\frac{\Box}{m^2}\right)^nf[\phi_i(x);x]\right]\,,
\end{align}
which reduces to the following form
when the spacetime-dependence of $f[\phi_i(x);x]$
is negligible compared to the mass $m$ of~$\sigma$:
\begin{align}
S_{\rm eff}&=\frac{1}{2m^2}\int d^4x\sqrt{-g}
\big(f[\phi_i(x);x]\big)^2\,.
\end{align}
Therefore,
in the cosmological perturbation around FRW backgrounds,
the kinetic term of $\sigma$ can be neglected
when the mass $m$ of $\sigma$ is much larger
than the Hubble parameter $H$ and
the mass $m_i$ of other fields $\phi_i$:
$m\gg H,m_i$.

\section{Numerical calculations of power spectrum}
\label{app_numerical}
To perform the integral~(\ref{I_ij}) numerically,
there are two technical obstacles~\cite{Chen:2009we,Chen:2012ge}:
spurious divergences at $x=0$ and oscillating behaviors at $x=\infty$.

\medskip
The integral~(\ref{I_ij}) contains two integrals,
$\int_0^\infty dx_1\, A_i(x_1)^\ast
\int_{x_1}^\infty dx_2\,A_j(x_2)$
and
$\int_0^\infty dx_1 \,\widetilde{A}_i(x_1)
\int_{x_1}^\infty dx_2\,A_j(x_2)$,
and each integral diverges for some parameter region.
However, we can show that such divergences cancel out
in the calculation of $\mathcal{C}_{ij}$'s.
For example, let us consider~$\mathcal{C}_{11}$ for real $0<\nu<3/2$.
The asymptotic behavior
of each integral in $\mathcal{I}_{11}$
around $x_1=x_2=0$
is given by
\begin{align}
\sim \int_0 dx_1\,x_1^{-1/2-\nu}\int_{x_1} dx_2\,x_2^{-1/2-\nu}\sim  \,0^{1-2\nu}\,,
\end{align}
which diverges for $\nu>1/2$.
We first notice that this kind of leading singularities cancel out
between two terms and the asymptotic behavior
of the total integral $\mathcal{I}_{11}$ is given by
\begin{align}
\sim i\int_0 dx_1\,x_1^{1/2-\nu}\int_{x_1} dx_2\,x_2^{-1/2-\nu}\sim i \,0^{2-2\nu}
\end{align}
up to a real constant number.  Although it is still singular for
$\nu>1$, such singular contribution is pure-imaginary and does not
contribute to ${\rm Re}[\,\mathcal{I}_{11}]$.  The higher order terms
are finite for $\nu<3/2$ and hence we conclude that
$\mathcal{C}_{11}={\rm Re}[\,\mathcal{I}_{11}]$ is finite.  In a similar
way, we can show that ${\rm Re[\,\mathcal{I}_{ij}+\mathcal{I}_{ji}]}$,
${\rm Im[\,\mathcal{I}_{12}-\mathcal{I}_{21}]}$, and ${\rm
Im[\,\mathcal{I}_{13}-\mathcal{I}_{31}]}$ are finite for $0<\nu<3/2$ and
$\nu=\text{ pure-imaginary}$, and therefore all the $\mathcal{C}_{ij}$'s
are finite.  To avoid this kind of spurious singularities in the
numerical calculation, we introduce a IR cut off $\epsilon_{\rm IR}$:
\begin{align}
\mathcal{I}_{ij}=\frac{\pi}{4}e^{\frac{i}{2}\pi(\nu-\nu^\ast)}
\int_{\epsilon_{\rm IR}}^\infty dx_1 \Big(A_i(x_1)^\ast-\widetilde{A}_i(x_1)\Big)
\int_{x_1}^\infty dx_2\,A_j(x_2)\,,
\end{align}
where we set $\epsilon_{\rm IR}=10^{-10}$ in our calculation.

\medskip
As is usual in the Feynman diagram calculation in the momentum
space, the integral $\mathcal{I}_{ij}$ oscillates at $x=\infty$ because
of the oscillating behavior of the mode functions $u_k$ and $v_k$, which
makes the numerical calculation difficult.
Following~\cite{Chen:2009we,Chen:2012ge}, we perform contour
deformations to avoid this kind of technical difficulties. 
Let us first
consider
the integral ${\rm Re}[\,\mathcal{I}_{ij}+\mathcal{I}_{ji}]$.
It is convenient to rewrite it as follows:
\begin{align}
\nonumber
{\rm Re}[\,\mathcal{I}_{ij}+\mathcal{I}_{ji}]&=
\frac{\pi}{4}e^{\frac{i}{2}\pi(\nu-\nu^\ast)}
{\rm Re}\left[\,
\int_{\epsilon_{\rm IR}}^\infty dx_1 \,A_i(x_1)^\ast
\int_{x_1}^\infty dx_2\,A_j(x_2)
+\int_{\epsilon_{\rm IR}}^\infty dx_1 \,A_j(x_1)
\int_{x_1}^\infty dx_2\,A_i(x_2)^\ast\right.
\\
\nonumber
&\qquad\qquad\qquad\qquad\left.-\int_{\epsilon_{\rm IR}}^\infty dx_1\,\widetilde{A}_i(x_1)
\int_{x_1}^\infty dx_2\,A_j(x_2)
-\int_{\epsilon_{\rm IR}}^\infty dx_1\,\widetilde{A}_j(x_1)
\int_{x_1}^\infty dx_2\,A_i(x_2)\,\right]\\
\nonumber
&=
\frac{\pi}{4}e^{\frac{i}{2}\pi(\nu-\nu^\ast)}
{\rm Re}\left[
\left(\int_{\epsilon_{\rm IR}}^\infty dx \,A_i(x)\right)^\ast
\int_{\epsilon_{\rm IR}}^\infty dy\,A_j(y)
\right.
\\
&\qquad\qquad\qquad\qquad\left.-\int_{\epsilon_{\rm IR}}^\infty dx_1\,\widetilde{A}_i(x_1)
\int_{x_1}^\infty dx_2\,A_j(x_2)
-\int_{\epsilon_{\rm IR}}^\infty dx_1\,\widetilde{A}_j(x_1)
\int_{x_1}^\infty dx_2\,A_i(x_2)\,\right]\,.
\end{align}
The first term in the bracket can be Wick rotated
without crossing any poles as
\begin{align}
\left(\int_0^\infty dx \,A_i(\epsilon_{\rm IR}+ix)\right)^\ast
\int_0^\infty dy\,A_j(\epsilon_{\rm IR}+iy)\,,
\end{align}
and the last two terms are Wick rotated as
\begin{align}
\int_0^\infty dx\,\widetilde{A}_i(\epsilon_{\rm IR}+ix)
\int_0^\infty dy\,A_j(\epsilon_{\rm IR}+ix+iy)
+\int_0^\infty dx\,\widetilde{A}_j(\epsilon_{\rm IR}+ix)
\int_0^\infty dy\,A_i(\epsilon_{\rm IR}+ix+iy)\,.
\end{align}
Then, we obtain the following expression of ${\rm Re}[\,\mathcal{I}_{ij}+\mathcal{I}_{ji}]$:
\begin{align}
{\rm Re}[\,\mathcal{I}_{ij}+\mathcal{I}_{ji}]=\frac{\pi}{4}e^{\frac{i}{2}\pi(\nu-\nu^\ast)}
{\rm Re}\left[\int_0^\infty dx \Big(A_i(\epsilon_{\rm IR}+ix)^\ast+\widetilde{A}_i(\epsilon_{\rm IR}+ix)\Big)
\int_0^\infty dy\,A_j(\epsilon_{\rm IR}+ix+iy)\right]+(i\leftrightarrow j)\,.
\end{align}
To avoid the singular behavior around $x=0$ discussed above, we further
modify the contour as follows:
\begin{align}
\nonumber
{\rm Re}[\,\mathcal{I}_{ij}+\mathcal{I}_{ji}]&=\frac{\pi}{4}e^{\frac{i}{2}\pi(\nu-\nu^\ast)}
{\rm Re}\left[\,i\int_0^1 dx \Big(A_i(\epsilon_{\rm IR}+x)^\ast+\widetilde{A}_i(\epsilon_{\rm IR}+x)\Big)
\int_0^\infty dy\,A_j(\epsilon_{\rm IR}+ix+iy)\right.\\
\label{plus_I_ij}
&\quad\left.+
\int_0^\infty dx \Big(A_i(\epsilon_{\rm IR}+1+ix)^\ast+\widetilde{A}_i(\epsilon_{\rm IR}+1+ix)\Big)
\int_0^\infty dy\,A_j(\epsilon_{\rm IR}+1+ix+iy)\right]+(i\leftrightarrow j)\,.
\end{align}
By performing similar contour deformations,
${\rm Im}[\,\mathcal{I}_{ij}-\mathcal{I}_{ji}]$ can be also expressed as follows:
\begin{align}
\nonumber
{\rm Im}[\,\mathcal{I}_{ij}-\mathcal{I}_{ji}]&=\frac{\pi}{4}e^{\frac{i}{2}\pi(\nu-\nu^\ast)}
{\rm Re}\left[\,\int_0^1 dx \Big(A_i(\epsilon_{\rm IR}+x)^\ast-\widetilde{A}_i(\epsilon_{\rm IR}+x)\Big)
\int_0^\infty dy\,A_j(\epsilon_{\rm IR}+ix+iy)\right]\\
\nonumber
&\quad+
\frac{\pi}{4}e^{\frac{i}{2}\pi(\nu-\nu^\ast)}
{\rm Im}\left[\int_0^\infty dx \Big(A_i(\epsilon_{\rm IR}+1+ix)^\ast+\widetilde{A}_i(\epsilon_{\rm IR}+1+ix)\Big)
\int_0^\infty dy\,A_j(\epsilon_{\rm IR}+1+ix+iy)\right]\\
\label{minus_I_ij}
&\quad-(i\leftrightarrow j)\,.
\end{align}The expressions (\ref{plus_I_ij}) and (\ref{minus_I_ij})
are used in our numerical calculations of the power spectrum.

\section{Analytical calculation of power spectrum for $c_\pi=c_\sigma$}
\label{app_analytic}
In this appendix
we calculate the power spectrum
for the case $r_s=c_\sigma/c_\pi=1$.  For this class of
models, we can analytically calculate the integrals~$\mathcal{I}_{ij}$'s
by extending the results in~\cite{Chen:2012ge}.  We first introduce a
function $\mathcal{A}(\ell,\nu,x)$ defined by
\begin{align}
\mathcal{A}(\ell,\nu,x)=x^{-\frac{1}{2}+\ell}e^{ i x}H_\nu^{(1)}(x)\,.
\end{align}
In terms of $\mathcal{A}$,
$A_i$ can be written as
\begin{align}
A_1(x)=\mathcal{A}(0,\nu,x)\,,
\quad
A_2(x)=\mathcal{A}(1,\nu-1,x)\,,
\quad
A_3(x)=i\mathcal{A}(1,\nu,x)\,,
\end{align}
and hence the $x$-integrals in the first term and the $x_2$-integral in
the second term of~(\ref{I_ij}) reduce to that of $\mathcal{A}$.
Similarly to the case in~\cite{Chen:2012ge}, the indefinite integral
$\mathcal{D}$ of~$\mathcal{A}$ can be expressed using hyper-geometric
functions as
\begin{align}
\nonumber
\mathcal{D}(\ell,\nu,x)&=\int dx\,\mathcal{A}(\ell,\nu\,x)\\
\nonumber
&=
\frac{2^\nu x^{\frac{1}{2}+\ell-\nu}\Gamma(\nu)}{i\pi(\frac{1}{2}+\ell-\nu)}
{}_2F_2\Big(\frac{1}{2}-\nu,\frac{1}{2}+\ell-\nu;\frac{3}{2}+\ell-\nu,1-2\nu;+ 2ix\Big)\\
&\quad+e^{-i\pi\nu}\frac{2^\nu x^{\frac{1}{2}+\ell+\nu}\Gamma(-\nu)}{i\pi(\frac{1}{2}+\ell+\nu)}
{}_2F_2\Big(\frac{1}{2}+\nu,\frac{1}{2}+\ell+\nu;\frac{3}{2}+\ell+\nu,1+2\nu;+ 2ix\Big)\,.
\end{align}
Then,
the integral $\mathcal{I}_{ij}$ can be written as
\begin{align}
\nonumber
\mathcal{I}_{ij}&=\frac{\pi}{8}e^{\frac{i}{2}\pi(\nu-\nu^\ast)}(-i)^{n_1-\tilde{n}_1+n_2-\tilde{n}_2}\Big(\mathcal{D}(n_1,\nu+n_2,\infty)-\mathcal{D}(n_1,\nu+n_2,0)\Big)^\ast
\Big(\mathcal{D}(\tilde{n}_1,\nu+\tilde{n}_2,\infty)
-\mathcal{D}(\tilde{n}_1,\nu+\tilde{n}_2,0)\Big)\\
\label{I_ij_first}
&\quad-\frac{\pi}{4}i^{\tilde{n}_1+\tilde{n}_2}e^{\frac{i}{2}\pi(\nu-\nu^\ast)}\int_0^\infty dx_1
\widetilde{A}_i(x_1)
\Big(\mathcal{D}(\tilde{n}_1,\nu+\tilde{n}_2,\infty)
-\mathcal{D}(\tilde{n}_1,\nu+\tilde{n}_2,x_1)\Big)\,,
\end{align}
where $(n_1,n_2),(\tilde{n}_1,\tilde{n}_2)=(0,0),(1,-1),(1,0)$
for $i,j=1,2,3$, respectively.
Here note that
$\mathcal{D}(\ell,\nu,x)$
gives a finite value
in the limit $x=\infty$ (see appendix~\ref{app_D} for the derivation):
\begin{align}
\label{D_inf}
\mathcal{D}(\ell,\nu,\infty)&=\frac{1}{i\sqrt{2\pi}}\frac{1}{\sin\pi\nu}
\Bigg(\frac{\Gamma(1/2+\ell-\nu)}{\Gamma(1/2-\ell-\nu)}-\frac{\Gamma(1/2+\ell+\nu)}{\Gamma(1/2-\ell+\nu)}\Bigg)
\Gamma(-\ell)(- 2i)^{-\ell}e^{- \frac{i}{2}\pi(\nu-1/2)}\\
&=\frac{\sqrt{\pi}(1-i)e^{-\frac{i}{2}\pi\nu}}{\cos \pi\nu}(-2i)^{-\ell}\frac{1}{\Gamma(\ell+1)}\frac{\Gamma(\frac{1}{2}+\ell+\nu)\Gamma(\frac{1}{2}+\ell-\nu)}{\Gamma(\frac{1}{2}+\nu)\Gamma(\frac{1}{2}-\nu)}\,,
\end{align}
where we used an $i\epsilon$-prescription to drop oscillatory terms
$\propto e^{2ix}$ at infinity $x\to\infty$. It should be also noted that
the asymptotic behavior of $\mathcal{D}(\ell,\nu,x)$ around $x=0$ is
given by
\begin{align}
\mathcal{D}(\ell,\nu,x)&=
\frac{2^\nu \Gamma(\nu)}{i\pi(\frac{1}{2}+\ell-\nu)}
\Big(x^{\frac{1}{2}+\ell-\nu}+\mathcal{O}(x^{\frac{3}{2}+\ell-\nu})\Big)
+e^{-i\pi\nu}\frac{2^\nu \Gamma(-\nu)}{i\pi(\frac{1}{2}+\ell+\nu)}
\Big(x^{\frac{1}{2}+\ell+\nu}+\mathcal{O}(x^{\frac{3}{2}+\ell+\nu})\Big)
\,,
\end{align}
which is singular for ${\rm Re}[1/2+\ell-\nu]<0$. 
However, as
mentioned in appendix~\ref{app_numerical}, this kind of singularities cancel out in the
calculation of~$\mathcal{C}_{ij}$'s and we drop them in the following
calculation.  We therefore rewrite~(\ref{I_ij_first}) as follows:
\begin{align}
\nonumber
\mathcal{I}_{ij}&=\frac{\pi}{8}e^{\frac{i}{2}\pi(\nu-\nu^\ast)}(-i)^{n_1-\tilde{n}_1+n_2-\tilde{n}_2}\Big(\mathcal{D}(n_1,\nu+n_2,\infty)\Big)^\ast
\mathcal{D}(\tilde{n}_1,\nu+\tilde{n}_2,\infty)
\\
\nonumber
&\quad-\frac{\pi}{4}i^{\tilde{n}_1+\tilde{n}_2}e^{\frac{i}{2}\pi(\nu-\nu^\ast)}\int_0^\infty dx_1
\widetilde{A}_i(x_1)
\Big(\mathcal{D}(\tilde{n}_1,\nu+\tilde{n}_2,\infty)
-\mathcal{D}(\tilde{n}_1,\nu+\tilde{n}_2,x_1)\Big)\\
\nonumber
&=\frac{\pi^2}{4}\frac{1}{\cos^2\pi\nu}
2^{-n_1-\tilde{n}_1}(-1)^{n_1-\tilde{n}_1+n_2-\tilde{n}_2}
\frac{(\frac{1}{2}+\nu^\ast+n_2)_{n_1}(\frac{1}{2}-\nu^\ast-n_2)_{n_1}}{\Gamma(1+n_1)}\frac{(\frac{1}{2}+\nu+\tilde{n}_2)_{\tilde{n}_1}(\frac{1}{2}-\nu-\tilde{n}_2)_{\tilde{n}_1}}{\Gamma(1+\tilde{n}_1)}
\\
\label{I_ij_second}
&\quad-\frac{\pi}{4}i^{\tilde{n}_1+\tilde{n}_2}e^{\frac{i}{2}\pi(\nu-\nu^\ast)}\int_0^\infty dx_1
\widetilde{A}_i(x_1)
\Big(\mathcal{D}(\tilde{n}_1,\nu+\tilde{n}_2,\infty)
-\mathcal{D}(\tilde{n}_1,\nu+\tilde{n}_2,x_1)\Big)\,,
\end{align}
with $\displaystyle(a)_m=\frac{\Gamma(a+m)}{\Gamma(a)}$.

\medskip We next perform the $x_1$-integral in~(\ref{I_ij_second}).  To
perform this kind of integrals, we use a trick of
resummation~\cite{Chen:2012ge}.  Expanding $\widetilde{\mathcal{A}}_i$
in $x$ and using the identity for Bessel functions,
\begin{align}
x^{-\nu}J_\nu(x)e^{ix}=\sum_{m=0}^\infty a_{m}^{\nu} x^m
\quad
{\rm with}
\quad
a_{m}^{\nu}=\frac{2^{m+\nu}i^m\,\Gamma(m+\nu+1/2)}{m! \sqrt{\pi}\,\Gamma(m+2\nu+1)}\,,
\end{align}
$\widetilde{\mathcal{A}}_i$ is rewritten as
\begin{align}
\nonumber
\widetilde{\mathcal{A}}_i
&=i^{n_1+n_2}x^{-\frac{1}{2}+n_1}e^{ix}\frac{-i}{\sin \pi(\nu^\ast+n_2)}\Big(e^{i\pi (\nu^\ast+n_2)}J_{\nu^\ast+n_2}(x)-J_{-(\nu^\ast+n_2)}(x)\Big)\\
&=\frac{-i^{1+n_1+n_2}}{\sin \pi\nu^\ast}\Big(e^{i\pi \nu^\ast}\sum_{m=0}^\infty a_m^{\nu^\ast+n_2}x^{-\frac{1}{2}+m+n_1+n_2+\nu^\ast}-(-1)^{n_2}\sum_{m=0}^\infty a_m^{-(\nu^\ast+n_2)}x^{-\frac{1}{2}+m+n_1-n_2-\nu^\ast}\Big)\,,
\end{align}
where again $(n_1,n_2)=(0,0),(1,-1),(1,0)$ for $i=1,2,3$, respectively.
Using this expression,
the $x_1$-integral in~(\ref{I_ij_second})
can be written as a sum of integrals
in the form
\begin{align}
\int_0^\infty dx \,x^{p}\left(\mathcal{D}(\ell,\nu,\infty)-\mathcal{D}(\ell,\nu,x)\right)\,.
\end{align}
Integrating by parts and
using $\partial_x\mathcal{D}(\ell,\nu,x)=\mathcal{A}(\ell,\nu,x)$
and $x^{1+p}\mathcal{A}(\ell,\nu,x)=\mathcal{A}(1+\ell+p,\nu,x)$,
we rewrite it as
\begin{align}
\nonumber
\int_0^\infty dx \,x^{p}\left(\mathcal{D}(\ell,\nu,\infty)-\mathcal{D}(\ell,\nu,x)\right)&
=\left[\frac{x^{1+p}}{1+p}\left(\mathcal{D}(\ell,\nu,\infty)-\mathcal{D}(\ell,\nu,x)\right)\right]_0^\infty
+\int_0^\infty dx \,\frac{1}{1+p}\mathcal{A}(1+\ell+p,\nu,x)\\
\nonumber
&=\frac{1}{1+p}\Big(\mathcal{D}(1+\ell+p,\nu,\infty)
+\left[x^{1+p}\mathcal{D}(\ell,\nu,x)
-\mathcal{D}(1+\ell+p,\nu,x)\right]_{x\to0}\Big)\\
\label{resum_int}
&=\frac{1}{1+p}\mathcal{D}(1+\ell+p,\nu,\infty)\,.
\end{align}
Here we again dropped the contribution from $\mathcal{D}$ at $x=0$,
which vanishes or cancels out in our calculation as mentioned earlier.
Then, the $x_1$-integral can be written as follows:
\begin{align}
\nonumber
&\quad-\frac{\pi}{4}i^{\tilde{n}_1+\tilde{n}_2}e^{\frac{i}{2}\pi(\nu-\nu^\ast)}\int_0^\infty dx_1
\widetilde{A}_i(x_1)
\Big(\mathcal{D}(\tilde{n}_1,\nu+\tilde{n}_2,\infty)
-\mathcal{D}(\tilde{n}_1,\nu+\tilde{n}_2,x_1)\Big)\\
\nonumber
&=\frac{\pi}{4}\frac{i^{1+n_1+\tilde{n}_1+\tilde{n}_2}}{\sin \pi\nu^\ast}\Bigg(
e^{\frac{i}{2}\pi(\nu+\nu^\ast+n_2)}\sum_{m=0}^\infty \frac{a_m^{\nu^\ast+n_2}}{\frac{1}{2}+m+n_1+n_2+\nu^\ast}\mathcal{D}(1/2+m+n_1+n_2+\nu^\ast+\tilde{n}_1,\nu+\tilde{n}_2,\infty)\\
&\quad\qquad\qquad\qquad\qquad-e^{\frac{i}{2}\pi(\nu-\nu^\ast-n_2)}\sum_{m=0}^\infty
\frac{a_m^{-(\nu^\ast+n_2)}}{\frac{1}{2}+m+n_1-n_2-\nu^\ast}
\mathcal{D}(1/2+m+n_1-n_2-\nu^\ast+\tilde{n}_1,\nu+\tilde{n}_2,\infty)\Big)\,.
\end{align}
After some lengthy calculations,
we obtain
\begin{align}
\nonumber
&\quad
\frac{\pi}{4}\frac{i^{1+n_1+\tilde{n}_1+\tilde{n}_2}}{\sin \pi\nu^\ast}
e^{\frac{i}{2}\pi(\nu\pm\nu^\ast\pm n_2)}
\frac{a_m^{\pm(\nu^\ast+n_2)}}{\frac{1}{2}+m+n_1\pm n_2\pm \nu^\ast}\mathcal{D}(1/2+m+n_1\pm n_2\pm \nu^\ast+\tilde{n}_1,\nu+\tilde{n}_2,\infty)\\
\nonumber
&=\frac{i\,e^{\pm i \pi\nu^\ast}}{\sin \pi\nu^\ast}\frac{2^{-n_1-\tilde{n}_1-2}(-1)^{m+\tilde{n}_2}}{(\frac{1}{2}+m+n_1\pm (\nu^\ast+ n_2))\left(\frac{1}{2}+m\pm(\nu^\ast+n_2)\right)_{1+n_1+\tilde{n}_1}}\\[1mm]
\label{to_be_re-summed}
&\quad\times\left\{\begin{array}{ll}(m+1)_{n_1\pm n_2+\tilde{n}_1\mp\tilde{n}_2}\,(-m-n_1\mp n_2-\tilde{n}_1\mp\tilde{n}_2\mp 2\nu^\ast)_{n_1\mp n_2+\tilde{n}_1\pm \tilde{n}_2}
&\quad\text{for} \,\,\,\,\text{real} \,\,\,\,\nu\,,  \\[2mm]
(m+1)_{n_1\pm n_2+\tilde{n}_1\pm\tilde{n}_2}\,(-m-n_1\mp n_2-\tilde{n}_1\pm\tilde{n}_2\mp 2\nu^\ast)_{n_1\mp n_2+\tilde{n}_1\mp \tilde{n}_2}
&\quad\text{for} \,\,\,\,\text{pure-imaginary} \,\,\,\,\nu\,,
\end{array}\right.
\end{align}
where $\displaystyle\Gamma(z)\Gamma(1-z)=\frac{\pi}{\sin \pi z}$ was
used. Finally, it is necessary to re-sum~(\ref{to_be_re-summed}) with
respect to $m$. In the case of ${\rm Re}[\,\mathcal{I}_{11}]$
$(n_i=\tilde{n}_i=0)$, for example, we perform the resummation as
follows:
\begin{align} \nonumber
&\quad-\frac{\pi}{4}e^{\frac{i}{2}\pi(\nu-\nu^\ast)}\int_0^\infty dx_1
\widetilde{A}_1(x_1) \Big(\mathcal{D}(0,\nu,\infty)
-\mathcal{D}(0,\nu,x_1)\Big)\\ \nonumber &=\frac{i}{4\sin
\pi\nu}\sum_{m=0}^\infty\Bigg[\frac{(-1)^me^{ i
\pi\nu}}{(m+\frac{1}{2}+\nu)^2} -\frac{(-1)^me^{- i
\pi\nu}}{(m+\frac{1}{2}-\nu)^2}\Bigg]\\ \nonumber &=\frac{i}{4\sin
\pi\nu}\Big[e^{ i \pi\nu}\Phi(-1,2,\tfrac{1}{2}+\nu)-e^{- i
\pi\nu}\Phi(-1,2,\tfrac{1}{2}-\nu)\Big]\\ \nonumber &=-\frac{i\,e^{ i
\pi\nu}}{16\sin \pi\nu}
\Bigg[\psi^{(1)}\Big(\frac{3}{4}+\frac{\nu}{2}\Big)-\psi^{(1)}\Big(\frac{1}{4}+\frac{\nu}{2}\Big)\Bigg]
+\frac{i\,e^{ -i \pi\nu}}{16\sin \pi\nu}
\Bigg[\psi^{(1)}\Big(\frac{3}{4}-\frac{\nu}{2}\Big)-\psi^{(1)}\Big(\frac{1}{4}-\frac{\nu}{2}\Big)\Bigg]\,.
\end{align}
Here $\Phi(z,s,\alpha)$ and $\psi^{(n)}(z)$ are the Lerch transcendent and the polygamma function, respectively:
\begin{align}
\Phi(z,s,\alpha)=\sum_{m=0}^\infty\frac{z^m}{(m+\alpha)^s}\,,
\quad
\psi^{(n)}(z)=(-1)^{n+1}n!\sum_{m=0}^\infty\frac{1}{(m+z)^{n+1}}\,,
\end{align}
which satisfy the following relation:
\begin{align}
\Phi(-1,n+1,z)=(-1)^n\,n!\,2^{-n-1}\Bigg[\psi^{(n)}\Big(\frac{z+1}{2}\Big)-\psi^{(n)}\Big(\frac{z}{2}\Big)\Bigg]\,.
\end{align}
We then obtain
\begin{align}
{\rm Re}[\,\mathcal{I}_{11}]=\frac{\pi^2}{4\cos^2 \pi\nu}+{\rm Re}\left[\frac{i\,e^{ -i \pi\nu}}{16\sin \pi\nu}
\Big[\psi^{(1)}\Big(\frac{3}{4}+\frac{\nu}{2}\Big)-\psi^{(1)}\Big(\frac{1}{4}+\frac{\nu}{2}\Big)\Big]
-\frac{i\,e^{ i \pi\nu}}{16\sin \pi\nu}
\Big[\psi^{(1)}\Big(\frac{3}{4}-\frac{\nu}{2}\Big)-\psi^{(1)}\Big(\frac{1}{4}-\frac{\nu}{2}\Big)\Big]\right]\,,
\end{align}
which reproduces the result in~\cite{Chen:2012ge}.
In a similar way,
we can obtain analytic expression for $\mathcal{I}_{ij}$'s,
and the results are summarized as follows:
{\allowdisplaybreaks
\begin{align}
\mathcal{I}_{22}&=\left\{
\begin{array}{ll}
&\displaystyle\frac{\pi^2|\nu-\frac{1}{2}|^2|\nu-\frac{3}{2}|^2}{16 \cos^2\pi\nu}+\frac{1}{16}-\frac{i}{128\sin\pi\nu}
\Big(e^{i\pi\nu}(2\nu -3)(2\nu -5)-e^{-i\pi\nu}(2\nu+1)(2\nu-1)\Big)
\\[2mm]
&\displaystyle+\frac{i}{256\sin\pi\nu}
(2\nu-1)^2(2\nu-3)^2\Big(e^{i\pi\nu}\Phi(-1,2,\tfrac{1}{2}+\nu)-e^{-i\pi\nu}\Phi(-1,2,\tfrac{5}{2}-\nu)\Big)\quad\text{for} \,\,\,\,\text{real} \,\,\,\,\nu\,,  \\[6mm]
&\displaystyle\frac{\pi^2|\nu-\frac{1}{2}|^2|\nu-\frac{3}{2}|^2}{16 \cos^2\pi\nu}+\frac{1}{16}
-\frac{i}{128\sin\pi\nu}\Big(e^{i\pi\nu}\frac{-3-82\nu+12\nu^2+72\nu^3}{3+2\nu}+e^{-i\pi\nu}\frac{-15-126\nu-36\nu^2+56\nu^3}{1+2\nu}\Big)
\\[2mm]
&\displaystyle+\frac{i}{8\sin\pi\nu}\nu(-3+4\nu^2)\Big(e^{i\pi\nu}
\,\Phi(-1,1,\tfrac{3}{2}+\nu)-e^{-i\pi\nu}\Phi(-1,1,-\tfrac{1}{2}-\nu)\Big)\\[2mm]
&\displaystyle+\frac{i}{256\sin\pi\nu}
(4\nu^2-1)(4\nu^2-9)\Big(e^{i\pi\nu}\Phi(-1,2,\tfrac{5}{2}+\nu)-e^{-i\pi\nu}\Phi(-1,2,\tfrac{1}{2}-\nu)\Big)\quad\text{for} \,\,\,\,\text{pure-imaginary} \,\,\,\,\nu\,,
\end{array}\right.\\
\nonumber
\mathcal{I}_{33}&=\frac{\pi^2|\nu+\frac{1}{2}|^2|\nu-\frac{1}{2}|^2}{16 \cos^2\pi\nu}
-\frac{1}{16} + 
 \frac{i}{128\sin\pi\nu} \Big(
e^{i\pi\nu}(2\nu-1)(2\nu-3)-e^{-i\pi\nu}(2\nu+1)(2\nu+3)\Big)
\\
&\quad-
 \frac{i}{256\sin\pi\nu} 
(2\nu+1)^2(2\nu-1)^2\Big(e^{i\pi\nu}\Phi(-1,2,\tfrac{3}{2}+ \nu)-e^{-i\pi\nu}\Phi(-1,2,\tfrac{3}{2}-\nu)\Big)\,,\\[2mm]
\nonumber
\mathcal{I}_{12}&=-\frac{\pi^2(\nu-\frac{1}{2})(\nu-\frac{3}{2})}{8\cos^2\pi\nu}
-\frac{i}{16\sin\pi\nu} \Big(
e^{i\pi\nu}(2\nu-1)+e^{-i\pi\nu}(2\nu+1)\Big)\\
\nonumber
&
\quad+\frac{i}{16\sin\pi\nu}(-3+4\nu^2) \Big(
e^{i\pi\nu}\Phi(-1,1,\tfrac{1}{2}+\nu)-e^{-i\pi\nu}\Phi(-1,1,\tfrac{1}{2}-\nu)\Big)\\
&
\quad-\frac{i}{32\sin\pi\nu}(-3+2\nu)(-1+2\nu) \Big(
e^{i\pi\nu}\Phi(-1,2,\tfrac{1}{2}+\nu)
-e^{-i\pi\nu}\Phi(-1,2,\tfrac{1}{2}-\nu)\Big)\,,\\
\nonumber
\mathcal{I}_{21}&=-\frac{\pi^2(\nu^\ast-\frac{1}{2})(\nu^\ast-\frac{3}{2})}{8\cos^2\pi\nu}+\frac{i(-1+4\nu^2)}{32\sin\pi\nu^\ast}\Big(e^{i\pi\nu^\ast}\Phi(-1,1,-\tfrac{1}{2}+\nu^\ast)-e^{-i\pi\nu^\ast}\Phi(-1,1,\tfrac{3}{2}-\nu^\ast)\Big)\\
\nonumber
&\quad-\frac{i(4\nu^2-5)}{32\sin\pi\nu^\ast}
\Big(e^{i\pi\nu^\ast}\Phi(-1,1,\tfrac{1}{2}+\nu^\ast)-e^{-i\pi\nu^\ast}\Phi(-1,1,\tfrac{5}{2}-\nu^\ast)\Big)\\
&\quad-\frac{i(-3+2\nu^\ast)(-1+2\nu^\ast)}{32\sin\pi\nu^\ast}\Big(e^{i\pi\nu^\ast}\Phi(-1,2,\tfrac{1}{2}+\nu^\ast)-e^{-i\pi\nu^\ast}\Phi(-1,2,\tfrac{5}{2}-\nu^\ast)\Big)\,,
\\[2mm]
\nonumber
\mathcal{I}_{13}&=\frac{\pi^2(\nu+\frac{1}{2})(\nu-\frac{1}{2})}{8 \cos^2\pi\nu}
+\frac{i}{16\sin\pi\nu}\Big( 
e^{i\pi\nu}(2\nu-1)+e^{-i\pi\nu}(2\nu+1)\Big)\\
\nonumber
&\quad-\frac{i(1+4\nu^2)}{16\sin\pi\nu}\Big( 
e^{i\pi\nu}\Phi(-1,1,\tfrac{1}{2}+\nu)-e^{-i\pi\nu}\Phi(-1,1,\tfrac{1}{2}-\nu)\Big)\\
&\quad+\frac{i(-1+4\nu^2)}{32\sin\pi\nu}\Big( 
e^{i\pi\nu}
\Phi(-1,2,\tfrac{1}{2}+\nu)-e^{-i\pi\nu}\Phi(-1,2,\tfrac{1}{2}-\nu)\Big)\,,\\[2mm]
\nonumber
\mathcal{I}_{31}&=\frac{\pi^2(\nu^\ast+\frac{1}{2})(\nu^\ast-\frac{1}{2})}{8 \cos^2\pi\nu}
+\frac{i(-1+4\nu^2)}{32\sin\pi\nu}\Big(e^{i\pi\nu}\Phi(-1,1,\tfrac{1}{2}+\nu)-e^{-i\pi\nu}\Phi(-1,1,\tfrac{1}{2}-\nu)\Big)\\
\nonumber
&\quad-\frac{i(3+4\nu^2)}{32\sin\pi\nu}\Big(e^{i\pi\nu}\Phi(-1,1,\tfrac{3}{2}+\nu)-e^{-i\pi\nu}\Phi(-1,1,\tfrac{3}{2}-\nu)\Big)\\
&\quad-\frac{i(-1+4\nu^2)}{32\sin\pi\nu}\Big(e^{i\pi\nu}\Phi(-1,2,\tfrac{3}{2}+\nu)-e^{-i\pi\nu}\Phi(-1,2,\tfrac{3}{2}-\nu)\Big)\,,\\[2mm]
\nonumber
\mathcal{I}_{23}&=-\frac{\pi^2|\nu-\frac{1}{2}|^2(\nu+\frac{1}{2})(\nu^\ast-\frac{3}{2})}{16 \cos^2\pi\nu}
+\frac{1}{16}-\frac{i}{128\sin\pi\nu^\ast}\Big(e^{i\pi\nu^\ast}(-11+12\nu^2)+e^{-i\pi\nu^\ast}\frac{-1 + 2 \nu^\ast}{-3+ 2\nu^\ast}(1-20\nu^\ast+20\nu^2)\Big)\,,\\
\nonumber
&\quad+\frac{i(-1+2\nu^\ast)(-1-4\nu^\ast+4\nu^2)}{32\sin\pi\nu^\ast}\Big(e^{i\pi\nu^\ast}\Phi(-1,1,-\tfrac{1}{2}+\nu^\ast)-e^{-i\pi\nu^\ast}\Phi(-1,1,\tfrac{3}{2}-\nu^\ast)\Big)\\
&\quad-\frac{i(-1+2\nu^\ast)^2(-3+2\nu^\ast)(1+2\nu^\ast)}{256\sin\pi\nu^\ast}\Big(e^{i\pi\nu^\ast}\Phi(-1,2,\tfrac{1}{2}+\nu^\ast)-e^{-i\pi\nu^\ast}\Phi(-1,2,\tfrac{5}{2}-\nu^\ast)\Big)\,,\\[2mm]
\nonumber
\mathcal{I}_{32}&=-\frac{\pi^2|\nu-\frac{1}{2}|^2(\nu-\frac{3}{2})(\nu^\ast+\frac{1}{2})}{16 \cos^2\pi\nu}
-\frac{1}{16}-\frac{i}{128\sin\pi\nu}
\Big(e^{i\pi\nu}\frac{-1+2\nu}{1+2\nu}(1-20\nu+20\nu^2)
+e^{-i\pi\nu}(1-24\nu+12\nu^2)\Big)\\
\nonumber
&+\frac{i(-1+2\nu)(-1-4\nu+4\nu^2)}{32\sin\pi\nu}\Big(e^{i\pi\nu}\Phi(-1,1,\tfrac{1}{2}+\nu)
-e^{-i\pi\nu}\Phi(-1,1,\tfrac{1}{2}-\nu)\Big)\\
&+\frac{i(-3+2\nu)(-1+2\nu)^2(1+2\nu)}{256\sin\pi\nu}\Big(e^{i\pi\nu}\Phi(-1,2,\tfrac{3}{2}+\nu)-e^{-i\pi\nu}\Phi(-1,2,\tfrac{3}{2}-\nu)\Big)\,.
\end{align}}
Here we used
$\displaystyle\sum_{m=0}^\infty(-1)^m=\frac{1}{1+1}=\frac{1}{2}$.
As we displayed in figure~\ref{fig:PS_m},
the analytic results obtained in this appendix
and the numerical results for $r_s=1$
well coincide with each other.

\subsection{Asymptotic behavior of $\mathcal{D}(\ell,\nu,x)$}
\label{app_D}
In this subsection we derive the asymptotic behavior (\ref{D_inf})
in the limit $x\to\infty$
of the function $\mathcal{D}(\ell,\nu,x)$:
\begin{align}
\nonumber
\mathcal{D}(\ell,\nu,x)&=
\frac{2^\nu x^{\frac{1}{2}+\ell-\nu}\Gamma(\nu)}{i\pi(\frac{1}{2}+\ell-\nu)}
{}_2F_2\Big(\frac{1}{2}-\nu,\frac{1}{2}+\ell-\nu;\frac{3}{2}+\ell-\nu,1-2\nu; 2ix\Big)\\
&\quad+e^{-i\pi\nu}\frac{2^{-\nu} x^{\frac{1}{2}+\ell+\nu}\Gamma(-\nu)}{i\pi(\frac{1}{2}+\ell+\nu)}
{}_2F_2\Big(\frac{1}{2}+\nu,\frac{1}{2}+\ell+\nu;\frac{3}{2}+\ell+\nu,1+2\nu; 2ix\Big)\,.
\end{align}
We use the following asymptotic expansion
of hypergeometric functions:
\begin{align}
\nonumber
{}_2F_2(a_1,a_2;b_1,b_2;z)&=\frac{\Gamma(b_1)\Gamma(b_2)}{\Gamma(a_1)\Gamma(a_2)}e^zz^{a_1+a_2-b_1-b_2}\sum_{k=0}^\infty c_kz^{-k}\\
\nonumber
&\quad+\frac{\Gamma(b_1)\Gamma(b_2)\Gamma(a_2-a_1)}{\Gamma(a_2)\Gamma(b_1-a_1)\Gamma(b_2-a_1)}(-z)^{-a_1}{}_3F_1(a_1,a_1-b_1+1,a_1-b_2+1;a_1-a_2+1;-1/z)\\
&\quad+\frac{\Gamma(b_1)\Gamma(b_2)\Gamma(a_1-a_2)}{\Gamma(a_1)\Gamma(b_1-a_2)\Gamma(b_2-a_2)}(-z)^{-a_2}{}_3F_1(a_2,a_2-b_1+1,a_2-b_2+1;a_2-a_1+1;-1/z)\,,
\end{align}
where $c_k$'s are numerical factors independent of $z$.
The first term gives an oscillating term $\sim e^{2ix}$,
which can be dropped by an $i\epsilon$-prescription.
Then, let us consider the contribution of the last two terms.
They are respectively in the form
\begin{align}
\nonumber
&\frac{1}{i\sqrt{2\pi} \ell}\frac{1}{\sin \pi\nu} x^{\ell}
e^{-\frac{i}{2}\pi(\nu-1/2)}{}_3F_1(1/2+\nu,1/2-\nu,-\ell;1-\ell; i/(2x))\\
\label{second_term}
&+\frac{1}{i\sqrt{2\pi}}\frac{1}{\sin\pi\nu}\frac{\Gamma(1/2+\ell-\nu)}{\Gamma(1/2-\ell-\nu)}\Gamma(-\ell)(- 2i)^{-\ell}e^{- \frac{i}{2}\pi(\nu-1/2)}{}_3F_1(1/2+\ell+\nu,1/2+\ell-\nu,0;1+\ell;i/(2x))\,,
\end{align}
and
\begin{align}
\nonumber
&-\frac{1}{i\sqrt{2\pi} \ell}\frac{1}{\sin \pi\nu} x^{\ell}
e^{-\frac{i}{2}\pi(\nu-1/2)}{}_3F_1(1/2+\nu,1/2-\nu,-\ell;1-\ell; i/(2x))\\
\label{third_term}
&-\frac{1}{i\sqrt{2\pi}}\frac{1}{\sin\pi\nu}\frac{\Gamma(1/2+\ell+\nu)}{\Gamma(1/2-\ell+\nu)}\Gamma(-\ell)(- 2i)^{-\ell}e^{- \frac{i}{2}\pi(\nu-1/2)}{}_3F_1(1/2+\ell+\nu,1/2+\ell-\nu,0;1+\ell;i/(2x))\,,
\end{align}
where note that hypergeometric functions ${}_pF_q(a_1,\ldots ,a_p;b_1,\ldots,b_q;z)$ are symmetric under the permutations of $a_i$'s and those of $b_i$'s, respectively.
The first term in (\ref{second_term}) and that in (\ref{third_term})
cancel out and we obtain
\begin{align}
\nonumber
\mathcal{D}(\ell,\nu,x)=&\frac{1}{i\sqrt{2\pi}}\frac{1}{\sin\pi\nu}\frac{\Gamma(1/2+\ell-\nu)}{\Gamma(1/2-\ell-\nu)}\Gamma(-\ell)(- 2i)^{-\ell}e^{- \frac{i}{2}\pi(\nu-1/2)}{}_3F_1(1/2+\ell+\nu,1/2+\ell-\nu,0;1+\ell;i/(2x))\\
&-\frac{1}{i\sqrt{2\pi}}\frac{1}{\sin\pi\nu}\frac{\Gamma(1/2+\ell+\nu)}{\Gamma(1/2-\ell+\nu)}\Gamma(-\ell)(- 2i)^{-\ell}e^{- \frac{i}{2}\pi(\nu-1/2)}{}_3F_1(1/2+\ell+\nu,1/2+\ell-\nu,0;1+\ell;i/(2x))\,,
\end{align}
where note that we did not use any approximation so far.
Finally, taking the limit $x\to\infty$,
we conclude that
\begin{align}
\label{D_asy_1}
\mathcal{D}(\ell,\nu,\infty)=\frac{1}{i\sqrt{2\pi}}\frac{1}{\sin\pi\nu}
\Bigg(\frac{\Gamma(1/2+\ell-\nu)}{\Gamma(1/2-\ell-\nu)}-\frac{\Gamma(1/2+\ell+\nu)}{\Gamma(1/2-\ell+\nu)}\Bigg)
\Gamma(-\ell)(- 2i)^{-\ell}e^{- \frac{i}{2}\pi(\nu-1/2)}\,,
\end{align}
where we used ${}_3F_1(a_1,a_2,a_3;b;0)=1$.
Using the identity
\begin{align}
\Gamma(z)\Gamma(1-z)=\frac{\pi}{\sin \pi z}\,,
\end{align}
we can also rewrite (\ref{D_asy_1}) as follows:
\begin{align}
\mathcal{D}(\ell,\nu,\infty)&=\frac{\sqrt{\pi}(1-i)e^{-\frac{i}{2}\pi\nu}}{\cos \pi\nu}(-2i)^{-\ell}\frac{1}{\Gamma(\ell+1)}\frac{\Gamma(\frac{1}{2}+\ell+\nu)\Gamma(\frac{1}{2}+\ell-\nu)}{\Gamma(\frac{1}{2}+\nu)\Gamma(\frac{1}{2}-\nu)}\,,
\end{align}
which reproduces the result in \cite{Chen:2012ge} for $\ell=0$.



\end{document}